\documentclass[a4paper,11pt]{article}
\pdfoutput=1

\usepackage[utf8]{inputenc}
\usepackage[margin=1in]{geometry}
\usepackage[T1]{fontenc}
\usepackage{booktabs,multirow}
\usepackage{bbold}
\usepackage{subcaption}
\usepackage{amsmath}
\usepackage{mathtools,amssymb,braket,dsfont}
\usepackage{url}
\usepackage{graphicx}
\usepackage{multicol}
\usepackage{multirow}
\usepackage{float}
\usepackage[dvipsnames]{xcolor}
\usepackage{slashed}
\usepackage[final]{hyperref}
\hypersetup{
	colorlinks=true,
	linkcolor=orange,
	citecolor=blue,
	filecolor=magenta,
	urlcolor=blue
}
\usepackage[numbers, sort&compress]{natbib}
\usepackage[utf8]{inputenc}
\usepackage[T1]{fontenc}
\usepackage{comment}
\newcommand{\beq}{\begin{equation}}
\newcommand{\eeq}{\end{equation}}
\newcommand{\bea}{\begin{eqnarray}}
\newcommand{\eea}{\end{eqnarray}}

\newcommand{\mgut}{M_\textsc{gut}}
\newcommand{\alphaU}{\alpha_\textsc{gut}}

\newcommand{\fref}[1]{Figure~\ref{#1}} 
\newcommand{\eref}[1]{Eq.\,(\ref{#1})}

\newcommand{\aref}[1]{Appendix~\ref{#1}}
\newcommand{\sref}[1]{Section~\ref{#1}}
\newcommand{\tref}[1]{Table~\ref{#1}}
\begin{document}

\mbox{}\vspace{1cm}

%=================
\begin{center}
	%{\Large \bf \boldmath Lepton Flavour Violation in minimal\\[2mm] Grand Unification from Type II seesaw} \\[1cm]
	 %{\LARGE \bf \boldmath Charged Lepton Flavour Violation\\[2mm] in minimal grand-unified type II seesaw models} \\[1cm]
	 	{\LARGE \bf \boldmath Lepton Flavour Violation in minimal\\[2mm] grand-unified type II seesaw models} \\[1cm]
	{\Large 
	Lorenzo Calibbi$\,^\dag$\footnote{calibbi@nankai.edu.cn} and
	Xiyuan Gao$\,^\star$\footnote{Xiyuan.Gao@campus.lmu.de} 
	}\\[20pt]
 	{%\small 
 	$^\dag$ School of Physics, Nankai University, Tianjin 300071, China  \\[5pt]}
 	{%\small 
 	 
 	$^\star$ Arnold-Sommerfeld-Center, Ludwig-Maximilians-Universität,\\ München 80333, Germany \\[5pt]}
 	 
% 	$^\star$ Fakultät für Physik, Ludwig-Maximilians-Universität München,\\
% 	D-80333 München, Germany  \\[5pt]}
% 	
\end{center}
\vspace*{1cm}
%=================

\begin{abstract}
\noindent
We revisit minimal non-supersymmetric models of $SU(5)$ Grand Unification with the type~II seesaw mechanism as the origin of neutrino masses. Imposing the requirement of gauge coupling unification and the proton lifetime bounds, we perform a Bayesian fit and obtain robust quantitative information on the mass scales of the beyond the Standard Model particles. We then study lepton-flavour-violating (LFV) processes induced by the type~II scalar triplet and its $SU(5)$ partners, showing that the interplay of upcoming searches for different LFV observables can provide additional information on the masses of the new particles, as well as non-trivial constraints on neutrino parameters.
\end{abstract}
\vspace*{0.5cm}

%=================

\thispagestyle{empty}

\newpage

\setcounter{tocdepth}{1}
\tableofcontents

\section{Introduction}
\setcounter{footnote}{0}

Grand Unification is a powerful guiding principle towards unveiling new physics beyond the Standard Model (SM). 
The coupling constants of the three SM interactions exhibit the tendency to unify to a common value at a very high energy scale $\mgut$. This has been long regarded as a hint for a Grand Unified Theory (GUT) where the SM gauge group $SU(3)_c\times SU(2)_L\times U(1)_Y$ is embedded in a simple gauge group, such as $SU(5)$ or $SO(10)$~\cite{Georgi:1974sy,Fritzsch:1974nn}. This paradigm is very appealing for a number of theoretical and phenomenological reasons\,---\,for a recent review, see~\cite{Croon:2019kpe}. In particular, GUTs give a rationale behind the otherwise unexplained quantum numbers of the SM fermions, thus accounting for the quantisation of the electric charge and the exact cancellation of gauge anomalies within each single generation of fermions. Furthermore, since quarks and leptons are embedded in common irreducible representations of the GUT group, interactions mediated by GUT gauge bosons unavoidably violate baryon and lepton number, making GUT models in principle testable by searches for proton decay.

For what concerns the following discussion, even more important is the observation that the contributions to the running of the gauge couplings due to the SM field content alone can not achieve a successful unification. Hence GUTs provide a strong motivation for the presence of new fields at intermediate or low-energy scales that can prompt gauge coupling unification. Other open problems of the SM, in particular the origin of neutrino masses, nicely lead to the same conclusion. It is therefore very tempting to look for the fields able to account for gauge coupling unification among those responsible for neutrino masses. However, that does not seem straightforward within the context of the simplest extensions of the SM addressing neutrino masses\,---\,Dirac neutrinos or Majorana neutrinos from type I seesaw\,---\,as they only involve singlet representations of the SM gauge group (the right-handed neutrinos) that (i)~do not affect the running of the gauge couplings themselves, and (ii)~are naturally embedded in GUT representations that do not comprise extra fields which could facilitate unification (as they are singlets within $SU(5)$ too, while they nicely fit the $\bf 16$ spinorial representation of $SO(10)$ together with the other SM fermions).
Thus, the programme of achieving a minimal connection between neutrino mass models and unification should rather focus on the other two types of seesaw mechanism or on radiative neutrino mass models.\footnote{For a comprehensive discussion of the low-energy phenomenology of the former, see~\cite{abada2007low}, while for a review on the latter we refer to~\cite{Cai:2017jrq}.} 

In this paper, we focus on what is perhaps the simplest possibility, type II seesaw~\cite{Magg:1980ut,Lazarides:1980nt,Mohapatra:1980yp,Schechter:1980gr}, that is, we are going to introduce a single scalar $SU(2)_L$ triplet and its $SU(5)$ partners contained in the $\bf 15$ representation. Requiring a successful gauge coupling unification and enforcing the proton decay bound set non-trivial constraints on the masses of these new particles, leading to interesting and potentially testable phenomenological consequences. This has been extensively studied in the literature~\cite{Dorsner:2005fq,Dorsner:2005ii,Dorsner:2006dj,Dorsner:2006hw,Dorsner:2007fy,FileviezPerez:2008dw,Antusch:2022afk}, with a particular focus on the possibility that some states are light enough to be within the reach of high-energy colliders, such as the Large Hadron Collider~(LHC).\footnote{For analogous studies within the context of type III seesaw, see~\cite{Bajc:2006ia,Dorsner:2006fx,Bajc:2007zf,FileviezPerez:2007bcw,Arhrib:2009mz,perez2018seesaw}, while a general discussion on gauge coupling unification due to intermediate-scale scalar fields can be found in Ref.~\cite{Olivas:2021nft}.} 

In the following, we revisit several variations of the $SU(5)$ embedding of type II seesaw (distinguished by the way employed to fix the ``wrong'' fermion mass relations predicted by the minimal Georgi-Glashow $SU(5)$ model~\cite{Georgi:1974sy}, see \sref{sec:models}) and extend the existing literature in multiple directions. We first perform a Bayesian fit to the gauge coupling unification requirement and the proton lifetime constraint, in order to obtain reliable quantitative information about the viable spectrum of the theory (\sref{sec:fit}) and compare it with direct searches for new physics at the LHC. 
In \sref{sec:lfv}, we move to what is the main focus of the paper: the study of charged lepton-flavour-violating (LFV) decays that are  induced by the fields contained in the $\bf 15$, which are unavoidable since the couplings of these fields to the SM fermions need to account for the observed neutrino masses and mixing and must thus be flavour-changing (and are to large extent known). Searches for LFV decays are among the most sensitive probes of new physics coupling to SM leptons. In particular, the ongoing experimental programme is capable to reach scales exceeding $10^7-10^8$~GeV~\cite{Calibbi:2017uvl}. We are going to study the potential of future experiments of testing type II seesaw GUT models and highlight how the interplay of different LFV observables can provide information on the masses of the new particles, as well as complementary constraints on the neutrino parameters that have not been measured yet.

%=================

\section{Minimal models of Grand Unification with type II seesaw}
\label{sec:models}
We start from the minimal non-supersymmetric $SU(5)$ Georgi-Glashow model~\cite{Georgi:1974sy}, with the SM fermions organised within the first two lowest-dimension $SU(5)$ representations ($\bf \overline{5}$ and $\bf 10$):
\begin{align}
\label{SMfermions}
    \psi_{\overline{\textbf{5}}}~= ~&(D_R)^c\,(\mathbf{\bar 3},{\bf 1},1/3) \oplus L_L \,({\bf 1},{\bf 2},-1/2)\,, \\
    \psi_{\textbf{10}}~ = ~&Q_L\,({\bf 3},{\bf 1},1/6) \oplus
    (U_R)^c \,(\mathbf{\bar 3}, {\bf 1}, -2/3) \oplus (E_R)^c\, ({\bf 1},{\bf 1},1)\,,
\end{align}
where the $SU(3)_c\times SU(2)_L\times U(1)_Y$ quantum numbers are shown in parenthesis. 

The scalar sector consists of the real field $\phi_\textbf{24}$ in a $\bf 24$-dimensional representation and the $\bf 5$-dimensional  $\phi_\textbf{5}$ (the latter one containing the SM Higgs doublet), whose vevs cause the two-step spontaneous breaking $SU(5) \,\xrightarrow{v_{24}} \,SU(3)_c\times SU(2)_L \times U(1)_Y \,\xrightarrow{v_{5}} \,SU(3)_c \times U(1)_{em}$. The SM gauge bosons are also contained in an adjoint representation of $SU(5)$, together with new ones (the vector leptoquarks $(\mathbf{3},{\bf 2},5/6)$ and $(\mathbf{\bar 3},{\bf 2},-5/6)$, typically denoted as $X^\mu$,~$Y^\mu$) that convert quarks into leptons, thus inducing proton decay. 
The mass of these latter fields is proportional to the GUT-breaking vev $v_{24}$ and, in the following, we are going to identify it 
with the unification scale $\mgut$ where the three gauge couplings meet (and employ such an assumption to assess the impact of the extra gauge bosons on the proton lifetime, see \sref{sec:fit}). In other words, we are assuming that $X^\mu$ and $Y^\mu$
do not contribute to the running of the gauge couplings below the unification scale. 
Similarly, we are also assuming that the mass of the colour triplet in $\phi_\textbf{5}$ is at the GUT scale or above. 
In fact, this field does not only give rise to additional, potentially dangerous, contributions to proton decay (that typically bound its mass to be $\gtrsim 10^{11}$~GeV) but also tends to spoil the successful unification of the gauge couplings~\cite{Dorsner:2005fq}.

On the other hand, we will allow (some of) the states belonging to the $SU(5)$-breaking Higgs field $\phi_\textbf{24}$ to have masses lighter than $\mgut$, so to trigger gauge coupling unification. Notice, however, that this can not be the case of the scalar states $(\mathbf{3},{\bf 2},5/6)$ and $(\mathbf{\bar 3},{\bf 2},-5/6)$ that are just the would-be Goldstone bosons from $SU(5)$ breaking and provide the longitudinal components of the GUT gauge bosons $X^\mu$, $Y^\mu$.

Within the minimal $SU(5)$ model, such as in the Standard Model, neutrinos are massless, which conflicts with the observation of neutrino oscillations. As anticipated in the introduction, we assume that neutrino masses arise from the seesaw mechanism of type II, which requires the introduction of a scalar triplet $\Delta$ with lepton-number-breaking interactions~\cite{Magg:1980ut,Lazarides:1980nt,Mohapatra:1980yp,Schechter:1980gr}.\footnote{Interestingly, type II seesaw can also successfully address leptogenesis and inflation~\cite{Barrie:2021mwi,Barrie:2022cub}.}
The simplest representation of $SU(5)$ where this field can fit in is the $\bf 15$ whose SM decomposition also contains a scalar leptoquark that, as customary, we denote as $\widetilde{R_2}$ (see e.g.~the review in Ref.~\cite{Dorsner:2016wpm}) and a scalar colour sextet $S$~\cite{Dorsner:2005fq}:
\begin{align}   
\phi_{\textbf{15}}& = \Delta\,({\bf 1},{\bf 3},1)\oplus \widetilde{R_2} \,({\bf 3},{\bf 2},1/6) \oplus S\,({\bf 6},{\bf 1},-2/3)\,.
\end{align}
The $SU(5)$ Lagrangian of the Yukawa sector thus reads:
\begin{align}
\label{eq:yuk}
-{\cal L}_{\text{Yukawa}} ~=~&
Y_u \epsilon_{ijklm}\overline{\psi_{\textbf{10}}^{ij}}(\psi_{\textbf{10}}^{kl})^c \phi_\textbf{5}^{m*}+
Y_{d\ell} \phi_\textbf{5}^{i} \overline{\psi_{\textbf{10}}^{ij}}(\psi_{\overline{\textbf{5}}}^{j})^c
+Y_{15} \overline{\psi_{\overline{\textbf{5}}}^i} \phi_{\textbf{15}}^{ij*} (\psi_{\overline{\textbf{5}}}^j)^c
+\text{h.c.}\,,\end{align}
where $i,j,k,l,m = 1-5$ are $SU(5)$ indices, $\epsilon_{ijklm}$ is a rank-5 totally antisymmetric tensor, $Y_u$, $Y_{d\ell}$, and $Y_{15}$ are $3\times3$ matrices of Yukawa couplings, whose flavour indices are not explicitly shown. 
The $Y_{15}$ terms give mass to neutrinos via type II seesaw in the usual way, see the discussion in~\sref{sec:lfv}.

As is well known, the second term in \eref{eq:yuk} predicts the following GUT-scale relations among lepton and down-type quark masses:
 \begin{align}
 \label{eq:relations}
     m_d = m_e\,,~ m_s = m_{\mu}\,,~m_b = m_{\tau}\,,\qquad [\textrm{minimal}~SU(5)]
 \end{align}
 which are notoriously at odds with the experimental measurements, even taking into account the renormalisation group running down to the electroweak scale~\cite{Huang:2020hdv,Antusch:2013jca}. Several ways to correct these relations have been proposed in the literature. In the following, we will review three popular choices and employ them to identify the possible minimal sets of fields that can provide phenomenologically viable fermion masses (including neutrino ones).
 
 %%%%%%%%% Model 1 %%%%%%%%%%
 
\paragraph{Model 1: non-renormalisable operators.}
The simplest way to fix the quark-lepton mass relations shown in \eref{eq:relations} is to add the following non-renormalisable operators~\cite{Ellis:1979fg,Dorsner:2005fq,Dorsner:2006hw}:

\begin{equation}
\begin{aligned}
\label{eq:NR}
-{\cal L}_{\text{Yukawa}} ~\supset~&
\frac{Y_u^\prime}{\Lambda}\epsilon_{ijklm}\overline{\psi_{\textbf{10}}^{ij}}(\psi_{\textbf{10}}^{kl})^c \phi_{\textbf{24}}^{mn}\phi_\textbf{5}^{n*}+
\frac{Y_u^{\prime\prime}}{\Lambda}\epsilon_{ijklm}\overline{\psi_{\textbf{10}}^{ij}}(\psi_{\textbf{10}}^{kn})^c \phi_{\textbf{24}}^{mn}\phi_\textbf{5}^{l*}\\
&+\frac{Y_{d\ell}^\prime}{\Lambda} \phi_\textbf{5}^{i} \overline{\psi_{\textbf{10}}^{ij}}\phi_{\textbf{24}}^{jk}(\psi_{\overline{\textbf{5}}}^{k})^c
+\frac{Y_{d\ell}^{\prime\prime}}{\Lambda} \phi_\textbf{5}^{i} \phi_{\textbf{24}}^{ij}\overline{\psi_{\textbf{10}}^{jk}}(\psi_{\overline{\textbf{5}}}^{k})^c
+\text{h.c.}\,,
\end{aligned}
\end{equation}
where $\Lambda \gg \mgut$, implying some new degrees of freedom beyond minimal $SU(5)$ (e.g.~at the Planck scale). 
From Eqs.~(\ref{eq:yuk},\,\ref{eq:NR}), one obtains the following Yukawa terms after $SU(5)$ breaking:

\begin{equation}
\begin{aligned}
-{\cal L}_{\text{Mass}}~\supset~&4\overline{Q_L}(Y_u+Y_u^T) \widetilde{H} U_R+\overline{Q_L}Y_{d\ell} H D_R+\overline{L_L}Y_{d\ell}^T H E_R\\
     &-\frac{6v_{24}}{\Lambda}\overline{Q_L}(Y'_u+Y_u'^T) \widetilde{H} U_R+\frac{v_{24}}{\Lambda}\overline{Q_L}Y_{d\ell}' H D_R-\frac{3v_{24}}{2\Lambda}\overline{L_L}Y_{d\ell}'^T H E_R\\
     &-\frac{v_{24}}{\Lambda}\overline{Q_L}(Y''_u-Y_u''^T) \widetilde{H} U_R-\frac{3v_{24}}{2\Lambda}\overline{Q_L}Y_{d\ell}'' H D_R-\frac{3v_{24}}{2\Lambda}\overline{L_L}Y_{d\ell}''^T H E_R
    +\text{h.c.} \,,
\end{aligned}
\end{equation}
which give for the fermion mass matrices
\begin{equation}
\begin{aligned}
\label{mass0}
&M_u=2\sqrt{2}v_5 \left(Y_u+Y_u^T-\frac{3v_{24}}{2\Lambda}Y_u^\prime-\frac{3v_{24}}{2\Lambda}Y_u^{\prime T}-\frac{v_{24}}{4\Lambda}Y_u^{\prime\prime}+\frac{v_{24}}{4\Lambda}Y_u^{\prime\prime T}\right),\\
&M_d=\frac{v_5}{\sqrt{2}} \left(Y_{d\ell}+\frac{v_{24}}{\Lambda}Y_{d\ell}^\prime-\frac{3 v_{24}}{2\Lambda}Y_{d\ell}^{\prime\prime}\right)\,,
\quad M_\ell=\frac{v_5}{\sqrt{2}} \left(Y_{d\ell}^T-\frac{3v_{24}}{2\Lambda}Y_{d\ell}^{\prime\,T}-\frac{3v_{24}}{2\Lambda}Y_{d\ell}^{\prime\prime\,T}\right)\,.    
\end{aligned}
\end{equation}
The presence of the additional contributions $\propto Y_{d\ell}^\prime,\,Y_{d\ell}^{\prime\prime}$ clearly breaks the minimal relations of \eref{eq:relations}, allowing to fit the observed fermion masses by suitably adjusting the entries of the matrices $Y_{d\ell}$, $Y_{d\ell}^\prime$ and $Y_{d\ell}^{\prime\prime}$. 

In summary, in the case of Model 1, the only states that, if lighter than $\mgut$, can possibly trigger gauge-coupling unification are those contained in $\phi_{\textbf{15}}$ and $\phi_{\textbf{24}}$~\cite{Dorsner:2005fq,Dorsner:2005ii}, which we display in the first block of \tref{tab:particles}.
The conditions required in order to have some of these fields much lighter than the GUT scale or, in other words, large mass splittings in the scalar sector (which will necessarily involve fine tunings), are discussed in Appendix~\ref{app:scalar-sector}.

\begin{table}[t!]
    \centering
    \renewcommand{\arraystretch}{1.2}
\begin{tabular}{|c|c| c c c| c c c |}
 \hline
\multicolumn{8}{|c|}{All models}\\ \hline 
 Field & $ SU(5)$ & $SU(3)_c$ & $SU(2)_L$ & $U(1)_Y $ & $b_3^I$ & $b_2^I$ &$ b_1^I$ \\ \hline 
  $\varrho_3$ & $\phi_\textbf{24} $& \bf 1 &  \bf 3 & 0 & 0 & 1/3 & 0\\
  $\varrho_8 $& $ \phi_\textbf{24}$ &  \bf 8 &  \bf 1 & 0 & 1/2 & 0 & 0\\
  \hline
  $\Delta$ & $\phi_{\textbf{15}}$ & \bf 1 & \bf 3 & 1 & 0 & 2/3 & 3/5\\
  $\widetilde{R_2}$ & $\phi_{\textbf{15}}$ & \bf 3 & \bf 2 & 1/6 & 1/3 & 1/2 & 1/30 \\
  $S$ & $\phi_{\textbf{15}}$ & \bf 6 & \bf 1 & -2/3 & 5/6 & 0 & 8/15\\ \hline\hline
  \multicolumn{8}{|c|}{Model 2}\\ \hline
  Field & $ SU(5)$ & $SU(3)_c$ & $SU(2)_L$ & $U(1)_Y $ & $b_3^I$ & $b_2^I$ &$ b_1^I$ \\ \hline 
  $\varphi_8$ & $\phi_\textbf{45}$ & \bf 8 &\bf 2 & 1/2 & 2 & 4/3 & 4/5 \\
  $  \varphi_{\overline{6}} $ & $\phi_\textbf{45}$ & $\mathbf{\bar 6}$ & \bf 1 & -1/3 & 5/6 & 0 & 2/15\\
  $\varphi_{3}^T$ & $\phi_\textbf{45}$ & \bf 3 &\bf  3 & -1/3 & 1/2 & 2 & 1/5\\
  $ \varphi_{3}^D$ & $\phi_\textbf{45}$ & $\mathbf{\bar 3}$ & \bf 2 & -7/6 & 1/3 & 1/2 & 49/30 \\
  $\varphi_{3}^S$ & $\phi_\textbf{45}$ & \bf 3 & \bf 1 & -1/3 & 1/6 & 0 & 1/15\\
  $ \varphi_{\overline{3}}^S$ & $\phi_\textbf{45}$ & $\mathbf{\bar 3}$ & \bf 1 & 4/3 & 1/6 & 0 & 16/15\\
  $H_2$ & $\phi_\textbf{45}$ & \bf 1 & \bf 2 & 1/2 & 0 & 1/6 & 1/10\\ \hline\hline
  \multicolumn{8}{|c|}{Model 3}\\ \hline
  Field & $ SU(5)$ & $SU(3)_c$ & $SU(2)_L$ & $U(1)_Y $ &$b_3^I$ & $b_2^I$ &$ b_1^I$ \\ \hline 
  $L_V$   & $ \psi^v_{\overline{\textbf{5}}}$ & \bf 1 & \bf 2 & -1/2 & 0 & 1/3 & 1/5\\
  $D^c_V$  & $\psi^v_{\overline{\textbf{5}}} $ & $\mathbf{\bar 3}$ & \bf 1 & 1/3 &  1/3 & 0 & 2/15\\
  $L_V^c$   & $ \psi^v_{{\textbf{5}}} $ & \bf 1 & \bf 2 & 1/2 & 0 & 1/3 & 1/5\\
  $D_V$  & $ \psi^v_{\textbf{5}} $ & \bf 3 & \bf 1 & -1/3 &  1/3 & 0 & 2/15\\\hline
  $Q_V$ & $\psi^v_{\textbf{10}} $ & \bf 3 & \bf 2 &  1/6 & 2/3 & 1 & 1/15\\
  $U^c_V$ & $\psi^v_{\textbf{10}} $ & $\mathbf{\bar 3}$  & \bf 1 & -2/3 & 1/3 & 0 & 8/15\\
  $E^c_V$ & $\psi^v_{\textbf{10}} $ & \bf 1 &\bf  1 & 1 & 0 & 0 & 2/5\\
  $Q^c_V$ & $\psi^v_{\overline{\textbf{10}} }$ & $\mathbf{\bar 3}$  & \bf 2 &  -1/6 & 2/3 & 1 & 1/15\\
  $U_V$ & $\psi^v_{\overline{\textbf{10}} }$ & \bf 3 & \bf 1 & 2/3 & 1/3 & 0 & 8/15\\
  $E_V$ & $ \psi^v_{\overline{\textbf{10}}} $ & \bf 1 & \bf 1 & -1 & 0 & 0 & 2/5\\
  \hline
 \end{tabular}
    \caption{New fields with mass possibly below $\mgut$, the corresponding group representations, and their contribution to the one-loop $\beta$ function coefficients of the SM gauge couplings.}
    \label{tab:particles}
\end{table}

%%%%%%%%% Model 2 %%%%%%%%%%

\paragraph{Model 2: scalar $\bf 45$.}
If one prefers to work within a renormalisable theory, the simplest choice is to add a $\bf 45$-dimensional scalar representation~\cite{Dorsner:2007fy}:
\begin{align}   
\phi_\textbf{45} =& \varphi_8({\bf 8},{\bf 2},1/2)\oplus \varphi_{\overline{6}} (\mathbf{\overline{6}},{\bf 1},-1/3)\oplus\varphi_{3}^T({\bf 3},{\bf 3},-1/3)\oplus \nonumber\\
& \varphi_{3}^D (\mathbf{\overline{3}}, {\bf 2}, -7/6) \oplus \varphi_{3}^S ({\bf 3},{\bf 1},-1/3) \oplus \varphi_{\overline{3}}^S (\mathbf{\overline{3}}, {\bf 1},4/3) \oplus H_2 ({\bf 1},{\bf 2},1/2)\,.
\end{align}
The additional terms in the Lagrangian of the Yukawa sector read:
\begin{equation}
\label{Yukawa}
    \begin{aligned}
-{\cal L}_{\text{Yukawa}} ~\supset ~ &
Y_u \epsilon_{ijklm}\overline{\psi_{\textbf{10}}^{ij}}(\psi_{\textbf{10}}^{kl})^c \phi_\textbf{5}^{m*}+
Y_{d\ell} \phi_\textbf{5}^{i} \overline{\psi_{\textbf{10}}^{ij}}(\psi_{\overline{\textbf{5}}}^{j})^c\\
&+Y_u^\prime \epsilon_{ijklm}\overline{\psi_{\textbf{10}}^{ij}}(\psi_{\textbf{10}}^{nk})^c \phi_\textbf{45}^{lmn*}+
Y_{d\ell}^\prime \phi_\textbf{45}^{ijk} \overline{\psi_{\textbf{10}}^{ij}}(\psi_{\overline{\textbf{5}}}^{k})^c
+\text{h.c.}\,.
    \end{aligned}
\end{equation}
Notice that $\phi_{\textbf{45}}$ is a rank 3 tensor, satisfying antisymmetric and traceless conditions:
$\phi_\textbf{45}^{ijk}=-\phi_\textbf{45}^{jik},$ $\sum_{j=1}^5\phi_\textbf{45}^{ijj}=0.$
So we can choose for its vev:
\begin{equation}
\langle\phi_\textbf{45}^{5ij}\rangle=-\langle\phi_\textbf{45}^{i5j}\rangle=\frac{1}{\sqrt{2}}v_{45} (4 \delta^{i4}\delta^{j4}-\delta^{ij}),\quad\quad (i,j = 1-4)\,,
\end{equation}
with other entries vanishing. The fermion mass terms then result:
\begin{equation}
\begin{aligned}
-{\cal L}_{\text{Mass}}~\supset~& 4\overline{Q_L}(Y_u+Y_u^T) \widetilde{H} U_R+\overline{Q_L}Y_{d\ell} H D_R+\overline{L_L}Y_{d\ell}^T H E_R\\
&-8\overline{Q_L}(Y_u^\prime-Y_u^{\prime\,T}) \widetilde{H_2} U_R-6\overline{Q_L}Y_{d\ell}^\prime H_2 D_R+2\overline{L_L}Y_{d\ell}^{\prime\,T} H_2 E_R
+\text{h.c.}\,,
\end{aligned}
\end{equation}
and one can get the following fermion mass matrices:
\begin{equation}
\label{Mass}
\begin{aligned}
&M_u=\frac{1}{\sqrt{2}} \left[4\left(Y_u+Y_u^T\right)v_5 -8 \left(Y_u^\prime-Y_u^{\prime\,T}\right)v_{45}^*\right]\,,\\
&M_d=\frac{1}{\sqrt{2}}\left(v_5 Y_{d\ell}+2v_{45}Y_{d\ell}^\prime\right)\,,
\quad M_\ell=\frac{1}{\sqrt{2}}\left(v_5Y_{d\ell}^T-6 v_{45}Y_{d\ell}^{\prime\,T}\right)\,.
\end{aligned}
\end{equation}

Again, it is apparent that the entries $M_u,M_d, M_{\ell}$ are all free parameters, such that the observed fermion masses and mixing can be easily fitted.\footnote{For a different approach, with the same field content as Model 2 (including the $\bf 45$) but employing non-renormalisable instead of renormalisable operators to correct \eref{eq:relations}, see Ref.~\cite{Antusch:2022afk}.} It is worth noting that at low energies this is just a two Higgs-doublet model~(2HDM) with $v_{45}$ of the order of the electroweak~(EW) scale, and $v_\text{EW}= \sqrt{v_5^2 + v_{45}^2} \approx 246$~GeV.

For this model, the extra fields possibly contributing to the running of the SM gauge couplings are those in $\phi_{\textbf{15}}$, $\phi_{\textbf{24}}$, and $\phi_{\textbf{45}}$, see \tref{tab:particles}. 
Details about masses and vevs of these scalar states can be found in Appendix~\ref{app:scalar-sector}.

%%%%%%%%% Model 3 %%%%%%%%%%

\paragraph{Model 3: vector-like fermions.}
The last possibility we consider is adding heavy fermions in vector-like representations of $SU(5)$ (and thus of the SM gauge group too)~\cite{Witten:1979nr,Dorsner:2014wva}, that is, the $\bf 5 \oplus \mathbf{\bar 5}$ representation: 
\begin{equation}
\begin{aligned}
\psi_{\overline{\textbf{5}}}^v&= D_V^c (\mathbf{\overline{3}},{\bf 1},1/3)\oplus L_V ({\bf 1}, {\bf 2},-1/2)\,,\\
\psi^v_{{\textbf{5}}} &= D_V (\mathbf{3},{\bf 1},-1/3)\oplus L^c_V ({\bf 1}, {\bf 2},1/2)\,,
\end{aligned}
\end{equation}
and/or $\bf 10 \oplus \mathbf{\overline{10}}$: 
\begin{equation}
\begin{aligned}
\psi_{\textbf{10}}^v&=Q_V ({\bf 3}, {\bf 2},1/6)\oplus U_V^c (\mathbf{\overline{3}},{\bf 1},-2/3)\oplus E_V^c ({\bf 1},{\bf 1},1)\,,\\
\psi_{\overline{\textbf{10}}}^v&=Q^c_V (\mathbf{\overline{3}}, {\bf 2},-1/6)\oplus U_V ({\bf 3},{\bf 1},2/3)\oplus E_V ({\bf 1},{\bf 1},-1)\,.
\end{aligned}
\end{equation}
These vector-like pairs of Weyl fermions combine into Dirac fermions that, with slight abuse of notation, we will also denote
as $\psi^v_{\overline{\textbf{5}}}$ and $\psi_{\textbf{10}}^v$.

The components of the above $SU(5)$ fields mix with SM leptons and quarks differently, thus correcting the mass relations in \eref{eq:relations}.
For example, with only one generation of vector-like fermions $\psi_{\overline{\textbf{5}}}^v$, the Lagrangian of the Yukawa sector becomes (in 4-component notation): 
\begin{equation}
\label{mass5}
\begin{aligned}
-{\cal L}_{\rm Yukawa} \supset &
\left(
  \begin{array}{cc}
    \overline{\psi_{\overline{\textbf{5}}}^\alpha} &
    \overline{\psi_{\overline{\textbf{5}}}^v} \\
  \end{array}
\right)
\left(
\begin{array}{cc}
  Y_{d\ell}^{\alpha\beta}\phi_\textbf{5}^* & M_{5}^\alpha+\lambda_{5}^\alpha\phi_\textbf{24} \\
  Y_{d\ell}^{\prime\,\beta} \phi_\textbf{5}^* & M_{5}^V+\lambda_{5}^V\phi_\textbf{24}
\end{array}
\right)
\left(
  \begin{array}{cc}
    (\psi_{\textbf{10}}^\beta)^c \\ \psi_{\overline{\textbf{5}}}^v  \\
  \end{array}
\right)+\text{h.c.}\\
\rightarrow & \left(
  \begin{array}{cc}
    \overline{(D^\alpha_R)^c} & \overline{D_V^c}\\
  \end{array}
\right)
\left(
\begin{array}{cc}
 \frac{1}{\sqrt{2}} v_5 Y_{d\ell}^{\alpha \beta}  & M_{5}^\alpha+\lambda_{5}^\alpha v_{24} \\
 \frac{1}{\sqrt{2}} v_5 Y_{d\ell}^{\prime\,\beta}   & M_{5}^V+ \lambda_{5}^V v_{24}
\end{array}
\right)
\left(
  \begin{array}{cc}
    (D^\beta_L)^c \\ D_V^c \\
  \end{array}
\right) + \\
& \left(
  \begin{array}{cc}
    \overline{E_L^\alpha} & \overline{E_V^\prime}\\
  \end{array}
\right)
\left(
\begin{array}{cc}
 \frac{1}{\sqrt{2}} v_5 Y_{d\ell}^{\alpha \beta}  & M_{5}^\alpha-\frac32\lambda_{5}^\alpha v_{24} \\
 \frac{1}{\sqrt{2}} v_5 Y_{d\ell}^{\prime \,\beta}   & M_{5}^V-\frac32 \lambda_{5}^V v_{24}
\end{array}
\right)
\left(
  \begin{array}{cc}
    E_R^\beta \\ E_V^\prime  \\
  \end{array}
\right),\\
\end{aligned}
\end{equation}
where  $\alpha$ and $\beta$ are flavour indices, $D_L$ and $E_L$ respectively denote the down-type quarks and charged leptons in the doublets $Q_L$ and $L_L$, and $E_V^\prime$ is the charged state in $L_V$.

As we can see, in addition to the standard interactions with SM fermions and Higgs fields, the vector-like fermion can also couple to chiral fermions (or itself) directly or via the $SU(5)$ adjoint scalar field $\phi_{\textbf{24}}$. So after spontaneous symmetry breaking, the two mass matrices for charged leptons and down-type quarks acquire six independent parameters: $(M_{5}^\alpha+\lambda_{5}^\alpha v_{24})/(M_{5}^V+ \lambda_{5}^V v_{24})$ and $(M_{5}^\alpha-\frac32\lambda_{5}^\alpha v_{24})/(M_{5}^V-\frac32 \lambda_{5}^V v_{24})$, the contribution from $\phi_\textbf{5}^*\overline{\psi_{\overline{\textbf{5}}}^v}(\psi_{\textbf{10}}^\beta)^c$ being negligible since $v_5\ll v_{24}, M_5$. For a detailed discussion, see Ref.~\cite{Babu:2012pb}. Therefore, one can correct the wrong quark-lepton mass relations with only one generation of vector-like fermions $\psi_{\overline{\textbf{5}}}^v$. 

The above result can be straightforwardly generalised to the $\psi_{\textbf{10}}^v$ case:.
\begin{equation}
\label{mass10}
\begin{aligned}
-{\cal L}_{\rm Yukawa} \supset &
\left(
  \begin{array}{cc}
    \overline{\psi_{\overline{\textbf{5}}}^\alpha} &
    \overline{\psi_{\textbf{10}}^v} \\
  \end{array}
\right)
\left(
\begin{array}{cc}
  Y_{d\ell}^{\alpha \beta}\phi_\textbf{5}^* & Y_{d\ell}^{\prime\,\alpha} \phi_\textbf{5}^*  \\
  M_{10}^\beta+\lambda_{10}^\beta\phi_\textbf{24} & M_{10}^V+\lambda_{10}^V\phi_\textbf{24}
\end{array}
\right)
\left(
  \begin{array}{cc}
    (\psi_{\textbf{10}}^\beta)^c \\ \psi_{\textbf{10}}^v  \\
  \end{array}
\right)+\text{h.c.} \\
\rightarrow & \left(
  \begin{array}{cc}
    \overline{(D^\alpha_R)^c} & \overline{D_V^\prime}\\
  \end{array}
\right)
\left(
\begin{array}{cc}
 \frac{1}{\sqrt{2}}v_5 Y_{d\ell}^{\alpha \beta} & 
 \frac{1}{\sqrt{2}}v_5 Y_{d\ell}^{\prime\,\alpha}   \\
  M_{10}^\beta+\frac14\lambda_{10}^\beta v_{24} & M_{10}^V+\frac14\lambda_{10}^Vv_{24}
\end{array}
\right)
\left(
  \begin{array}{cc}
    (D^\beta_L)^c \\ D_V^\prime  \\
  \end{array}
\right) + \\
& \left(
  \begin{array}{cc}
    \overline{E_L^\alpha} & \overline{E_V}\\
  \end{array}
\right)
\left(
\begin{array}{cc}
 \frac{1}{\sqrt{2}}v_5 Y_{d\ell}^{\alpha \beta} 
 & \frac{1}{\sqrt{2}}v_5 Y_{d\ell}^{\prime\,\alpha}   \\
  M_{10}^\beta+\frac32\lambda_{10}^b v_{24} & M_{10}^V+\frac32\lambda_{10}^V v_{24}
\end{array}
\right)
\left(
  \begin{array}{cc}
    E_R^\beta \\ E_V  \\
  \end{array}
\right),\\
\end{aligned}
\end{equation}
where $D_V^\prime$ is the $Q=-1/3$ state in $Q_V$.

In this scenario, the states belonging to the vector-like fermions $\psi_{\overline{\textbf{5}}}^v$ or $\psi_{\textbf{10}}^v$ would also contribute to the running of the gauge couplings, alongside the scalar fields in $\phi_{\bf 5}$ and $\phi_{\bf 15}$, see \tref{tab:particles}.

%==============================================

\section{Fit of the mass spectrum of minimal models}
\label{sec:fit}
In this section, we present the results of a Bayesian analysis aimed at constraining the mass spectrum of the models introduced above.
In principle, the physical masses of the new particles displayed in \tref{tab:particles} could range from $m_Z$ to $M_{\textsc{gut}}$ (or above).
However, the parameter space is tightly constrained because (i)~the three SM gauge coupling constants of a realistic GUT model must converge at a high-energy scale, and (ii)~such scale must be large enough not to cause unacceptably fast proton decay rates. 

%%%%%%%%%%%%%%

\paragraph{Gauge coupling unification.}  
Solving the renormalisation group equations (RGEs) of the SM gauge couplings (taking into account the effect of the new intermediate-scale fields), one can impose the unification of the three constants $\alpha_i \equiv g^2_i/4\pi$ to a common value $\alphaU$ at a scale $\mgut$. At one loop, this provides the three following equations~\cite{giveon19915}:
\begin{equation}
\label{eq:eff}
\alphaU^{-1}= \alpha_{i}^{-1}(m_Z) - \frac{b_i^{\rm eff}}{2 \pi} \ln\left(\frac{\mgut}{m_Z}\right),\quad b_i^{\rm eff} \equiv b^\textsc{sm}_{i}+ \sum_I b_i^I r_{I}\,,\quad
r_{I} \equiv \frac{\ln(\mgut/M_I)}{\ln(\mgut/m_Z)}\subset [0,1]\,,
\end{equation}
where $i=1,2,3$ labels the three gauge interactions, $b^\textsc{sm}_{i}$ are the one-loop $\beta$-function coefficients, $(b^\textsc{sm}_{3},b^\textsc{sm}_{2},b^\textsc{sm}_{1}) = (-7,-19/6,41/10)$, due to the SM field content, and the index $I$ runs over the new fields with mass $M_I < \mgut$, whose contributions to the $\beta$ functions are denoted as $b_i^I$. The latter quantities just depend on the quantum numbers of the fields under $SU(3)_c\times SU(2)_L\times U(1)_Y$, see e.g.~\cite{Machacek:1983tz}, and are listed in the last three columns of Table~\ref{tab:particles}. 

Eliminating $\alphaU$ and $\ln(\mgut/m_Z)$ in \eref{eq:eff}, one can get a constraint on the mass spectrum from gauge coupling unification, in terms of experimentally measured quantities~\cite{giveon19915}:
\begin{equation}
\label{Btest}
\frac{b_2^{\rm eff}-b_3^{\rm eff}}{b_1^{\rm eff}-b_2^{\rm eff}}= \frac{\alpha_2^{-1}(m_Z)-\alpha_3^{-1}(m_Z)}{\alpha_1^{-1}(m_Z)-\alpha_2^{-1}(m_Z)}
=\frac{5\sin^2\theta_w-5\alpha_{em}/\alpha_s}{3-8\sin^2\theta_w}
%\equiv B_\text{test} 
%=0.718 \pm 0.004\,,
=0.717 \pm 0.002\,,
\end{equation}
where $\alpha_{em}^{-1}=127.952\pm0.009$, $\alpha_s\equiv \alpha_3(m_Z) =0.1179\pm 0.0009$, $\sin^2\theta_w=0.23121\pm0.00004$ are, respectively, the electromagnetic coupling constant, the strong coupling constant, and the weak mixing angle at the electroweak scale $m_Z$~\cite{Zyla:2020zbs}, and the GUT normalisation $g_1 = \sqrt{5/3}\,g^\prime$ of the hypercharge coupling has been employed. 

For a given set of intermediate fields that satisfy \eref{Btest}, one can then employ the equations with $i=1,2$ in (\ref{eq:eff}) to obtain the following expression for the GUT scale:
\begin{equation}
\label{M_GUT}
\ln\left(\frac{\mgut}{m_Z}\right)= \frac{6\pi-16\pi \sin^2\theta_w}{5\alpha_{em}(b_1^{\rm eff}-b_2^{\rm eff})}\,. 
\end{equation}

Notice that Eqs.~(\ref{eq:eff}-\ref{M_GUT}) neglect the fact that the above-quoted values of $b^\textsc{sm}_i$ include the contributions of top quarks, hence it would be correct to consider the running above the top mass scale $m_t$, that is, to employ $\alpha^{-1}_i(m_t)$ in the formulae and substitute $m_t\to m_Z$ elsewhere. 
However, the numerical impact would be negligible: using the central values for $\alpha^{-1}_i(m_t)$ calculated in Ref.~\cite{Huang:2020hdv}, we find that the quantity in \eref{Btest} is shifted to $\approx 0.719$, well within the experimental uncertainty quoted above. Similarly, the effect of the substitution $m_t\to m_Z$ in the logarithms of \eref{eq:eff} is tiny. We expect a larger numerical deviation from the above unification requirement if two-loop RGEs are considered. Such an effect is typically of the same order of magnitude of unknown\,---\,in our context\,---\,threshold effects from mass splittings of the states at $M_I$ and $\mgut$ (see e.g.~the analytical discussion in Ref.~\cite{Alciati:2005ur}).  Since, for simplicity, we refrain from modelling the uncertainties due to unknown thresholds, we are going to neglect two-loop corrections as well.

%%%%%%%%%%%%%%

\paragraph{Proton lifetime.}
 As mentioned in the previous section, the extra $SU(5)$ gauge bosons $X_{\mu}$ and $Y_{\mu}$ can convert quarks and leptons into each other, and thus mediate proton decay. 
 
 \begin{table}[b!]
\renewcommand{\arraystretch}{1.3}
    \centering
    \begin{tabular}{c c c }
   \hline
 %   Mode    & 90\% CL limit (years) & Ref. \\
    Mode    & Limit (years) & Ref. \\
    \hline
    $p \rightarrow \pi^0 e^+$ & $ >2.4\times 10^{34}$ &  \cite{Super-Kamiokande:2020wjk} \\
    $p \rightarrow \pi^0 \mu^+$ & $ >1.6 \times 10^{34}$ &  \cite{Super-Kamiokande:2020wjk} \\ \hline
    $p \rightarrow K^0 e^+$ & $ >1.0\times 10^{33}$ &  \cite{Super-Kamiokande:2005lev} \\
    $p \rightarrow K^0 \mu^+$ & $ >3.6 \times 10^{33}$ &  \cite{Super-Kamiokande:2022egr} \\ \hline
    $p \rightarrow \pi^+ \overline{\nu}$ &  $> 3.9\times 10^{32}$ & \cite{Super-Kamiokande:2013rwg} \\
    $p \rightarrow K^+ \overline{\nu}$ &  $> 5.9\times 10^{33}$ & \cite{Super-Kamiokande:2014otb} \\
    \hline
    \end{tabular}
    \caption{90\% CL limits from proton decay searches on $\tau(p\to X)\equiv 1/\Gamma(p\to X)$.}
    \label{tab:pdecay}
\end{table}

The current best limits on the proton lifetime were set in 2020 by SuperKamiokande~(SK) searching for $p\to \pi^0 e^+$ and $p\to \pi^0 \mu^+$, see \tref{tab:pdecay}.
%$\tau(p \rightarrow \pi^0 e^+)\equiv {1}/{\Gamma (p \rightarrow \pi^0 e^+)}>2.4\times 10^{34}~\text{years}$,
%$\tau(p \rightarrow \pi^0 \mu^+)\equiv {1}/{\Gamma (p \rightarrow \pi^0 \mu^+)}>1.6\times 10^{34}~\text{years}$~\cite{Super-Kamiokande:2020wjk}.
The contribution of the $SU(5)$ gauge bosons to this kind of decay modes (and the analogous ones into neutral kaons) reads~\cite{Nath:2006ut,perez2018seesaw}
\begin{align}
\label{eq:p-to-e}
    \Gamma(p\rightarrow \pi^0 \ell_i^+)~=~& \frac{\pi\,m_p\, \alphaU^2}{2 \mgut^4} A^2 
  \bigg\{
   \left|(V_1)_{11} (V_3)_{1i} \langle \pi^0 | (ud)_R u_L |p\rangle \right|^2 \\
& +\left|\left[(V_1)_{11} (V_2)_{i1} + (V_1 V_\textsc{ckm}^*)_{11} (V_2V_\textsc{ckm}^T)_{i1} \right]
\langle \pi^0 | (ud)_L u_L |p\rangle \right|^2 \nonumber \bigg\}\,,\\
\label{eq:p-to-ke}
\Gamma(p\rightarrow K^0 \ell_i^+)~=~& \frac{\pi\,m_p\, \alphaU^2}{2 \mgut^4} \left(1-\frac{m_K^2}{m_p}\right)^2 A^2 
  \bigg\{
   \left|(V_1)_{11} (V_3)_{2i} \langle K^0 | (us)_R u_L |p\rangle \right|^2 \\
& +\left|\left[(V_1)_{11} (V_2)_{i2} + (V_1 V_\textsc{ckm}^*)_{12} (V_2V_\textsc{ckm}^T)_{i1} \right]
\langle K^0 | (us)_L u_L |p\rangle \right|^2 \nonumber  \bigg\}\,,
\end{align}
where we identified the mass of the mediators $X_{\mu}$ and $Y_{\mu}$ with $\mgut$ and  $A$ is a renormalisation factor accounting for the running of the baryon-number violating operators from the GUT scale to $m_p$ (cf.~\aref{app-pdecay} for details and for the numerical values of the hadronic matrix elements). Furthermore, $V_\textsc{ckm}$ is the CKM mixing matrix and the other matrices are defined in terms of the biunitary rotations that diagonalise the fermion masses ($V_f^\dag M_f V_f' = M_f^\textrm{diag}$) as follows
\begin{equation}
\label{NewMixing}
\begin{aligned}
V_1 \equiv V_u'^\dagger V_u^*\,,\quad V_2\equiv V_\ell'^\dagger V_d^*\,,\quad V_3\equiv V_\ell^\dagger V_d'^*\,.
\end{aligned}
\end{equation}

Within minimal $SU(5)$, all the above matrices equal $\mathbb{1}$ and the decay width in \eref{eq:p-to-e} only depends
on known CKM angles. This is not anymore the case in presence of the more general mass matrices considered in the previous section that can correctly account for the observed fermion mass relations. Thus, in the models we are considering, $p\rightarrow \pi^0 \ell_i^+$ depends on the unknown (and, within the SM, unobservable) right-handed rotations $V^\prime_f$ through the combinations in \eref{NewMixing}.
It is therefore possible that non-trivial (and somewhat tuned) flavour structures of $M_f$ conspire to suppress the $p$-decay rates in these channels~\cite{Dorsner:2004xa,Nath:2006ut}.

On the other hand, the decay modes involving neutrinos are subject to weaker constraints (cf.~\tref{tab:pdecay})
%($\tau(p \rightarrow K^+ \overline{\nu}) > 5.9\times 10^{33}~\text{years}$~\cite{Super-Kamiokande:2014otb},
%$\tau(p \rightarrow \pi^+ \overline{\nu}) > 3.9\times 10^{32}~\text{years}$~\cite{Super-Kamiokande:2013rwg}) 
but are theoretically more robust. In fact, it has been noted that summing over the (experimentally unobservable) anti-neutrino flavours 
makes the dependence on the PMNS mixing drop and leads to a much cleaner theoretical prediction for these channels than for $p\rightarrow \pi^0 \ell_i^+$~\cite{Dorsner:2004xa,perez2018seesaw}:
\begin{align}
\label{eq:p-to-pi}
\Gamma(p\rightarrow \pi^+ \overline{\nu})~=~& \frac{\pi\,m_p\, \alphaU^2}{2 \mgut^4} A^2 |(V_1 V_\textsc{ckm})_{11} \langle \pi^+ | (du)_R d_L |p\rangle|^2\,,\\
\Gamma(p\rightarrow K^+ \overline{\nu})~ =~ & \frac{\pi\,m_p \,\alphaU^2}{2 \mgut^4}\left(1-\frac{m_K^2}{m_p}\right)^2 
A^2 \nonumber \bigg\{
\left|(V_1 V_\textsc{ckm})_{11} \langle K^+ | (us)_R d_L |p\rangle \right|^2 \nonumber\\
&+ \left|(V_1 V_\textsc{ckm})_{12} \langle K^+ | (ud)_R s_L |p\rangle \right|^2  \bigg\}\,.
\label{eq:p-to-K}
\end{align}
As we can see, the only residual dependence on the fermion flavour structure is encoded in $V_1$, a matrix that equals the identity if the up-quark mass matrix $M_u$ is symmetric. In our models, this occurs if the contribution $\propto Y_u^{\prime\prime}$ is subdominant in
\eref{mass0} (Model 1), that $\propto Y_u^{\prime}$ is negligible in \eref{Mass} (Model 2), and only $\bf 5 + \mathbf{\bar 5}$ vector-like fermions are introduced (Model 3). 
Furthermore, even for a non-symmetric $M_u$, the results obtained setting $V_1 \to \mathbb{1}$ in Eqs.~(\ref{eq:p-to-pi},\,\ref{eq:p-to-K}) are still a very
good approximation if $V_u'$ has got a hierarchical structure akin to that observed in the left-handed sector.

Finally, let us notice that, besides the vector bosons $X_{\mu}$ and $Y_{\mu}$, scalar particles such as the colour triplet in $\phi_\mathbf{5}$ and $\varphi_{3}^T$, $\varphi_{3}^S$, $\varphi_{\overline{3}}^S$ (cf.~\tref{tab:particles}) also endanger proton stability.\footnote{Our scalar fields could induce both $B-L$ conserving and violating processes, see~\aref{app-pdecay} for details.} When considering minimal scenarios, it is therefore reasonable to set their masses at the GUT scale directly. However, when the relevant Yukawa couplings are small, these fields could also be lighter than $\mgut$ by several orders of magnitude. We will comment about their possible impact on our fit below.

%%%%%%%%%%%%%%

\paragraph{Fitting procedure.}
The mass spectrum is calculated by means of the following steps. Firstly, we sample uniformly the initial parameters in \eref{eq:eff}, $\{r_I \subset [0,1]\}$, and enforce the unification constraint of \eref{Btest}. Next, we calculate the resulting GUT scale according to \eref{M_GUT}, and use it in \eref{eq:eff} to obtain the masses of the new particles, $\{M_I\}$, and the unified coupling $\alphaU$. 
This information can be converted into a prediction for the proton decay rates, once additional assumptions on the flavour structure of
the mixing in \eref{NewMixing} are made (that we will discuss below, when presenting our results). Finally, applying the relevant SK bounds on proton decay reported in \tref{tab:pdecay}, we get probability distributions for the spectrum of the new particles and the proton lifetime.

We start considering the simplest models (that is, minimal in terms of field content) that are compatible with all phenomenological observations related to neutrino and fermion masses and proton stability, thus reducing the number of free parameters in our fit. First, let us notice that the particles whose contribution to the $U(1)_Y$ $\beta$-function coefficient is larger than the $SU(2)_L$ one should better not contribute much to the running of the gauge couplings, as their effect is to decrease $\mgut$ and thus endanger proton stability (according to \eref{M_GUT}, $(b_1^{\rm eff}-b_2^{\rm eff})$ should be as small as possible for the sake of a large $\mgut$). Therefore, we start setting the masses of such fields at $\mgut$. Similarly, we do not consider at first scalars that directly mediate proton decay, as we mentioned above. Under these assumptions, our parameter space is rather limited; we will discuss below how the fit is affected by enlarging it. All models have as free parameters the masses of the $SU(2)_L$ triplet $\varrho_3$ and the colour octet $\varrho_8$ from the GUT Higgs $\bf 24$, and those of the seesaw triplet $\Delta$ and the leptoquark $\widetilde{R_2}$ from the $\bf 15$, cf.~\tref{tab:particles}. In addition, Model 2 features the masses of the colour octet and the second Higgs doublet in the $\bf 45$, $\varphi_8$ and $H_2$, and Model 3 the vector-like fermions $L_V+ L^c_V$ and $Q_V+  Q^c_V$.

%%%%%%%%%%%%%%
%
\begin{figure}[t]
    \centering
    \includegraphics[width=1.\textwidth]{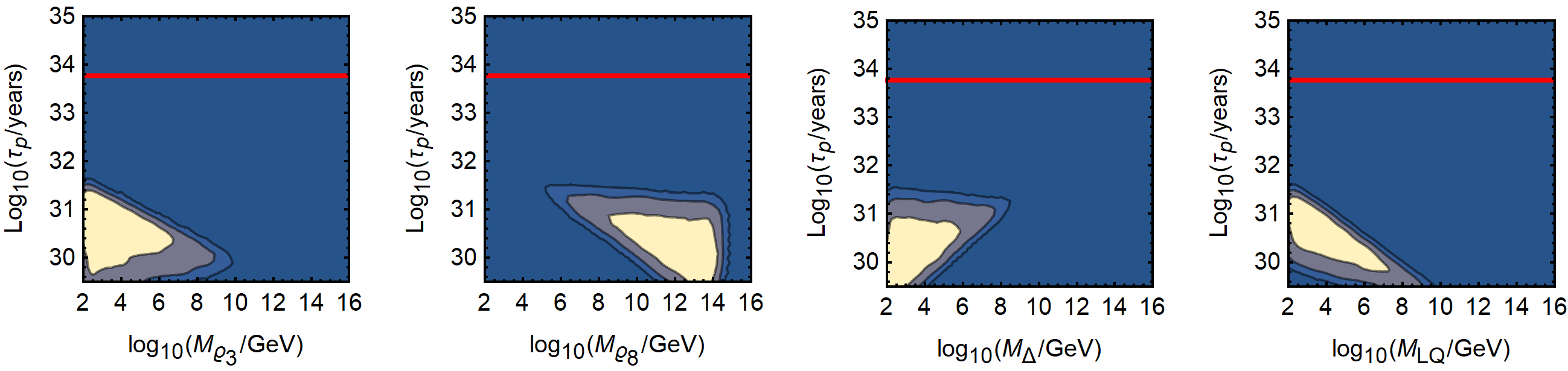}
    \caption{Result of the fit for Model 1 (non-renormalisable operators) shown on planes displaying the extra field masses and the proton lifetime from $p\rightarrow K^+ \overline{\nu}$ (the most constraining of the theoretically clean decay modes). The red line depicts the corresponding SK lower limit, $5.9\times 10^{33}~\text{years}$~\cite{Super-Kamiokande:2014otb}. Regions favoured by the fit at 1$\sigma$, 2$\sigma$ and $3\sigma$ are highlighted.}
    \label{Non-re}
\end{figure}

\subsection{Model 1}
\label{sec:model1}
As discussed above, the minimal setup of this model just comprises 4 parameters. In \fref{Non-re}, we show the result of the fit in terms of these parameters and the resulting proton lifetime. The latter was estimated based only on the theoretically clean mode $p\rightarrow K^+ \overline{\nu}$, assuming that $p\rightarrow \pi^0 \ell_i^+$ can be somewhat suppressed by the flavour structure of the fermion masses. For the calculation, we have taken for the mixing matrix $V_1 = \mathbb{1}$, cf.~\eref{NewMixing}, hence the plots illustrate to a very good approximation both the case of an (approximately) symmetric up-quark mass matrix, as well as a hierarchical structure of the right-handed mixing. For this fit, we did not impose the experimental proton decay limits. In fact, as we can see, the region favoured by the fit corresponds to a proton lifetime more than two orders of magnitude smaller than the present SK bound. Therefore, this setup is excluded, barring very fine-tuned flavour structures of that Yukawa matrices such that $p\rightarrow \pi^0 \ell_i^+$, $p\rightarrow \pi^+\overline{\nu}$, and $p\rightarrow K^+ \overline{\nu}$ be all simultaneously suppressed. 
The reason why this model is so strongly disfavoured is that there are too few parameters to achieve a high $\mgut$. Enlarging the parameter space by including more states from $\phi_{\bf 24}$ and $\phi_{\bf 15}$ with $M_I < \mgut$ would not improve the situation: as discussed above,
the presence of these other fields would, in fact, tend to further lower $\mgut$ and/or introduce new sources of $p$ decay.

\begin{figure}[t!]
\centering
\includegraphics[width=1.\textwidth]{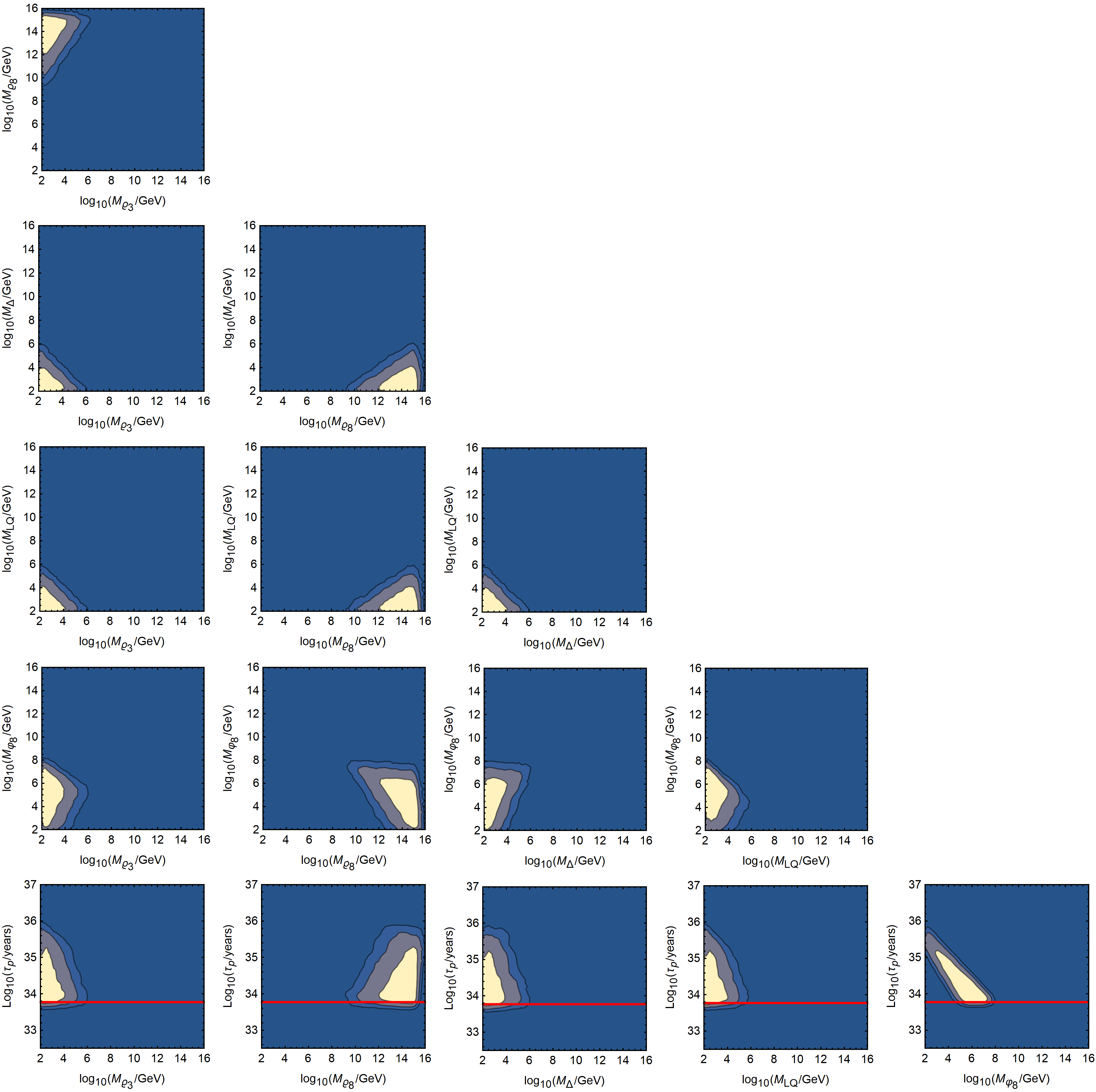} 
\caption{Model 2 (scalar $\bf 45$): result of the fit for the minimal  6-parameter setup. 
The proton lifetime $\tau_p$ vs.~the mass parameters and the two-dimensional correlation plots for the particle masses are shown. $\tau_p$
was calculated considering the clean $p\rightarrow K^++\overline{\nu}$ decay modes. Colours as in \fref{Non-re}.}
\label{fig:minimal-fit}
\end{figure}

\subsection{Model 2}
\label{sec:model2}
In Model 2, we have in addition $\varphi_8$ and $H_2$ (both from $\phi_{\bf 45}$) that, as argued above, can contribute to gauge coupling unification without endangering proton stability. We start considering only the effect of the colour octet and $SU(2)_L$ doublet $\varphi_8$\,---\,alongside the fields contained in $\phi_{\bf 24}$ and $\phi_{\bf 15}$ that we included in the fit of Model 1\,---\,while set the mass of the second Higgs doublet $H_2$ equal to $\mgut$. 
The outcome of this 5-parameter fit is shown in \fref{fig:minimal-fit}. As for Model 1, proton lifetime was calculated considering $p\rightarrow K^+ \overline{\nu}$ with $V_1 = \mathbb{1}$ and conservatively assuming that flavour mixing can suppress $p\rightarrow \pi^0 \ell_i^+$ to a sufficient extent. Here, in contrast to Model 1, we are imposing the SK bound as a constraint of the fit.
As we can see from the last row of \fref{fig:minimal-fit}, the effect of the colour octet $\varphi_8$ is to raise the GUT scale to such an extent that, at the 1$\sigma$\,(3$\sigma$) level, a proton lifetime up to about $10^{35\,(36)}$~years can be easily achieved. This requires the octet to live at an intermediate to low scale ($\lesssim 10^8$~GeV). In fact, the proton lifetime is anti-correlated to the octet mass (cf.~the bottom-right plot of the figure), as first observed in Ref.~\cite{Dorsner:2006dj}. 
The plots in \fref{fig:minimal-fit} also show that a good fit requires that the $Y=0$ triplet $\varrho_3$ from $\phi_{\bf 24}$ as well as the seesaw triplet $\Delta$ and the scalar leptoquark $\widetilde{R_2}$ from $\phi_{\bf 15}$ should all be rather light ($\lesssim 10$~TeV, at $1\sigma$).

This scenario could be regarded as a `minimal predictive grand-unified type II seesaw model'. Indeed, it is `minimal' and 'predictive' due to the following reasons: 
\begin{itemize}
    \item Only bosonic fields are added to the SM (and to minimal $SU(5)$), no additional (vector-like) fermions are required. 
    Furthermore, both the scalar $\bf 15$ and $\bf 45$ representations are contained in a single $SO(10)$ representation of dimension $\bf 126$.
    \item All of the 5 new particles considered in \fref{fig:minimal-fit} are necessary for a successful gauge coupling unification. As shown below, $\varphi_8$ can not be replaced with $H_2$ as the latter field does not raise $\mgut$ so much. In addition, even the `scalar gluon' $\varrho_8$, which does not contribute to the running of $\alpha_{1,2}$, is also crucial, because it can help balance $(b_2^{\rm eff}-b_3^{\rm eff})$ and $(b_1^{\rm eff}-b_2^{\rm eff})$ in~\eref{Btest}, when the latter quantity increases. 
    \item No fine tuning in the Yukawa sector is required and all the flavour mixing angles could have `natural' and generic values.
    \item As we have seen, light fields are predicted, in particular the weak triplet $\Delta$ and the leptoquark $\widetilde{R_2}$, which could therefore induce large LFV effects, as we are going to discuss in the next section.
    \item The $Y=0$ triplet $\varrho_3$  is also required to be light. Interestingly, this field can be responsible for the shift of the $W$ boson mass that the recent result of the CDF collaboration~\cite{CDF:2022hxs} seems to indicate: the anomaly can be accommodated with $M_{\varrho_3} \approx 10$~TeV if the triplet-Higgs doublet trilinear coupling (in our context $\phi_{\bf 5}\phi_{\bf 5}\phi_{\bf 24}$) is of order $M_{\varrho_3}$~\cite{DiLuzio:2022xns}.\footnote{For discussions of the CDF anomaly with a $Y=0$ triplet in the context of GUTs see \cite{Evans:2022dgq,Senjanovic:2022zwy}.}
    \item The anti-correlation between the mass of the octet $\varphi_8$ and the proton lifetime has important phenomenological implications: if proton decay will be further constrained by next-generation large-volume detectors, such as JUNO~\cite{JUNO:2015sjr}, DUNE~\cite{DUNE:2015lol}, Hyper-Kamiokande~\cite{Hyper-Kamiokande:2018ofw}, $\varphi_8$ could be an accessible target for future runs of the LHC or the proposed high-energy hadron colliders. Vice versa, if colliders further constrain the $\varphi_8$ mass, this would favour a proton lifetime possibly within the reach of future experiments.
\end{itemize}

\begin{figure}[t!]
    \centering
    \includegraphics[width=1.\textwidth]{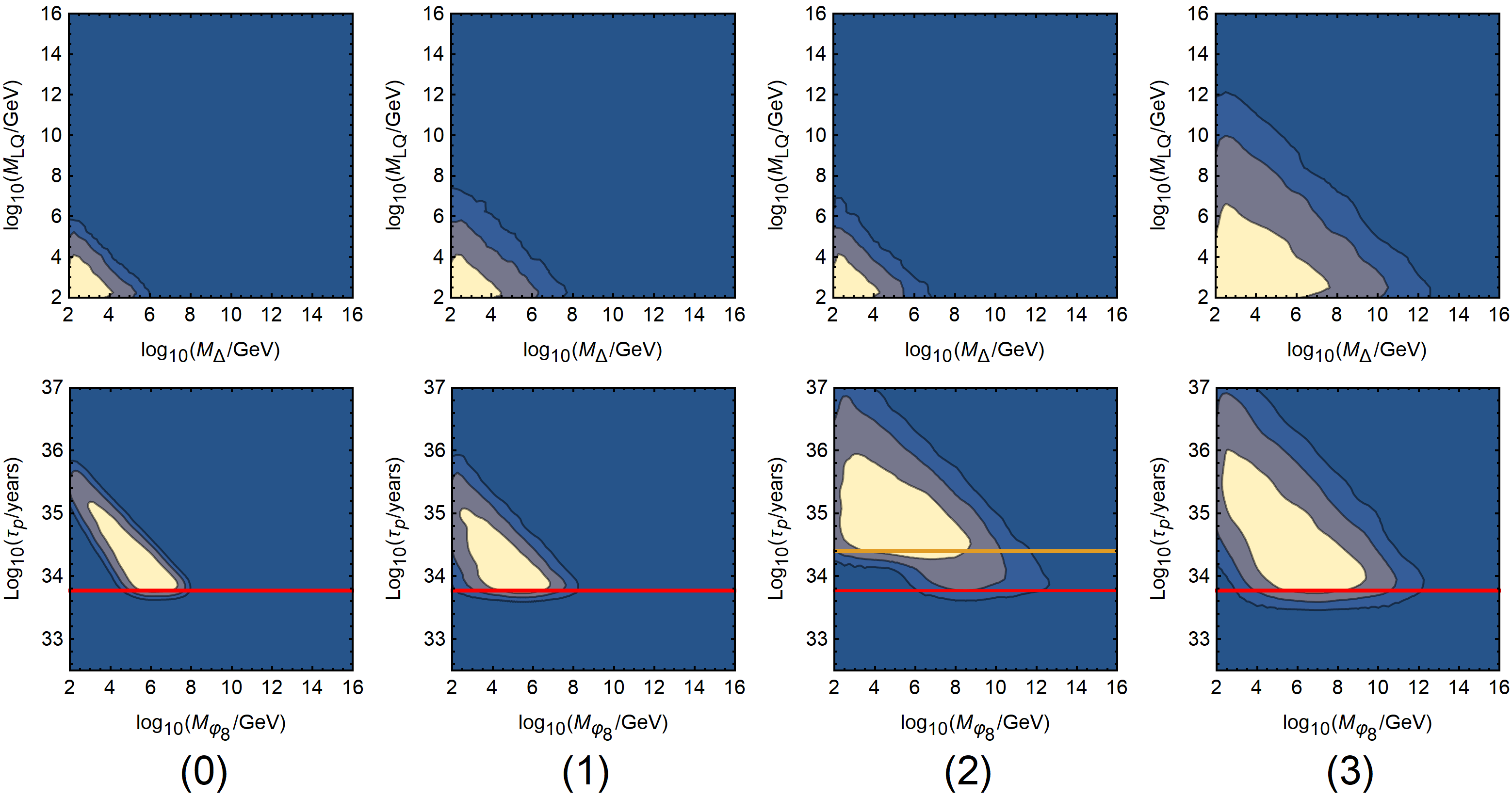}
    \caption{Impact on $M_\Delta$, $M_{LQ}$, and the $M_{\varphi_8}-\tau_p$ correlation of relaxing the assumptions of the Model 2 fit. From left to right: (0)~minimal setup as in \fref{fig:minimal-fit}, (1)~2HDM, (2)~generic flavour mixing, (3)~$p$-decay mediators. See the text for details.}
    \label{fig:comparison}
\end{figure}

As the above-discussed `minimal grand-unified type II seesaw model' has so many interesting phenomenological implications, it is important to discuss how robust the latter are. In other words, if we introduce more parameters, will the favoured masses of LFV mediators and the $M_{\varphi_8}-\tau_p$ correlation change much? We can consider three paths for a next-to-minimal extension of the minimal scenario: 
\begin{enumerate}
    \item[(1)] Set all the other particles ($S$ from $\phi_\textrm{\bf 15}$ and $\varphi_{\overline{6}}$, $\varphi_3^D$, $H_2$ from $\phi_\textrm{\bf 45}$) except for the proton decay mediators lighter than the GUT scale, such that they can also contribute to the RGEs. Due to the presence of the second Higgs doublet\,---\,the only among these fields that, as argued above, can have a positive impact on gauge coupling unification\,---\,we label this scenario 2HDM (two Higgs doublet model).
    \item[(2)] Allow arbitrary flavour mixing, that is, include the mixing angles in the matrices~(\ref{NewMixing}) among the free parameters to fit.\footnote{In this case, we can impose the bounds from all the proton decay channels, $p\rightarrow  \pi^0/K^0+e^+/\mu^+$ and $p\rightarrow \pi^+/K^+ + \overline{\nu}$, as we have all information to calculate the rates in Eqs.~(\ref{eq:p-to-e}-\ref{eq:p-to-K}).} This may suppress proton decay rates and relax the mass constraints on the light particles. In fact, according to \eref{Yukawa}, $M_u$ receives an anti-symmetric contribution such that $V_1$ is now a general unitary matrix.
    \item[(3)] Take the scalar proton decay mediators (the colour triplet $H_T$ in $\phi_{\bf 5}$, the $\phi_{\bf 45}$ fields $\varphi_{3}^T$, $\phi_{\bf 45}$ and $\varphi_{3}^S$) lighter than $\mgut$. We let their masses free  to range from $\mgut$ down to about $10^{13}$ GeV, which is the order of magnitude of the bounds from $p$-decay searches if the values of the couplings of these fields to SM fermions are in the ballpark of the SM Yukawa couplings~\cite{Nath:2006ut}.
\end{enumerate}

\begin{table}[t!]
\renewcommand{\arraystretch}{1.3}
    \centering
    \begin{tabular}{|c c c|}
 %   \hline
 %       & $M_{\Delta}$ (TeV) & $M_{LQ}$ (TeV) \\
    \hline
     \multicolumn{1}{|l}{(0) minimal fit}  & $M_{\Delta}$ (TeV) & $M_{LQ}$ (TeV) \\ \hline
    $1\sigma$~UL  & 1.6  & 1.5 \\
    $2\sigma$~UL  &  39  &  33\\
    $3\sigma$~UL  &  449  & 335\\
    \hline\hline
   \multicolumn{1}{|l}{(1) 2HDM}   & $M_{\Delta}$ (TeV) & $M_{LQ}$ (TeV) \\ \hline
    $1\sigma$~UL  & 2.2  & 1.6 \\
    $2\sigma$~UL  & 136 & 73 \\
    $3\sigma$~UL  &  $6.4\times10^3$  & $4.8\times10^3$\\
    \hline\hline
   \multicolumn{1}{|l}{ (2) flavour mix.}   & $M_{\Delta}$ (TeV) & $M_{LQ}$ (TeV) \\ \hline
    $1\sigma$~UL  & 2.0  & 1.5 \\
    $2\sigma$~UL  & 61  &  48 \\
    $3\sigma$~UL  &  $1.6\times10^3$  & $1.5\times10^3$\\
    \hline\hline
%    + $p$-decay mediators $\gtrsim10^{13}$ GeV & & \\
%    \multicolumn{1}{|l}{(3) $p$-decay med.}    & $M_{\Delta}$ (TeV) & $M_{LQ}$ (TeV) \\ \hline
%    $1\sigma$~UL  & 89  & 30 \\
%    $2\sigma$~UL &  $1.6\times10^5$ & $4.3\times10^4$\\
%    $3\sigma$~UL  &  $1.0\times10^8$  & $4.1\times10^7$\\
%    \hline
%    +Proton Decay Mediators $\gtrsim10^{11}$ GeV  & & \\
    \multicolumn{1}{|l}{(3) $p$-decay med.}    & $M_{\Delta}$ (TeV) & $M_{LQ}$ (TeV) \\ \hline
    $1\sigma$~UL  & $134$  & $33$ \\
    $2\sigma$~UL  & $1.5\times10^{5}$ & $4.9\times10^4$ \\
    $3\sigma$~UL  &  $1.8\times10^{8}$  & $8.5\times10^{7}$\\
    \hline
    \end{tabular}
    \caption{
    Model 2: 1$\sigma$, 2$\sigma$, and $3\sigma$ upper limits (UL) of the marginalised 1D probability distributions of the masses of the $\phi_{\bf 15}$ fields $\Delta$ and $\widetilde{R_2}$ for the minimal setup and the next-to-minimal fits described in the main text.}
    \label{tab:comparison}
\end{table}

\fref{fig:comparison} and \tref{tab:comparison} show the impact on the fit of the above relaxed assumptions. 
While the 3$\sigma$ upper bounds soar in the next-to-minimal scenarios, the 1$\sigma$-favoured regions remain at the TeV scale, with the exception of case (3), where the fit is substantially relaxed. 
In fact, among all these new scalars, only $\varphi_3^T$ and $H_2$ from $\varphi_{\textbf{45}}$ satisfy $b_1^I<b_2^I$ and thus can play a role in relaxing the mass bounds in the next-to-minimal scenarios,\footnote{The impact of $H_2$ is however very mild, as shown by the results for case (1) in \fref{fig:comparison} and \tref{tab:comparison}. Notice in particular that a light $\varphi_8$ is still required to achieve successful unification.} while the other fields considered in scenarios (1) and (3) can only decrease $\mgut$ and thus in fact tighten these mass bounds. Similarly, relaxing the flavour structure of the fermion mass matrices can loosen the $p$-decay constraints and have a significant impact on the $M_{\varphi_8}-\tau_p$ correlation (cf.~the column (2) of \fref{fig:comparison} where we plot $\min[\tau(p \to \pi^0 e^+), \tau(p \to K^+ \overline \nu)]$ and show both experimental limits) but does not affect much the prediction for the masses of the seesaw triplet and the leptoquark.

Searches for the production of new-physics particles at the LHC have already started to test this model. Constraints on the mass of $\Delta$ can be obtained searching, in particular, for the electroweak production of the doubly-charged states $pp\to \Delta^{++} \Delta^{--}$
followed by decays into same-sign leptons or $W$ bosons, $\Delta^{++} \to \ell^+_i \ell^+_j$ and $\Delta^{++} \to W^+ W^+$.
The former mode\,---\,which dominates if the vev of $\Delta$ induced by electroweak symmetry breaking is small, $\langle\Delta\rangle \lesssim 10^{-4}$~GeV, see e.g.~\cite{Ashanujjaman:2021txz}\,---\,provides a cleaner signature that leads to a stronger limit, $M_\Delta \gtrsim 800$~GeV~\cite{ATLAS:2017xqs}. If on the contrary the decay into $W$ dominates, the current lower limit on $M_\Delta$ is about 350~GeV~\cite{ATLAS:2021jol}.

Bounds on leptoquarks are even more stringent, as $\widetilde{R_2}$ can be copiously produced via strong interactions. The state with $Q=2/3$ decays fully visibly into (right-handed) down-type quarks and charged leptons, $\widetilde{R_2}^{2/3} \to \ell_i^+ d_j$. As we will discuss in the next section, the flavour structure of the $\widetilde{R_2}$ couplings (like that of $\Delta$) is dictated by the neutrino mass matrix, that is, by the large PMNS mixing angles. Therefore, $\widetilde{R_2}$ tends to decay `democratically' into all combinations of quark and lepton flavours, resulting in a large yield for the signal $e/\mu\,+$\,jet and a limit $M_{LQ}\gtrsim 1.6-1.8$~TeV~\cite{ATLAS:2020dsk}. 

The phenomenology of the colour-octet isospin doublet $\varrho_8$ has been extensively studied in the context of minimal extensions of the SM scalar sector, starting from Ref.~\cite{manohar2006flavor}.
After production via strong interactions, both states in $\varrho_8$ (charged and neutral) would decay into quark pairs through the couplings $Y_u^\prime$ and $Y_{d\ell}^\prime$ in \eref{Yukawa}. If these matrices feature a flavour hierarchy resembling that of the SM Yukawa couplings, decays into third generation quarks ($\varrho_8^0 \to t\bar t$, $\varrho_8^+ \to t\bar b$) will dominate. In such a case, the current LHC bounds are estimated to be in the 800$-$1000~GeV range~\cite{darme2018cornering,Miralles:2019uzg,Cacciapaglia:2020vyf}. However, notice that only sizeable couplings to first and second generation down-type quarks are strictly required in order to correct the relations in \eref{eq:relations}. If such couplings dominate, the particles in the octet would mostly decay into two light jets and, thus, be subject to much more stringent constraints from searches for heavy di-jet resonances, corresponding to a lower bound of about 4~TeV~\cite{CMS:2018mgb}.

%%%%%%%%%%%%%%%%%%%%%%%%%%%%%%%%%%%%%%%%%%%%%%

\subsection{Model 3}
\label{sec:model3}
\begin{figure}[t]
    \centering
    \includegraphics[width=1.\textwidth]{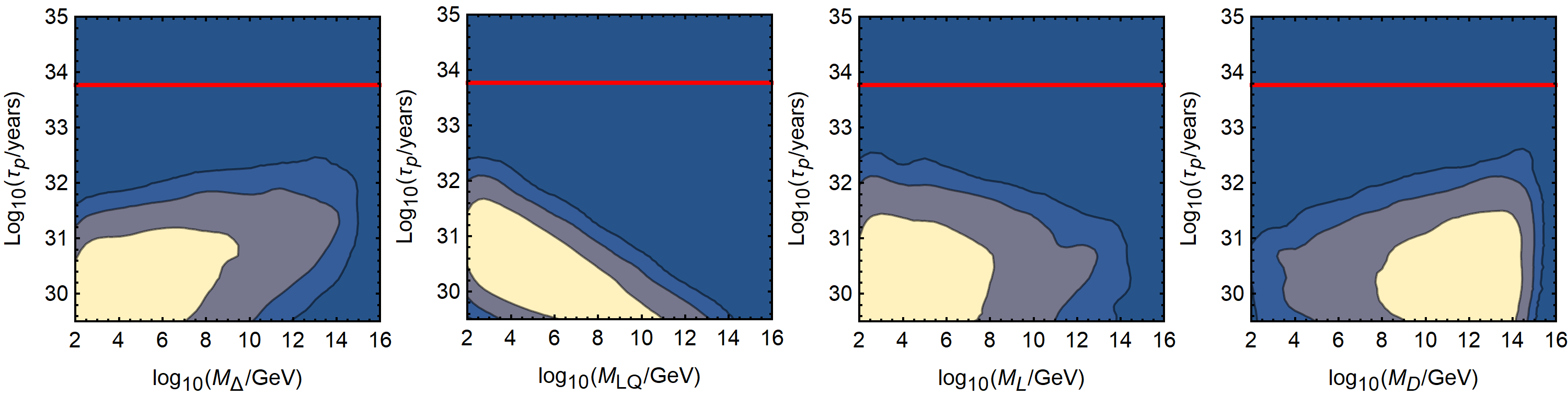}\\
    \includegraphics[width=.75\textwidth]{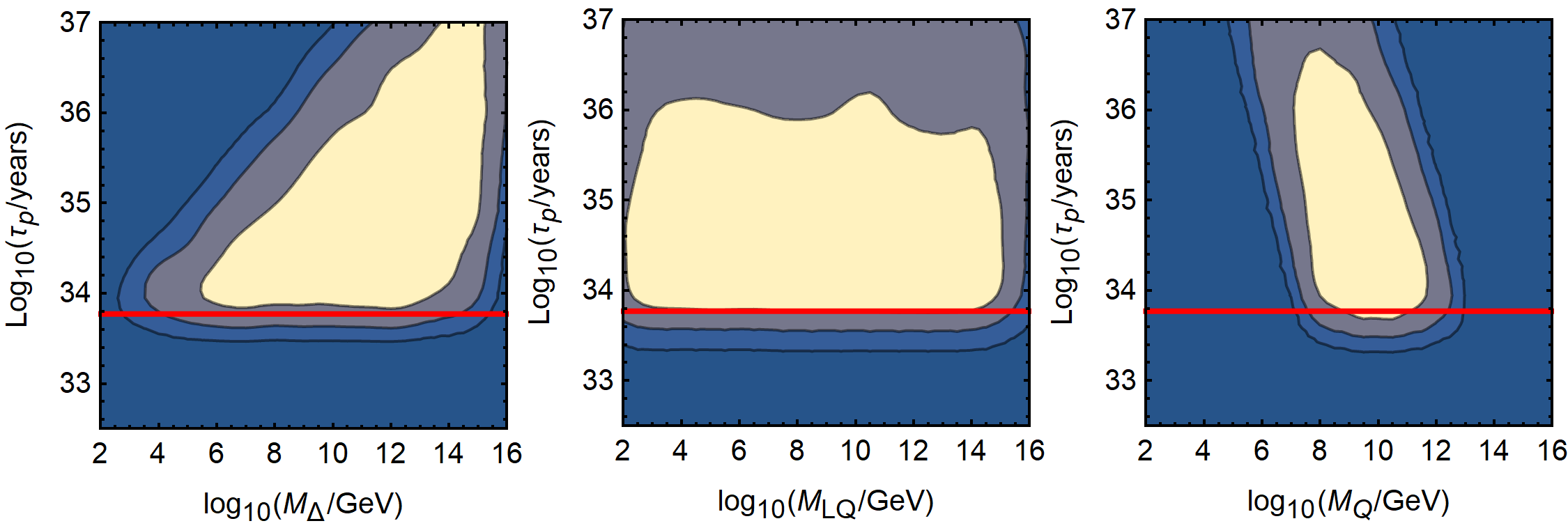}
    \caption{Results of the fit for Model 3, considering the minimal field content of \fref{Non-re} plus one generation of vector-like fermions in the $\mathbf{5} + \mathbf{\overline{5}}$ representation (first row) or one generation in the  $\mathbf{10} + \mathbf{\overline{10}}$ (second row).}
    \label{fig:model3}
\end{figure}
In the case of Model 3, where only vector-like fermions are added to the minimal $SU(5)$ field content, the regions of the parameter space favoured by the fit are very different. As discussed above, correct fermion mass relations and unification can be achieved by the usual fields in $\phi_\mathbf{24}$ and $\phi_\mathbf{15}$ plus vector-like leptons $L_V +  L^c_V$ (that is, introducing a fermionic $\mathbf{5} + \mathbf{\bar 5}$), but the impact of these latter field on $\mgut$ is limited. As a consequence, one generation of vector-like leptons is far insufficient to raise $\mgut$ with respect to Model 1 (\fref{Non-re}) at a level compatible with the $p$-decay bounds.
This is shown in the first row of \fref{fig:model3}. Only multiple $\mathbf{5} + \mathbf{\bar 5}$ generations could evade the limits on $p$-decay
without relying on tuning in the fermion mixing. We checked that at least five generations are needed. 
On the other hand, if one introduces fermions in the $\mathbf{10} + \mathbf{\overline{10}}$, only one generation of $Q_V + Q^c_V$ is enough to achieve unification at a large enough $\mgut$. In the latter case, the $\phi_{\bf 15}$ fields $\Delta$ and $\widetilde{R_2}$ do not even need to be light. This is explicitly shown in the second row of \fref{fig:model3}. As we can see, the central values for $M_{\Delta}$ and $M_{LQ}$ could be much higher compared to the results we shown for Model~2 (cf.~\tref{tab:comparison}) and their 1$\sigma$ favoured ranges span almost all scales between $m_Z$ and $\mgut$. Hence Model 3 (with a single $\mathbf{10} + \mathbf{\overline{10}}$), while being perfectly viable, completely lacks the predictivity and the interesting phenomenological features of Model~2. A similar conclusion would hold also for the case of multiple ($\ge5$) $\mathbf{5} + \mathbf{\bar 5}$ generations, as shown by the first row of \fref{fig:model3}.

%%%%%%%%%%%%%%%%%%%%%%%%%%%%%%%%%%%%%%%%%%%%%%%%%%%%%%

\section{Type II seesaw fields and Lepton Flavour Violation}
\label{sec:lfv}
In this section, we focus on the low-energy phenomenology of the fields in $\phi_{\textbf{15}}$ associated to the generation of neutrino masses.
In particular, the seesaw triplet $\Delta$ and the scalar leptoquark $\widetilde{R_2}$ unavoidably mediate LFV interactions, as we are discussing in the following. Furthermore, within the most successful (and predictive) of the models analysed above (see \sref{sec:model2}), gauge coupling unification requires them to be rather light, $\lesssim \mathcal{O}(10)$~TeV, which makes searches for LFV processes the most promising experimental handle to test type II seesaw unification. 

The couplings to leptons of these fields follow from the third term in \eref{eq:yuk}:
\begin{equation}
\label{eq:typeIIint}
-{\cal L}_{\text{Yukawa}} \supset
Y_{15}^{\alpha\beta}\, \overline{\psi_{\overline{\textbf{5}}}}_{\,\alpha} \phi_{\textbf{15}}^* \psi_{\overline{\textbf{5}}\,\beta}^c+\text{h.c.} %\\& 
~\rightarrow ~ Y_{\Delta}^{\alpha\beta}\, \overline{L_L}_{\,\alpha} \Delta i \sigma_2 L_{L\,\beta}^c
+ Y_{LQ}^{\alpha\beta}\, \overline{D_{R}}_{\,\alpha} \widetilde{R_2} L_{L\,\beta}  +\text{h.c.}\,,
\end{equation}
where $\alpha$ and $\beta$ are flavour indices and we work in the basis where the charged-lepton and down-quark mass matrices are flavour diagonal. The conventions we adopt for the decomposition of the $SU(2)_L$ representations are
\begin{align}
\label{SU2sym}
    \Delta =\left(
\begin{array}{cc}
    \Delta^-/\sqrt{2} & \Delta^{0} \\
    \Delta^{--}  & -\Delta^-/\sqrt{2}
\end{array}
\right)\,, \quad\quad \widetilde{R_2}^T= \left(\widetilde{R_2}^{2/3},\, \widetilde{R_2}^{-1/3}\right)\,.
\end{align}

At the GUT scale the triplet and leptoquark Yukawa matrices in \eref{eq:typeIIint} match to $Y_{15}$ as 
\begin{equation}
\label{eq:yukGUT}
Y_\Delta = Y_{LQ}/\sqrt2 = Y_{15}\,, \quad\quad \text{[GUT scale]\,.}    
\end{equation}
At lower scales, they are renormalised according to the RGEs reported in \aref{app:LQrge}, resulting in the TeV-scale relation 
\begin{equation}
\label{eq:yukTeV}
Y_{LQ} \approx 2.1\,Y_\Delta\,,  \quad\quad \text{[TeV scale]\,.}    
\end{equation}
What makes this framework predictive is that the flavour structure of both matrices is related to the observed neutrino masses and mixing.
Neutrino mass terms arise from the explicit breaking of the lepton number that is a consequence of the couplings of the triplet to leptons in \eref{eq:typeIIint} in combination with the following scalar potential term:
\begin{equation}
\label{eq:DeltaL}
-{\cal L}~ \supset ~ \mu\,\phi_{\bf 5} \phi_{\bf 15}^* \phi_{\bf 5}  +\text{h.c.}
~\rightarrow ~ \mu_\Delta\, H^T i \sigma_2 \Delta  H  +\text{h.c.}\,.
\end{equation}
The resulting Majorana neutrino mass matrix reads
\begin{equation}
\label{eq:mnu}
m_\nu = \sqrt2 \,Y_\Delta v_\Delta =  Y_\Delta  \,\frac{\mu_\Delta\, v^2}{M_\Delta^2}\,,
\end{equation}
where $v_\Delta$ is the vev the triplet acquires upon electroweak symmetry breaking, that is, $\langle \Delta^0 \rangle = v_\Delta/\sqrt2$, and $v = v_5$.\footnote{This is equal to $v_\text{EW}\approx 246$~GeV for models without a second Higgs doublet. On the contrary, if $H_2\subset  \phi_{45}$ exists, one has $v_5 = v_\text{EW}\cos\beta$ with $\tan\beta \equiv v_{45}/v_{5}$ being a free parameter. If this is the case, the bounds on the couplings discussed below have to be rescaled by an $\mathcal{O}$(1) factor, while the rest of the phenomenological discussion does not change.}
\eref{eq:mnu} shows that the flavour structure of the matrix $Y_\Delta$\,---\,and consequently of $Y_{LQ}$ too\,---\,is the same as that of the neutrino mass matrix. 
In other words, in the charged-lepton mass basis, $Y_\Delta$ unavoidably features off-diagonal LFV entries dictated by the (large) mixing angles of the PMNS matrix, following from 
\begin{align}
\label{diag}
m_\nu^{\rm diag} = U_\textsc{pmns}^T \, m_\nu\, U_\textsc{pmns}\,,
\end{align}
where $m_\nu^{\rm diag}$ is the diagonal matrix of the neutrino mass eigenvalues $(m_1,m_2,m_3)$ and $U_\textsc{pmns}$ is the PMNS mixing matrix (cf.~\aref{app:nupar} for details).
Notice however that the absolute size of the couplings in $Y_\Delta$ and $Y_{LQ}$ is not uniquely determined (even for a given $M_\Delta$) because of the dependence of $m_\nu$ on the lepton-breaking dimensionful parameter $\mu_\Delta$. In particular, for a small enough $\mu_\Delta$, the observed values of the neutrino masses can be reproduced even with a light triplet and $\sim \mathcal{O}(1)$ couplings in $Y_\Delta$\,---\,which greatly enhances the LFV effects, as we will show below. On the other hand, the Yukawa couplings could still be extremely small if $\mu_\Delta$ is sizeable.\footnote{One can obtain the loose lower bound $|Y_\Delta^{\alpha\beta}| \gtrsim 10^{-12}$ from the electroweak-fit constraint on $\Delta\rho$, which requires $v_\Delta \lesssim 1$~GeV (see e.g.~\cite{Kanemura:2022ahw}) in \eref{eq:mnu}.}
Ratios of rates of different LFV processes overcome this source of uncertainty and, as discussed below, can provide further constraints on the spectrum of the model, in particular on the ratio $M_{LQ}/M_\Delta$. This opportunity, in combination with the unification requirements on the particle masses and the fact that, following from \eref{eq:typeIIint}, the leptoquark couplings $Y_{LQ}$ are flavour symmetric and linked to the neutrino mass matrix, makes the LFV phenomenology of $SU(5)$ type-II seesaw models much more predictive than the generic setups previously studied, e.g.~in the model-independent analyses of Refs.~\cite{abada2007low,Dorsner:2016wpm}.

\begin{table}[t]
\renewcommand{\arraystretch}{1.3}
    \centering
    \begin{tabular}{l c c }  
    \hline
    Observable & 90\% CL upper limit & Future sensitivity \\
    \hline
    ${\rm BR}(\mu^+\to e^+\gamma)$ & $4.2\times 10^{-13}$~\cite{TheMEG:2016wtm} & $ 6 \times 10^{-14}$~\cite{Baldini:2018nnn} \\
    ${\rm BR}(\mu^+\to e^+e^-e^+)$ & $1.0\times 10^{-12}$~\cite{Bellgardt:1987du} & $10^{-16}$~\cite{Blondel:2013ia} \\
    ${\rm CR}(\mu^-\, N\to e^- \,N)$ & $7.0\times 10^{-13}$ ($N=$\,Au)~\cite{Bertl:2006up} & $6\times 10^{-17}$ ($N=$\,Al)~\cite{Bartoszek:2014mya,Kuno:2013mha} \\
    \hline
    ${\rm BR}(K_L\to \mu^\pm e^\mp)$ & $4.7\times 10^{-12}$~\cite{BNL:1998apv} & $\sim 10^{-12}$~\cite{Goudzovski:2022vbt} \\
    ${\rm BR}(K_L\to \pi^0 \mu^+ e^-)$ & $7.6\times 10^{-11}$~\cite{KTeV:2007cvy} & $\sim 10^{-12}$~\cite{Goudzovski:2022vbt} \\
    ${\rm BR}(K^+\to \pi^+ \mu^+ e^-)$ & $1.3\times 10^{-11}$~\cite{Sher:2005sp} & $\sim 10^{-12}$~\cite{Goudzovski:2022vbt} \\
    ${\rm BR}(K^+\to \pi^+ \mu^- e^+)$ & $5.2\times 10^{-10}$~\cite{Appel:2000tc} & $\sim 10^{-12}$~\cite{Goudzovski:2022vbt} \\
    \hline
    \end{tabular}
    \caption{Current experimental bounds and future expected sensitivities on the LFV processes relevant for our analysis.}
    \label{tab:lfv}
\end{table}

\subsection{LFV observables}

Both the triplet $\Delta$ and the leptoquark $\widetilde{R_2}$ induce LFV processes already at the tree level. Here we focus on $\mu-e$ flavour violation that is subject to the best limits at present and has the most promising experimental prospects, see e.g.~\cite{Calibbi:2017uvl}.
Present bounds and future expected sensitivities on the processes we are interested in are reported in \tref{tab:lfv}.

A tree-level exchange of the triplet mediates $\mu\to eee$~\cite{abada2007low}:
\begin{align}   
    \text{BR}(\mu\rightarrow eee) = \frac{1}{4 G_F^2M_\Delta^4} \left|Y_{\Delta}^{21}\right|^2  \left|Y_{\Delta}^{11}\right|^2\,
\end{align}
where $G_F$ is the Fermi constant. 

The leptoquark $\widetilde{R_2}$ can induce at tree level $\mu \to e$ conversion in atomic nuclei, with a conversion rate given by~\cite{FileviezPerez:2008dw,Dorsner:2016wpm}:
\begin{align}
    \text{CR}(\mu\, N\rightarrow e\, N)&= \frac{m_\mu^5}{4\Gamma_{\text{capt}}\, M_{LQ}^4} \left(V^{(p)}+2 V^{(n)}\right)^2 \left|Y_{LQ}^{21}\right|^2  \left|Y_{LQ}^{11}\right|^2 \,,
\end{align}
which is as usual normalised by the capture rate $\Gamma_\text{capt}$ of muons by the nucleus $N$. $V^{(p)}$ and $V^{(n)}$ are overlap integrals between
muon and electron wave functions and nucleons density distributions~\cite{Kitano:2002mt}. The most recent evaluation of these quantities can be found in Ref.~\cite{Heeck:2022wer}.\footnote{The present best limit on $\mu \to e$~conversion was obtained on gold and the upcoming experiments plan to employ aluminium targets, see \tref{tab:lfv}. Thus we are using the following input for our analysis~\cite{Heeck:2022wer}:
$V^{(p)}(\text{Au})~=~0.0866\,,V^{(n)}(\text{Au})=0.129$ and $\Gamma_{\text{capt}}(\text{Au})~=~13.07\times10^6~{\rm s}^{-1}$;
$V^{(p)}(\text{Al})~=~0.0165\,,V^{(n)}(\text{Al})~=~0.0178$ and $\Gamma_{\text{capt}}(\text{Al})~=~0.7054\times10^6~{\rm s}^{-1}$. The capture rates were taken from \cite{Kitano:2002mt}.
}

The leptoquark also contributes at tree level to LFV decays of mesons, in particular the tightly constrained neutral kaon decay
$K_L\rightarrow \mu e$, whose branching ratio reads~\cite{FileviezPerez:2008dw,Dorsner:2016wpm}:
\begin{align}
\label{eq:KL}
    \text{BR}(K_L\rightarrow \mu e)&= \frac{m_K\tau_{K_L}}{256\pi}\frac{m_\mu^2 f_{K_L}^2 }{M_{LQ}^4}\left(1-\frac{m_{\mu}^2}{m_{K_L}^2}\right)^2
    \left|Y_{LQ}^{12}Y_{LQ}^{12\,*}+Y_{LQ}^{11} Y_{LQ }^{22\,*}\right|^2\,,
\end{align}
where $f_{K_L} \simeq 160$~MeV and $\tau_{K_L} = 5.116\times 10^{-8}$\,s are $K_L$ decay constant and lifetime~\cite{Zyla:2020zbs}.
Semileptonic kaon decays are also induced. Following \cite{bevcirevic2016lepton}, we find
\begin{align}
    \frac{d}{d q^2}\text{BR}(K \rightarrow \pi \mu e) ~  = ~& \frac{(m_{\mu}^2-q^2)^2 \tau_{K} \lambda^{\frac{1}{2}}(\sqrt{q^2},m_{K},m_{\pi})}{12288 \pi^3 \, m_{K}^3 M_{LQ}^4 q^6} \mathcal{Y} \\
    &\times \left[3 |f_0(q^2)|^2 (m_{K}^2-m_{\pi}^2 )^2 m_{\mu}^2+|f_+(q^2)|^2 ( m_{\mu}^2 +2 q^2)\lambda(\sqrt{q^2},m_{K},m_{\pi}) \right]\,,
    \nonumber
\end{align}
where $q^2 = (p_\mu + p_e)^2$ (with $m_\mu^2\lesssim q^2 \le (m_K-m_\pi)^2$), $\lambda(a,b,c) \equiv [a^2 -(b - c)^2][a^2 - (b + c)^2]$, and the form factors are about the same for $K^+$ and $K_L$ (up to percent level corrections~\cite{Cirigliano:2011ny}) and depend very weakly on $q^2$~\cite{KTeV:2004ozu}. For our numerical study, we employ
$f_0(q^2)\simeq f_+(q^2)\simeq f_0(0)\simeq f_+(0)\simeq0.9677$~\cite{FlavourLatticeAveragingGroup:2019iem}.
The above expression depends on the following combinations of couplings: 
%
\begin{comment}
$\mathcal{Y} = |Y_{LQ}^{12}|^4$ for $K^+\rightarrow \pi^+\mu^+ e^-$, $\mathcal{Y} = |Y_{LQ}^{11}Y_{LQ}^{22}|^2$ for $K^+\rightarrow \pi^+\mu^- e^+$, $\mathcal{Y} = \frac{1}{2}|Y_{LQ}^{12}Y_{LQ}^{12\,*}+Y_{LQ}^{11} Y_{LQ }^{22\,*}|^2 $ for $K_L\rightarrow \pi^0\mu^+ e^-$.  
\end{comment}
%
\begin{equation}
\label{eq:3body}
 \mathcal{Y} = ~ \left\{\begin{array}{ll}
      |Y_{LQ}^{12}|^4   &   [K^+\rightarrow \pi^+\mu^+ e^-]\,, \\[4pt]
       |Y_{LQ}^{11}Y_{LQ}^{22\,*}|^2  & [K^+\rightarrow \pi^+\mu^- e^+]\,, \\[4pt]
       \frac{1}{2}|Y_{LQ}^{12}Y_{LQ}^{12\,*}+Y_{LQ}^{11} Y_{LQ }^{22\,*}|^2 & [K_L\rightarrow \pi^0\mu^+ e^-]\,.
 \end{array}
 \right. 
\end{equation}

As one can see from \eref{eq:KL}, $K_L\rightarrow \pi^0\mu^+ e^-$ has the same dependence as $K_L\rightarrow \mu e$, and one numerically finds
$\text{BR}(K_L \rightarrow \pi^0 \mu^+ e^-) \approx 0.04 \times \text{BR}(K_L\rightarrow \mu^+ e^-)$, hence it can not provide additional information. On the contrary, both $K^+\rightarrow \pi^+\mu^+ e^-$ and $K^+\rightarrow \pi^+\mu^- e^+$ have a cleaner dependence on the entries of $Y_{LQ}$, hence they can help to study its flavour structure, as we will see below.

In principle, both the triplet and the leptoquark also induce $\mu\to e\gamma$ at one loop. However, the contribution from loops involving down-type quarks and $\widetilde{R_2}$ is strongly suppressed\,---\,the terms in the amplitude being $\propto (m_{d,s,b}/M_{LQ})^2$\,---\,thus the branching ratio is very well approximated by the contribution of the type II seesaw triplet alone~\cite{abada2007low}:
\begin{align}   
    \text{BR}(\mu\rightarrow e\gamma) = \frac{\alpha}{48\pi \,G_F^2M_\Delta^4} \frac{25}{64} \left|\sum_\beta Y_\Delta^{2\beta\,*} Y_\Delta^{1\beta} \right|^2\,.
\end{align}

\subsection{Numerical analysis}
\label{sec:results}

Numerically, the above formulae give
\begin{equation}
    \begin{aligned}
    \text{BR}(\mu\rightarrow eee)\, &\simeq~ 1.1\times 10^{-12} \left( \frac{10~\textrm{TeV}}{M_\Delta}\right)^4 \left(\frac{\left|Y_{\Delta}^{21}\right|^2  \left|Y_{\Delta}^{11}\right|^2}{0.05^4}\right)\,, \\% \nonumber\\
    \text{BR}(\mu\rightarrow e\gamma)\, & \simeq~ 3.6\times 10^{-13} \left( \frac{10~\textrm{TeV}}{M_\Delta}\right)^4 \left(\frac{\left|\sum_\beta Y_\Delta^{2\beta\,*} Y_\Delta^{1\beta} \right|^2}{0.4^4}\right)\,, \\%\nonumber\\
    \text{CR}(\mu \,\text{Au}\rightarrow e \,\text{Au}) \, & \simeq~ 
    2.4 \times \text{CR}(\mu \,\text{Al}\rightarrow e \,\text{Al})\, \simeq~
    7.3\times 10^{-13} \left( \frac{10~\textrm{TeV}}{M_{LQ}}\right)^4 \left(\frac{\left|Y_{LQ}^{21}\right|^2  \left|Y_{LQ}^{11}\right|^2}{0.02^4}\right)\,, \\%\nonumber\\
    \text{BR}(K_L\rightarrow \mu e)\, & \simeq~ 3.2\times 10^{-12}\left( \frac{10~\textrm{TeV}}{M_{LQ}}\right)^4\left(\frac{\left|Y_{LQ}^{12}Y_{LQ}^{12\,*}+Y_{LQ}^{11} Y_{LQ }^{22\,*}\right|^2}{0.04^4}\right)\,, \\%\nonumber\\
     \text{BR}(K^+\rightarrow \pi^+\mu^+ e^-)\, & \simeq~  1.2   \times 10^{-11} 
     \left( \frac{10~\textrm{TeV}}{M_{LQ}}\right)^4 \left(\frac{\left|Y_{LQ}^{21}\right|^4}{0.15^4}\right)\,, \\
        \text{BR}(K^+\rightarrow \pi^+\mu^- e^+)\, & \simeq~   6.2   \times 10^{-10} 
     \left( \frac{10~\textrm{TeV}}{M_{LQ}}\right)^4 \left(\frac{\left|Y_{LQ}^{11}Y_{LQ}^{22\,*}\right|^2}{0.4^4}\right)\,.
     \label{eq:LFVest}
    \end{aligned}
\end{equation}

\begin{figure}[t!]
  \centering
  \includegraphics[width=0.55\textwidth]{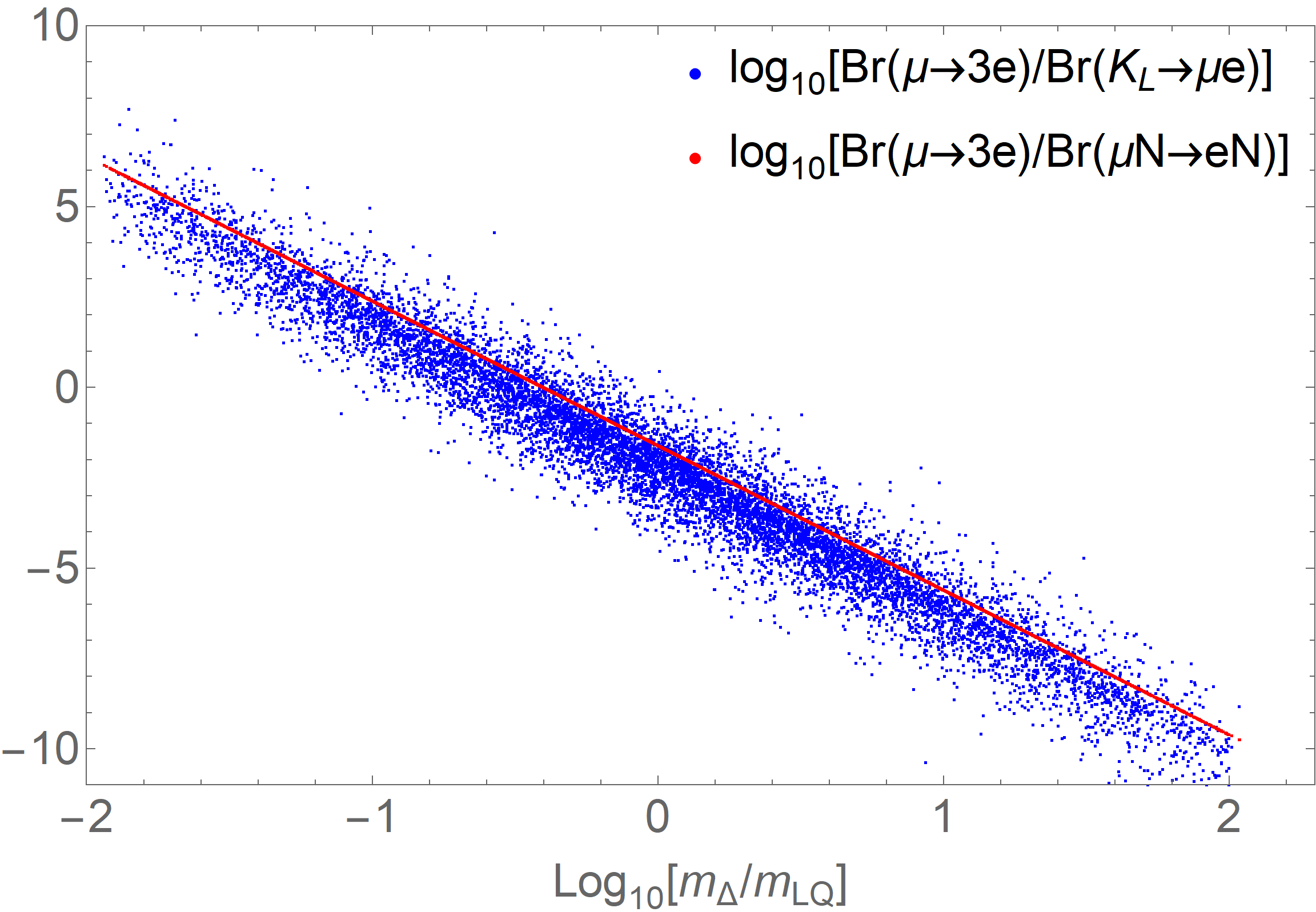}
  \caption{ Ratios $\text{BR}(\mu\rightarrow eee)/ \text{CR}(\mu\,\text{Al} \rightarrow e\,\text{Al})$ (red points) and 
   $\text{BR}(\mu\rightarrow eee)/ \text{BR}(K_L\rightarrow \mu e)$ (blue points) as functions of $M_\Delta / M_{LQ}$. See the text for details.
  }\label{fig:LFV-mass-ratio}
\end{figure}

These results, in combination with the experimental limits in \tref{tab:lfv}, show that for the spectrum favoured by our Model 2 fit in \sref{sec:model2} ($M_\Delta,\,M_{LQ}\lesssim 10$~TeV) the relevant couplings are approximately constrained to be $|Y^{\alpha\beta}_\Delta|\lesssim 0.05$, $|Y^{\alpha\beta}_{LQ}|\lesssim 0.02$, with $\mu \to eee$ and $\mu \,N\rightarrow e \,N$ providing the most stringent bounds, which translate into lower limits on the masses as stringent as $M_\Delta~\gtrsim~200$~TeV, $M_{LQ}~\gtrsim~500$~TeV, for $\mathcal{O}(1)$ couplings. \tref{tab:lfv} and \eref{eq:LFVest} show that upcoming experiments will improve these limits by about one order of magnitude.

Even if not so constraining, $\mu\to e \gamma$, $K_L\to \mu e$ and $K^+\rightarrow \pi^+\mu e$ depend on different combinations of the couplings, hence ratios of the branching ratios of different modes can provide information on the flavour structure of $Y_\Delta$ and $Y_{LQ}$, that is, on the flavour structure of the neutrino mass matrix, \eref{eq:mnu}, as we will discuss below.

%%%%%%%%%%%%%%%%%%%%%%%%%%%%

The matrices $Y_\Delta$ and $Y_{LQ}$ are related by the GUT boundary condition, \eref{eq:yukGUT}. Moreover, as discussed in \aref{app:LQrge},  the RGE running does not affect their flavour structure, only the overall normalisation. Therefore, we see from \eref{eq:LFVest} that the 
ratio between the rates of $\mu \to eee$ and $\mu \to e$ conversion in nuclei only depends on $M_{LQ}/M_{\Delta}$:
\begin{equation}
\label{eq:meconv-ratio}
\text{BR}(\mu\rightarrow eee) \,\simeq 0.0021 \,\left( \frac{M_{LQ}}{M_\Delta}\right)^4   \text{CR}(\mu \,\text{Au}\rightarrow e \,\text{Au}) \,\simeq    0.0049 \,\left( \frac{M_{LQ}}{M_\Delta}\right)^4  \text{CR}(\mu \,\text{Al}\rightarrow e \,\text{Al}) \,,
\end{equation}
where we employed the TeV-scale relation \eref{eq:yukTeV}. It is then clear that measurements (or constraints) of different LFV processes can provide  non-trivial information on the mass spectrum of the theory, to be combined with the constraints from gauge coupling unification and proton decay discussed in the previous section. This is also depicted in \fref{fig:LFV-mass-ratio}, where we plot the ratio $\text{BR}(\mu\rightarrow eee)/ \text{CR}(\mu\,\text{Al} \rightarrow e\,\text{Al})$ (red points) as a function of $M_\Delta / M_{LQ}$, varying the mass parameters within the 1$\sigma$-favoured region of the Model 2 fit reported in \sref{sec:model2}.

\begin{figure}[t!]
  \centering
  \includegraphics[width=0.48\textwidth]{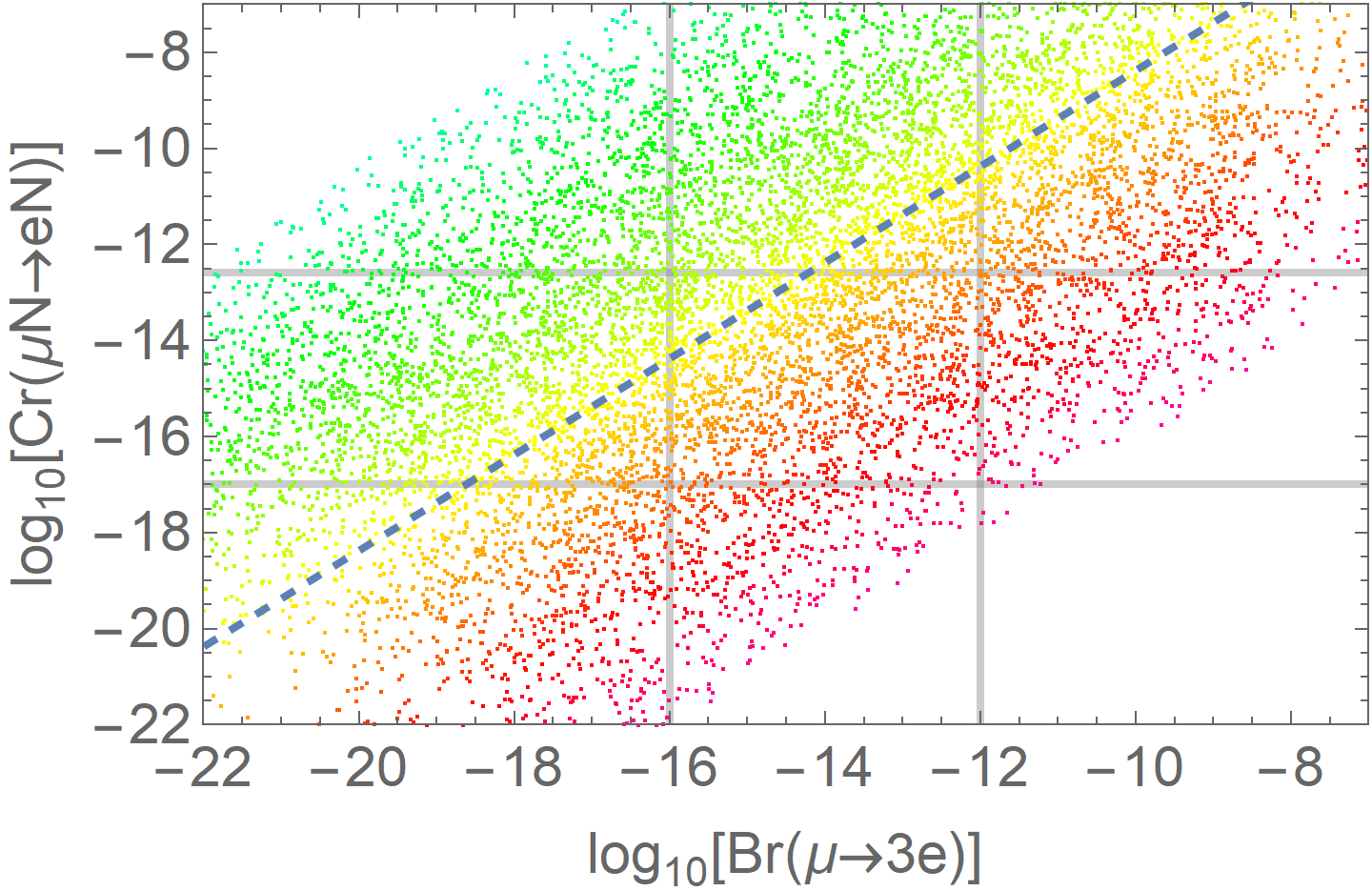}
  \hfill
  \includegraphics[width=0.48\textwidth]{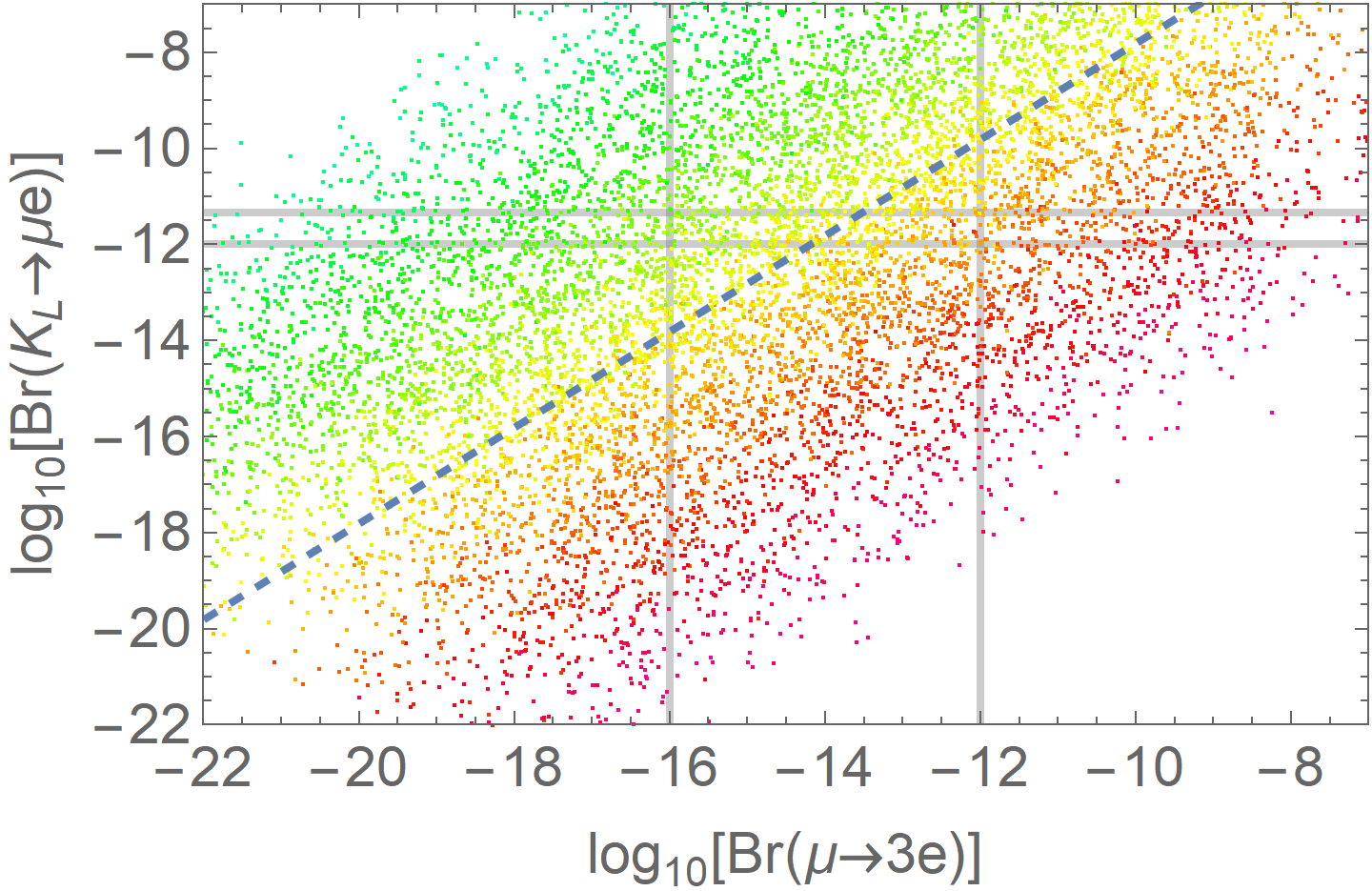}\\
  \vspace{0.5cm}
  \includegraphics[width=0.6\textwidth]{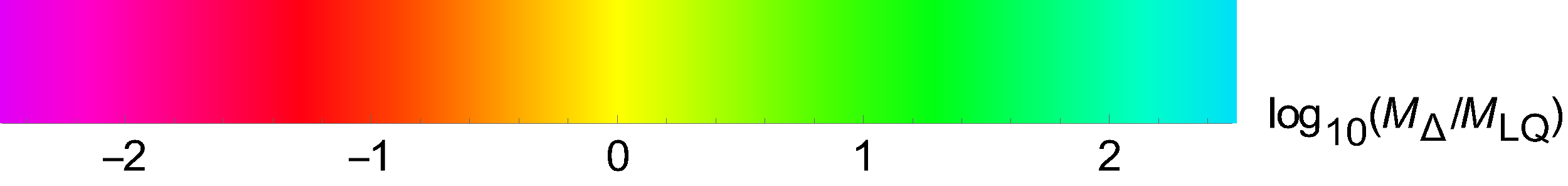}
    \caption{$\text{CR}(\mu\,\text{Al} \rightarrow e\,\text{Al})$ vs~$\text{BR}(\mu\rightarrow eee)$ (left panel) and 
   $\text{BR}(K_L\rightarrow \mu e)$ vs~$\text{BR}(\mu\rightarrow eee)$ (right panel) for the same variation of the parameters as in \fref{fig:LFV-mass-ratio} (see text for details). The colour of the points denotes the value of $M_\Delta / M_{LQ}$, as indicated under the plots, and the  dashed line corresponds to $M_\Delta = M_{LQ}$ and to setting the combinations of couplings appearing in \eref{eq:LFVest}
   to their fitted central values. The gray lines indicate the present and future experimental limits as in \tref{tab:lfv}. The present bound on $\mu\to e$ conversion was rescaled according to \eref{eq:meconv-ratio}.}
   \label{LFV processes1}
\end{figure}
\begin{figure}[t!]
  \centering
  \includegraphics[width=0.48\textwidth]{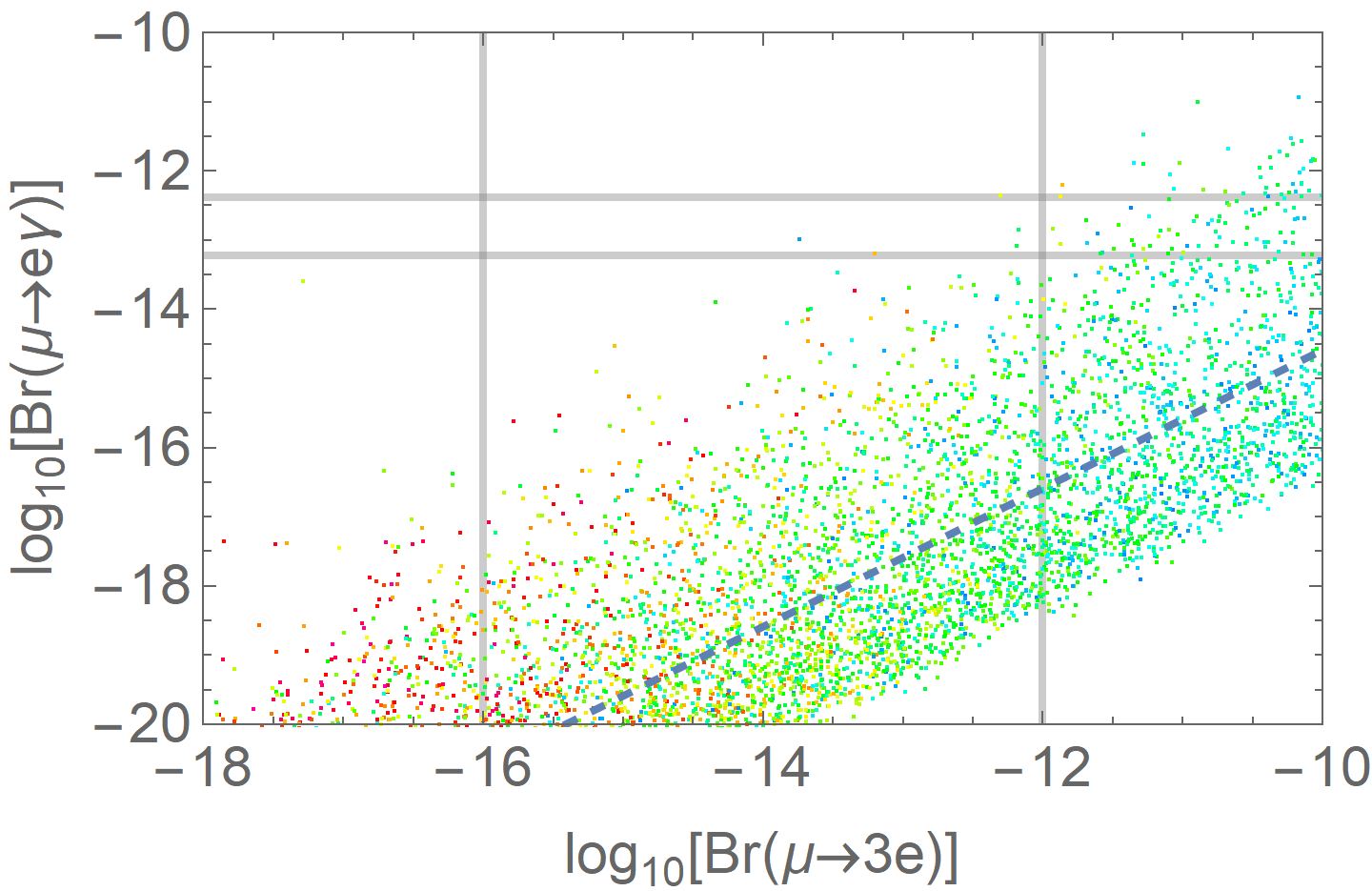}\hfill
  \includegraphics[width=0.48\textwidth]{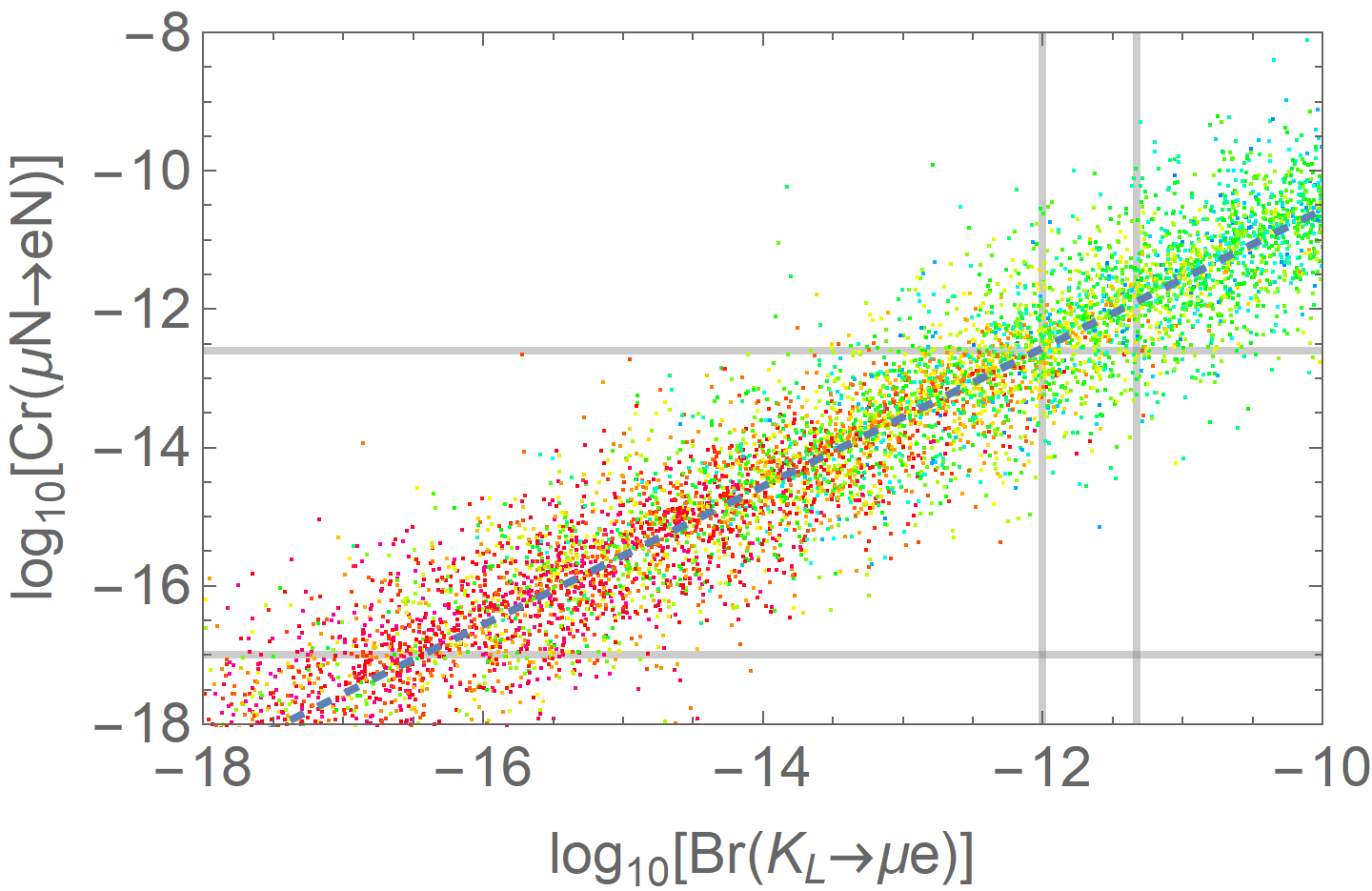}\\
  \includegraphics[width=0.48\textwidth]{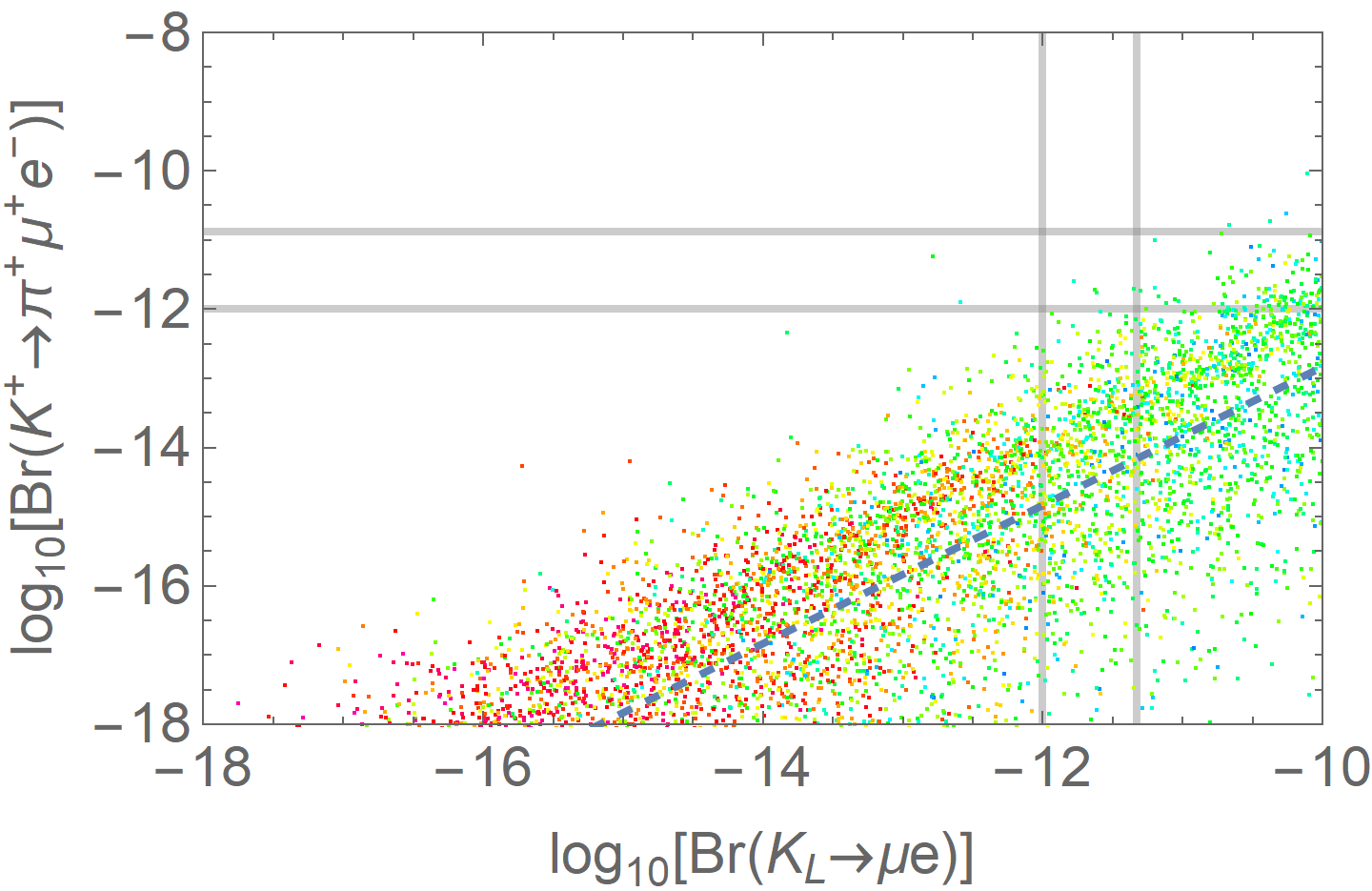}\hfill
  \includegraphics[width=0.48\textwidth]{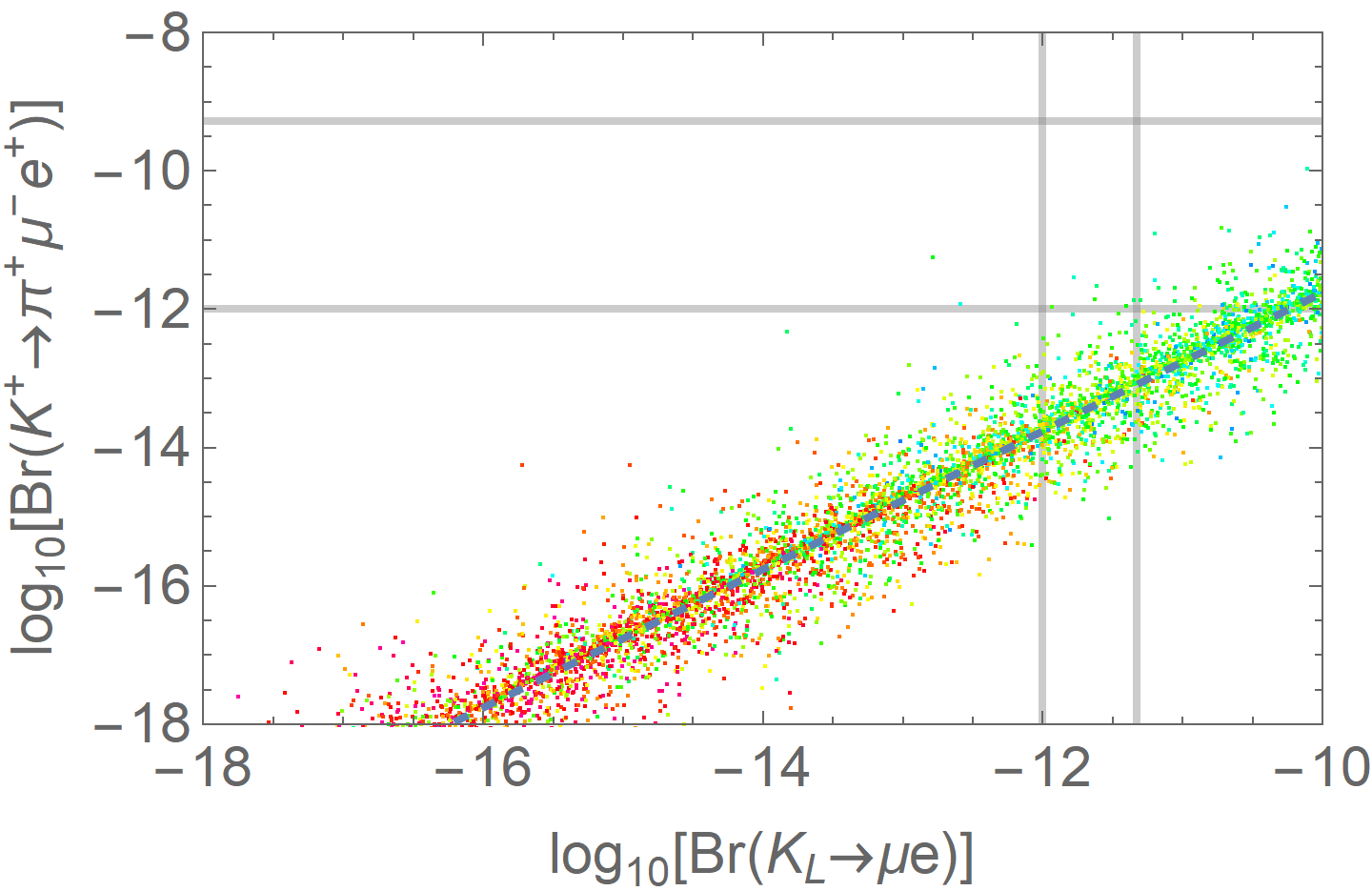}\\
  \vspace{0.5cm}
  \includegraphics[width=0.6\textwidth]{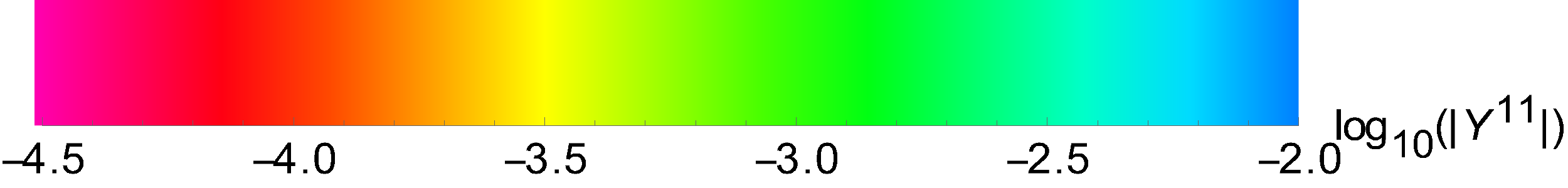} 
  \caption{Correlations between processes induced by the exchange of the same field ($\Delta$ or $\widetilde{R_2}$)
  for the same variation of the parameters as in Figures~\ref{fig:LFV-mass-ratio} and \ref{LFV processes1}. The colour of the points denotes the value of  $\log_{10}(|Y_{\Delta,LQ}^{11}|)$, as indicated under the plots. 
  %The  dashed line corresponds to setting the combinations of couplings appearing in \eref{eq:LFVest} to their fitted central values, and the gray lines indicate the present and future experimental limits as in \tref{tab:lfv}. 
  Lines as in \fref{LFV processes1}.
  }
  \label{LFV processes2}
  \end{figure}

The figure also displays $\text{BR}(\mu\rightarrow eee)/ \text{BR}(K_L\rightarrow \mu e)$ (blue points). In the latter case, the correlation is much less pronounced since the two processes depend on different combinations of the coupling matrices (see \eref{eq:LFVest}), which are in turn affected by the uncertainty stemming from the neutrino parameters in \eref{eq:mnu}. To produce \fref{fig:LFV-mass-ratio}, we employ the fits provided in Refs.~\cite{Esteban:2020cvm,nufit} for the mixing angles and neutrino mass differences, while the poorly-constrained Dirac phase and the unknown Majorana phases of the PMNS have been uniformly varied within $[0,2\pi)$, and we scanned the value of the lowest neutrino mass (assuming normal hierarchy) in the range $0.001\,\text{eV} \le m_1 \le 0.1\,\text{eV}$. The logarithm of the absolute strength of the coupling was varied uniformly in the range $-4.5 \le \log_{10}(|Y_{\Delta}^{11}|) \le -2$. 

The same choice of parameters has been employed to generate the plots of \fref{LFV processes1}, where the rates of $\mu \to eee$, $\mu\,\textrm{Al}\to e\,\textrm{Al}$ and $K_L \to \mu e$ are compared to the present bounds and future experimental sensitivities reported in \tref{tab:lfv}.  
As we can see, a large portion of the parameter space is already excluded and substantially more is within the sensitivity of the upcoming experiments, in particular Mu3e~\cite{Blondel:2013ia} and Mu2e/COMET~\cite{Bartoszek:2014mya,Kuno:2013mha}. Therefore, unless the overall size of the Yukawa couplings is considerably smaller than the range we considered, the spectrum of the model favoured by gauge coupling unification will likely provide positive LFV signals and, as Figures~\ref{fig:LFV-mass-ratio} and \ref{LFV processes1} show, such measurements would pinpoint the mass ratio $M_{LQ}/M_{\Delta}$ (besides measuring $|Y_{\Delta,\,LQ}^{21}|  |Y_{\Delta,\,LQ}^{11}| / M_{\Delta,\,LQ}^2$, that is, the coefficients of the LFV operators induced by a triplet or a leptoquark exchange). 

Of course, cleaner correlations are observed when considering pairs of processes induced by the same field, as shown in \fref{LFV processes2}. The first plot displays $\mu\to eee$ and $\mu\to e\gamma$, that is, processes due to the triplet $\Delta$.  The other three panels depict processes that are mediated by the leptoquark $\widetilde{R_2}$. These plots also show how present and future experimental bounds can constrain the overall value of the Yukawa couplings. 

The spread of the points in  \fref{LFV processes2} follows from the different combinations of the couplings relevant for different processes, as illustrated in \eref{eq:LFVest}, and thus is entirely due to the present uncertainty on the neutrino parameters in \eref{eq:mnu}. 
This is a clear indication that measuring the rates of different LFV modes mediated by the same state from $\phi_{\bf 15}$ would provide 
precious information on the neutrino parameters beyond that that is currently available from the observation of neutrino oscillations and other neutrino experiments. However, the prospects of this programme do not seem very good in the case of the processes induced by $\Delta$: the first panel of \fref{LFV processes2} indeed shows that it is unlikely to observe $\mu \to e\gamma$ given the present constraint on $\mu \to eee$, which is a general feature of type II seesaw models irrespective of their possible GUT embedding.\footnote{This conclusion can be relaxed in specific cases where $m_\nu$ (and thus $Y_\Delta$) features texture zeroes (see e.g.~\cite{Zhou:2015qua} for an assessment of such a possibility), for instance as a consequence of a flavour symmetry. In this kind of scenarios, one may envisage the possibility that either $Y_{\Delta}^{21} = 0$ or $Y_{\Delta}^{11} = 0$ at some high-energy scale related to new flavour dynamics and that the vanishing entry is only radiatively generated by running the matrix down to $M_\Delta$ through the RGEs shown in \aref{app:LQrge}. From \eref{eq:LFVest} we see that this would suppress $\mu\to eee$ and make $\mu\to e\gamma$ comparatively more constraining. In the following, we do not further entertain a situation of this kind. 
} 
Similarly, charged kaon decays (as shown in the second row of the figure) are not as promising as $K_L \to e\mu$.
This latter mode, in combination with $\mu\to e$ in nuclei (see the second plot of \fref{LFV processes2}), seems instead to offer a suitable option to probe the flavour structure of $Y_{LQ}$, and thus of $m_\nu$\,---\,especially if future experiments will be able to probe it substantially below the $10^{-12}$ level. Needless to say, a direct connection of these leptoquark-induced processes to the neutrino sector is  possible only in presence of an underlying GUT structure as the one we are considering here (and it would be a crucial indication thereof). 

\begin{figure}[t!]
    \centering
    \includegraphics[width=1.\textwidth]{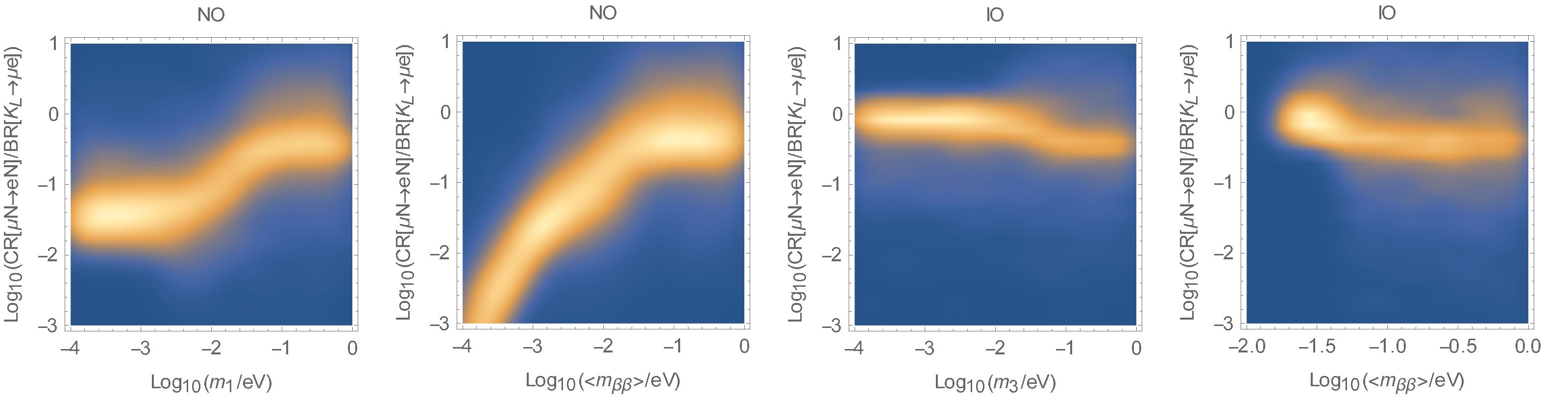}
    \caption{Ratio $\text{CR}(\mu \,\textrm{Al}\rightarrow e\,\,\textrm{Al})/\text{BR}(K_L\rightarrow \mu e)$ as a function of the lowest neutrino mass and the effective Majorana neutrino mass obtained by marginalising over the other neutrino parameters. Left: normal ordering (NO). Right: inverted ordering (IO).}
    \label{fig:m1}
\end{figure}

We start studying the dependence of the ratio $\text{CR}(\mu \,\textrm{Al}\rightarrow e\,\,\textrm{Al})/\text{BR}(K_L\rightarrow \mu e)$ on the parameters of the neutrino mass matrix $m_\nu$.
In \aref{app:nupar}, we show the standard paramerisation that we employ for the PMNS matrix appearing in \eref{eq:mnu} and the dependence of this ratio of LFV rates on the 9 parameters of the neutrino sector. 
In particular, \fref{allparameters} shows that measuring or constraining $\text{CR}(\mu \,\textrm{Al}\rightarrow e\,\,\textrm{Al})/\text{BR}(K_L\rightarrow \mu e)$ would not provide useful information on the oscillation parameters, that is, the PMNS mixing angles, the neutrino mass splittings, and the Dirac CP-violating phase. In contrast, this ratio is very sensitive to parameters that are so far unknown: the two Majorana phases $\alpha_{21}$ and $\alpha_{31}$, and the absolute neutrino mass $m_\textrm{min}$,\footnote{As customary, $m_\textrm{min} = m_1$ for the normal mass ordering (NO), $m_1 < m_2 < m_3$, and $m_\textrm{min} = m_3$, for the inverted mass ordering (IO), $m_3 < m_1 < m_2$.}
hence we focus here on these interesting quantities.

\begin{figure}[t!]
    \centering
    \includegraphics[width=1.\textwidth]{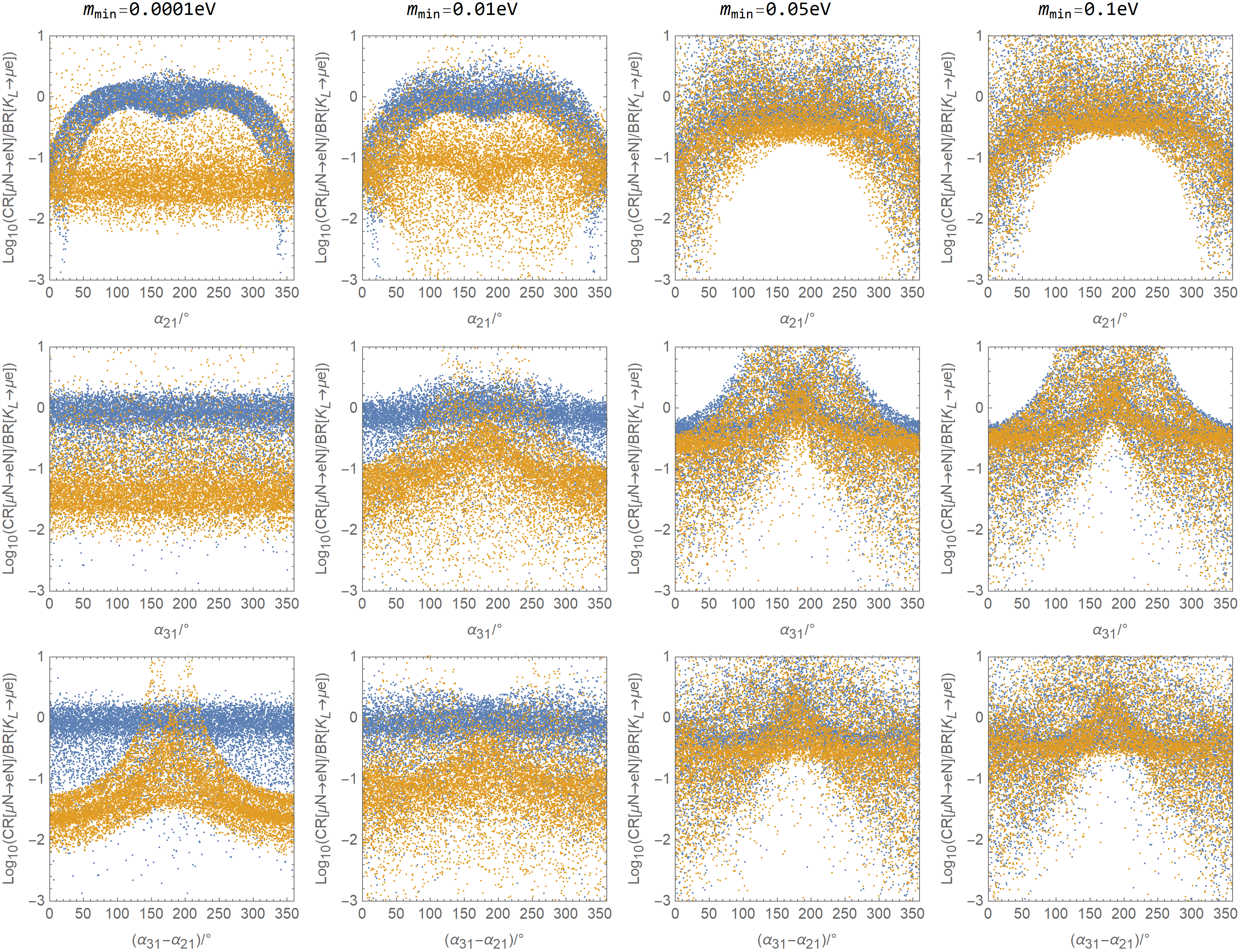}
    \caption{Ratio $\text{CR}(\mu \,\textrm{Al}\rightarrow e\,\,\textrm{Al})/\text{BR}(K_L\rightarrow \mu e)$ as a function of the PMNS Majorana phases $\alpha_{21}$ (first row), $\alpha_{31}$ (second row), and the combination $(\alpha_{31}-\alpha_{21})$ (third row) obtained by marginalising over the other neutrino parameters~\cite{Esteban:2020cvm} for different values of the lowest neutrino mass $m_\textrm{min}$. The NO case is denoted by orange points, the IO case by blue points.}
    \label{fig:Maj1}
\end{figure}

In \fref{fig:m1}, we display the dependence of $\text{CR}(\mu \,\textrm{Al}\rightarrow e\,\,\textrm{Al})/\text{BR}(K_L\rightarrow \mu e)$ on $m_\textrm{min}$ and the effective Majorana neutrino mass $\langle m_{\beta\beta} \rangle \equiv | \sum_{i} U_{ei}^2 m_i|$\,---\,where $U_{ei}$ are the first-row elements of the PMNS matrix in~\eref{PMNS}\,---\,marginalised over the other neutrino parameters. 
This shows that, in the context of our GUT models, measuring $K_L\rightarrow \mu e$ with a rate more than 10 times larger than $\mu\to e$ conversion in nuclei would strongly disfavour the inverted ordering and, more importantly, point to a light absolute mass and effective mass, $m_1,\,\langle m_{\beta\beta} \rangle \lesssim 10^{-2}$~eV, a situation rather challenging for other experimental probes such as searches for neutrinoless double-beta decays~\cite{Agostini:2022zub}.

\begin{figure}[t!]
    \centering
    \includegraphics[width=1.\textwidth]{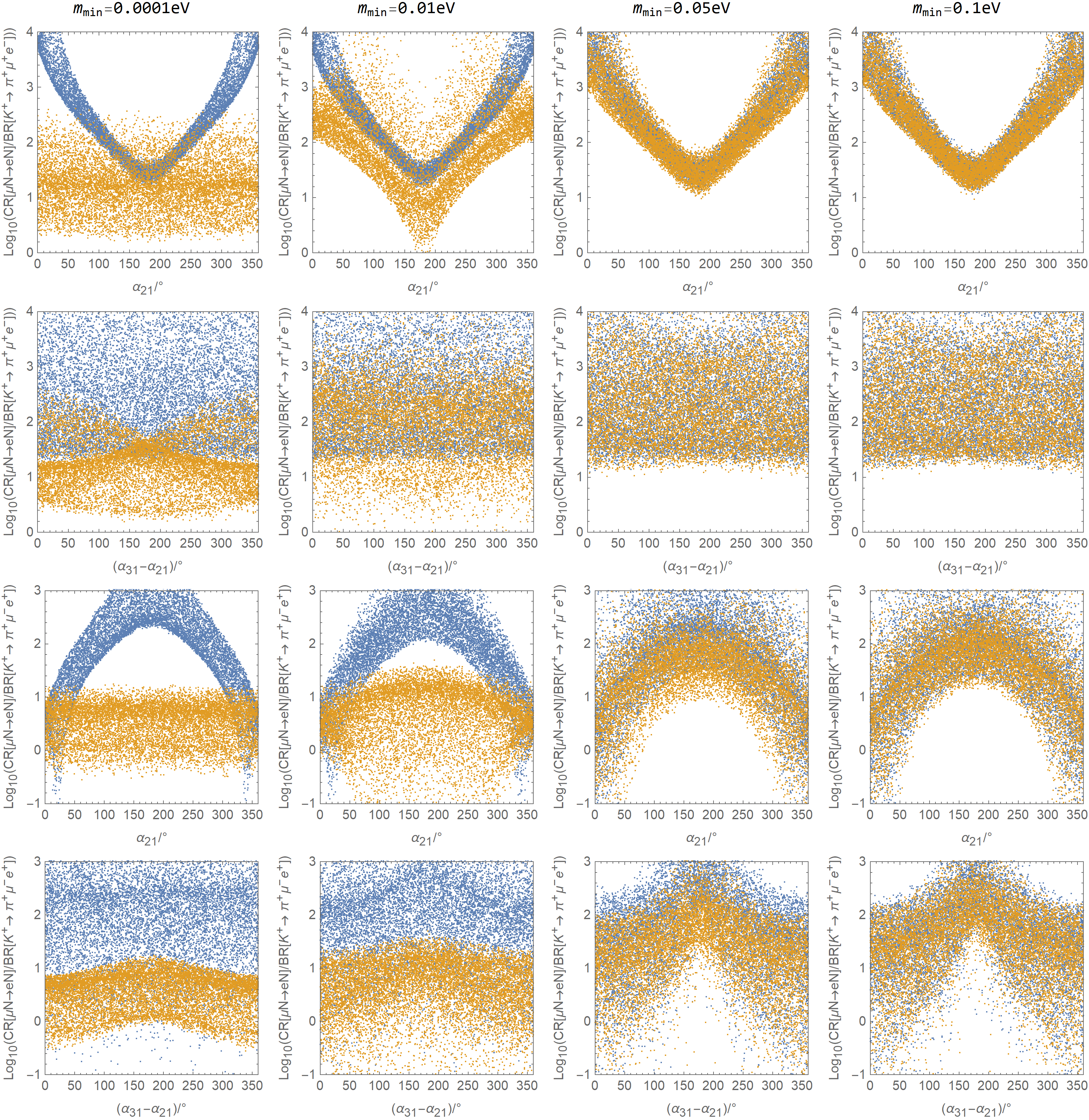}
    \caption{Same as \fref{fig:Maj1} for $\text{CR}(\mu \,\textrm{Al}\rightarrow e\,\,\textrm{Al})/\text{BR}(K^+\rightarrow \pi^+ \mu^+ e^-)$ (first and second row) and $\text{CR}(\mu \,\textrm{Al}\rightarrow e\,\,\textrm{Al})/\text{BR}(K^+\rightarrow \pi^+ \mu^- e^+)$ (third and fourth row). } 
    \label{Fig:Maj2}
\end{figure}

In \fref{fig:Maj1}, we plot $\text{CR}(\mu \,\textrm{Al}\rightarrow e\,\,\textrm{Al})/\text{BR}(K_L\rightarrow \mu e)$ as a function of the Majorana phases for different values of $m_\textrm{min}$. 
These plots show how, within an underlying GUT structure, the comparison of these two LFV processes can shed light on the unknown Majorana phases, especially in presence of a future determination of (or a more stringent constraint on) $m_\text{min}$.
As an example, we can see that, for a relatively sizeable $m_\text{min}$, 
$\text{CR}(\mu \,\textrm{Al}\rightarrow e\,\,\textrm{Al}) \ll \text{BR}(K_L\rightarrow \mu e)$ would require $\alpha_{21}$ to be quite close to 0. On the contrary, $\text{CR}(\mu \,\textrm{Al}\rightarrow e\,\,\textrm{Al}) \gg \text{BR}(K_L\rightarrow \mu e)$ would point to values of $\alpha_{31}$ not far from $\pi$. One can also see one of the phases becoming unphysical in the opposite limit $m_\text{min} \to 0$. 

Following from \eref{eq:LFVest}, the results of Figures~\ref{fig:m1} and~\ref{fig:Maj1} can be traced back to the change in the relative size of $Y_{LQ}^{11}$, $Y_{LQ}^{12}$, and $Y_{LQ}^{22}$ for different values of $m_\textrm{min}$, $\alpha_{21}$ and $\alpha_{31}$. 
One can hence expect to obtain an even better sensitivity on these parameters by comparing $\mu\to e$ conversion in nuclei with charged kaon LFV modes, since the latter processes feature a simpler dependence on the couplings than $K_L$, without in particular interference terms, cf.~\eref{eq:3body}.
This is indeed shown by \fref{Fig:Maj2} where similar results for
\mbox{$\text{CR}(\mu \,\textrm{Al}\rightarrow e\,\,\textrm{Al})/\text{BR}(K^+\rightarrow \pi^+ \mu^+ e^-)$} and 
\mbox{$\text{CR}(\mu \,\textrm{Al}\rightarrow e\,\,\textrm{Al})/\text{BR}(K^+\rightarrow \pi^+ \mu^- e^+)$} are displayed. As one can see, the complementarity of $K^+\rightarrow \pi^+ \mu^+ e^-$ and $K^+\rightarrow \pi^+ \mu^- e^+$ in constraining $\alpha_{21}$ is particularly pronounced. However, fully exploiting the interplay of different kaon LFV modes would require a future search campaign able to reach sensitivities substantially below $10^{-12}$, as shown by the second row of \fref{LFV processes2}.

%\newpage

%%%%%%%%%%%%%%%%%%%%%%%%%%%%%%%

\section{Summary and conclusions}
\label{sec:conc}

In this article, we have revisited a class of $SU(5)$ GUT models with minimal field contents (see \tref{tab:particles}) that allow for successful unification of the gauge couplings and account for the origin of neutrino masses via type II seesaw. In \sref{sec:models}, we classified our models based on how realistic fermion masses are achieved, studied their spectrum compatible with unification and $p$-decay constraints in \sref{sec:fit}, and finally discussed in detail their observable consequences in terms of LFV decays in \sref{sec:lfv}. 

The main findings of our study can be summarised as follows. 
\begin{itemize}
    \item The minimal $SU(5)$ setup with non-renormalisable interactions (``Model 1'') is excluded by proton decay searches, barring the case of fine cancellations triggered by a very peculiar flavour structure of the Yukawa couplings, hence it is strongly disfavoured, see \sref{sec:model1}.
    \item For what concerns models featuring vector-like matter (``Model 3''), we separately considered the case of a single $\bf 5 \oplus \mathbf{\overline{5}}$ fermionic representation and that with a single $\bf 10 \oplus \mathbf{\overline{10}}$. The former case is also strongly disfavoured by proton decay but it may become viable if multiple generations\,---\,at least 5\,---\,of vector-like fermions are introduced. The latter case is instead viable in its simplest form. However, the model is not predictive as the constraints on its spectrum are very loose and, in particular, no field is required to be light for the sake of unification and proton decay, cf.~\sref{sec:model3}.
    \item  The model with an additional scalar $\bf 45$ and renormalisable interactions (``Model 2'') can successfully achieve unification with a long enough proton lifetime and, especially in its minimal realisations, features very interesting predictions, as discussed at length in~\sref{sec:model2}. Several fields are required to be light (that is, not much above the TeV scale), in particular the type II seesaw fields in the $\bf 15$ representation that mediate LFV interactions. 
   \item The couplings of these fields (the seesaw triplet $\Delta$ and its $SU(5)$ partner, the scalar leptoquark $\widetilde{R_2}$) to SM fermions are linked to one another by the $SU(5)$ structure and thus their LFV effects are related. From this it follows that measuring \mbox{$\text{BR}(\mu\rightarrow eee)/ \text{CR}(\mu\,N \rightarrow e\,N)$} would pinpoint the mass ratio $M_\Delta/ M_{LQ}$, see \fref{fig:LFV-mass-ratio}. Such a measurement (or constraint, in case only one of the two LFV processes is observed) could be then confronted with unification and $p$-decay constraints (as well as information from collider searches) in order to see if a consistent picture emerge.
    \item Instead, ratios of processes mediated by the same field (the most promising being $\mu \to e$ conversion in nuclei and $K_L \to \mu e$, both due to the leptoquark) provide information on the flavour structure of the couplings and thus directly on the neutrino mass matrix (in the charged-lepton mass basis), as both matrices $Y_{\Delta}$ and $Y_{LQ}$ are proportional to $m_\nu$. We showed that ratios of different LFV branching ratios can be particularly sensitive to the neutrino parameters that can not be measured through oscillation experiments, namely the Majorana phases and the absolute mass (see Figures~\ref{fig:m1}-\ref{allparameters2}).
   \item While some of our results apply to more general extensions of the SM featuring the triplet~$\Delta$ (e.g.~to a generic type II seesaw) or the leptoquark $\widetilde{R_2}$, the connection between the processes induced by these two fields obviously requires the presence of a GUT. 
   Our results show that measuring the rates of several LFV modes may allow to collect enough evidence of such a connection and thus of an underlying GUT structure.
\end{itemize}

%%%%%%%%%%%%%%

\paragraph{Acknowledgments.}
LC is partially supported by the National Natural Science Foundation of China under the Grant No.~12035008.

%\newpage

%%%%%%%%%%%%%%%%%

\appendix

\section{Scalar potential and scalar mass spectrum}
\label{app:scalar-sector}

We discuss here the feasibility of the scalar mass spectra characterised by large mass splittings among the states belonging to the same $SU(5)$ representation that, according to the fit in \sref{sec:fit}, facilitate gauge coupling unification, and also if it is possible to achieve vev hierarchies in agreement with phenomenological requirements.
In general, while the superpotential within supersymmetric $SU(5)$ GUTs is strongly constrained by holomorphicity and renormalisability, that is not the case for a non-supersymmetric theory. In our scenarios, there are far more free parameters in the scalar potential, providing no fixed relationships among the masses of the new scalars and thus allowing (at the price of fine tunings) the large mass splittings assumed in \sref{sec:fit}.

\subsection{\texorpdfstring{$\phi_{\textbf{24}}$}{}}
In our models, the dominant terms of the scalar potential dictating the $\phi_{\textbf{24}}$ components masses and the vacuum expectation value $v_{24}$ read: 
\begin{equation}
\label{V24}
    V_{\textbf{24}}= -\frac{1}{2}m_{\textbf{24}}^2 \text{Tr}[\phi_{\textbf{24}}^2] 
    + \sqrt{\frac{10}{3}}\mu_{\textbf{24}} \text{Tr}[\phi_{\textbf{24}}^3]
    +\frac{1}{8}\lambda_1 \text{Tr}[\phi_{\textbf{24}}^2] \text{Tr}[\phi_{\textbf{24}}^2]
    +\frac{15}{2}\lambda_2\text{Tr}[\phi_{\textbf{24}}^4].
\end{equation}
The interaction terms with $\phi_{\textbf{5}}$, $\phi_{\textbf{15}}$, and $\phi_{\textbf{45}}$ are neglected here, as  $v_5,\, v_{15}(\equiv v_\Delta),\,v_{45}\ll m_{\textbf{24}}, v_{24} $ and the corresponding couplings are strongly suppressed after spontaneous symmetry breaking. \eref{V24} is in fact the same as the scalar potential in minimal $SU(5)$~\cite{Guth:1981uk,Kumericki:2017sfc}. Hence, requiring $\mu_{\textbf{24}}\xrightarrow{}-4\lambda_2v_{24}$, one can get:\footnote{For the components of $\phi_{\textbf{24}}$, we adopt here the same conventions as in Ref.~\cite{Kumericki:2017sfc}.}
\begin{equation}
    \begin{aligned}
    m_{\varrho_1}^2 &= (\lambda_1+25 \lambda_2) v_{24}^2\,, \quad
    m_{\varrho_8}^2 = 25\lambda_2 v_{24}^2\,, \quad
    m_{\varrho_3}^2 \xrightarrow{} 0\,,\\
    v_{\textsc{gut}}^2 &\equiv v_{24}^2 = \frac{2 }{\lambda_1+6 \lambda_2}m_{\textbf{24}}^2\,.
    \end{aligned}
\end{equation}
These expression imply that the hierarchy $m_{\varrho_3}\ll m_{\varrho_8}\ll v_{24}$ is achievable, as required by the results of our fit. Furthermore, one can check that, in this case, $\frac{\partial^2 V_{\textbf{24}}}{\partial \phi_{\textbf{24}}^2} = \frac{2\lambda_1+20\lambda_2}{\lambda_1+6\lambda_2}\, m_{\textbf{24}}^2$ can be positive, so that $v_{24}$ is really a local minimum. 

\subsection{\texorpdfstring{$\phi_{\textbf{5}}$}{} and \texorpdfstring{$\phi_{\textbf{45}}$}{}}

Both $\phi_{\textbf{5}}$ and $\phi_{\textbf{45}}$ contain a SM-like Higgs doublet (respectively, $H$ and $H_2$) and, due to ${\bf 45} \otimes \bar{\bf 5}  \supset {\bf 24}$ the $H$-$H_2$ mixing term also exists. At low energies, this is a generic 2HDM where the masses of the heavy states are all free parameters, see e.g.~\cite{Branco:2011iw}.
For instance, the mass terms for the two neutral CP-even states are given by: 
\begin{equation}
    V(h_1, h_2) = \frac{1}{2}(m_{11}^2h_1^2+m_{22}^2h_2^2-2 m_{12}^2 h_1 h_2) +  \text{quadratic terms}\,,
\end{equation}
where $m_{ij}$ are in general all at the GUT scale, if one does not invoke fine tuning. Then, the two local minima lie on:
\begin{equation}
    \begin{aligned}
    v_{5}^2 &\sim   \mathcal{O}(\lambda^{-1})\times  (m_{11}^2-m_{12}^2 \tan\beta)\,,\\
    v_{45}^2 &\sim  \mathcal{O}(\lambda^{-1})  \times  (m_{22}^2-m_{12}^2 \cot\beta)\,, \\
    \end{aligned}
\end{equation}
 where $\tan\beta=v_{45}/v_{5}$ and all the quadratic coupling strengths are assumed to be at $\mathcal{O}(\lambda)$ for simplicity. As $v_{45}^2+v_{5}^2 = v_{\text{EW}}^2$ and $v_{45}\sim v_{5}$, one gets the following mass matrix for $h_1, h_2$:
\begin{equation}
    M_{h_1 h_2}=
    \left(
    \begin{array}{cc}
         \tan\beta & -1 \\
        -1 & \cot\beta
    \end{array}
    \right)m_{12}^2
    + 
    \left(
    \begin{array}{cc}
        \lambda_{11} & \lambda_{12} \\
        \lambda_{21} & \lambda_{22}
    \end{array}
    \right)
    v_\text{EW}^2\,.
\end{equation}
The contribution from the quadratic terms (due to EW-symmetry breaking) is taken into account in $\lambda_{ij} v_\text{EW}^2$, where $\lambda_{ij}$ are in general all $\mathcal{O}(\lambda)$. $M_{h_1 h_2}$ has two eigenvalues: $m_{12}^2/(\cos\beta\sin\beta)\sim v_\textsc{gut}^2 $ and $\mathcal{O}(\lambda) v_\text{EW}^2$, corresponding to the squared mass of $h_2$ and $h_1$ individually. Therefore, due to the mixing, one can get the desired hierarchy between $m_{h_2}$ and its vev $v_{45}$. 

Strictly speaking, $\varphi_{3}^S$ in $\phi_{\textbf{45}}$ could mix with the scalar triplet in $\phi_{\textbf{5}}$, but the mixing angle is independent of $\beta$ so that both mass eigenstates can still be at GUT scale. This is because $v_{24}$ is large and the cubic and quadratic interaction terms with $\phi_{\textbf{24}}$ should not be neglected here, providing more free parameters.\footnote{The coupling to $\phi_{\textbf{15}}$ is not dominant as custodial symmetry requires $v_{15}\ll v_{EW}$.} Due to the same reason, the masses of the components of $\phi_{\textbf{5}}$ and $\phi_{\textbf{45}}$ are all independent, which means that a light $\varphi_8$, as required by the fit, is realisable without extending the model. For the explicit expressions, we refer to Section 4.1 of Ref.~\cite{Kumericki:2017sfc}.

\subsection{\texorpdfstring{$\phi_{\textbf{15}}$}{}}

The dominant scalar potential terms for the masses of the components of $\phi_{\textbf{15}}$ are: 
\begin{equation}
\begin{aligned}
    V_{\textbf{15}}= &-\frac{1}{2} m_{\textbf{15}}^2 \text{Tr}[\phi_{\textbf{15}}\phi_{\textbf{15}}^*]
    +\mu_{\textbf{15}} \text{Tr}[\phi_{\textbf{15}}\phi_{\textbf{15}}^*\phi_{\textbf{24}}] +\\
    &b_1\text{Tr}[\phi_{\textbf{15}}\phi_{\textbf{15}}^*]\text{Tr}[\phi_{\textbf{24}}\phi_{\textbf{24}}]
    +30b_2\text{Tr}[\phi_{\textbf{15}}\phi_{\textbf{15}}^*\phi_{\textbf{24}}\phi_{\textbf{24}}]
    +15b_3\text{Tr}[\phi_{\textbf{15}}\phi_{\textbf{24}}\phi_{\textbf{15}}^*\phi_{\textbf{24}}]\,,
\end{aligned}
\end{equation}
Again, when $m_{\Delta},m_{\widetilde{R_2}}\gg v_\text{EW}$, the interactions with $\phi_{\textbf{5}}$, $\phi_{\textbf{45}}$ or the $\phi_{\textbf{15}}$ quadratic couplings can be neglected. After $SU(5)$ breaking, the mass spectrum reads:\footnote{$
\phi_{\textbf{15}}=
\left(
\begin{array}{cc}
    S & \widetilde{R_2}/\sqrt{2} \\
    \widetilde{R_2}/\sqrt{2}  & i\sigma_2\Delta
\end{array}
\right)
$, with $\Delta$ and $\widetilde{R_2}$ shown in~\eref{SU2sym}.}
\begin{equation}
    \begin{aligned}
    m_{\Delta}^2&= -m_{\textbf{15}}^2+6 \mu_{\textbf{15}} v_{24}+ 2 b_1 v_{24}^2+18 b_2 v_{24}^2+9 b_3 v_{24}^2\,,\\
    m_{\widetilde{R_2}}^2&= -m_{\textbf{15}}^2+\mu_{\textbf{15}} v_{24}+2 b_1 v_{24}^2+13 b_2 v_{24}^2-6 b_3 v_{24}^2\,,\\
    m_{S}^2&= -m_{\textbf{15}}^2-4 \mu_{\textbf{15}} v_{24}+2 b_1 v_{24}^2+8 b_2 v_{24}^2+4 
    b_3 v_{24}^2\,.\\
    \end{aligned}
\end{equation}
Resorting to fine tuning, $0< m_{\Delta}^2\sim m_{\widetilde{R_2}}^2 \ll m_{S}^2\sim v_{24}^2$ is possible, as assumed in the rest of the paper. Furthermore, the cubic term in~\eref{eq:DeltaL} gives a non-zero vacuum expectation value: $v_{\Delta}\sim \mu  v_5^2/(\sqrt{2}m_{\Delta}^2)$, which is small (suppressed by $m_{\Delta}$) as desired.

\color{black}

%%%%%%%%%%%%%%%%%%%%%%%%%%%%%%%%%%%%%%

\section{More details on proton decay}
\label{app-pdecay}

\subsection{Renormalisation of the baryon-number-violating operators}
The renormalisation factor appearing in Eqs.~(\ref{eq:p-to-e}-\ref{eq:p-to-K}) is given by $A=A_{LD} A_{SD}$, where $A_{LD}$ and $A_{SD}$ account for the long-distance and short-distance running of the baryon-number-violating operators, respectively, see e.g.~\cite{Nath:2006ut}.
The former one corresponds to the QCD running from $m_t$ to the proton mass scale:
\begin{align}
  A_{LD} = \left( \frac{\alpha_3 (m_p)}{\alpha_3 (m_c)} \right)^{2/9}
     \left( \frac{\alpha_3 (m_c)}{\alpha_3 (m_b)} \right)^{6/25}
     \left( \frac{\alpha_3 (m_b)}{\alpha_3 (m_t)} \right)^{6/23} ~ \approx ~ 1.5\,.
\end{align}
The short-distance contribution encodes the renormalisation of the operators from the GUT scale down to $m_t$. This can be given in terms of the running of the gauge couplings, that is, in terms of the SM $\beta$-function coefficients plus the contribution of the extra fields:
\begin{align} \label{eq:ASD}
  A_{SD} = &
    \left( \frac{\alpha_1 (m_t)}{\alpha_1 (M_I)} \right)^{\! - \frac{23}{30 b^\textsc{sm}_1}}
    \left( \frac{\alpha_2 (m_t)}{\alpha_2 (M_I)} \right)^{\! - \frac{3}{2 b^\textsc{sm}_2}}
    \left( \frac{\alpha_3 (m_t)}{\alpha_3 (M_I)} \right)^{\! - \frac{4}{3 b^\textsc{sm}_3}}  \nonumber \\
&   \times  \left( \frac{\alpha_1 (M_I)}{\alphaU} \right)^{\! - \frac{23}{30 (b^\textsc{sm}_1 + \Delta b_1)}}
    \left( \frac{\alpha_2 (M_I)}{\alphaU} \right)^{\! - \frac{3}{2 (b^\textsc{sm}_2+ \Delta b_2)}}
    \left( \frac{\alpha_3 (M_I)}{\alphaU} \right)^{\! - \frac{4}{3 (b^\textsc{sm}_3+ \Delta b_3)}},
\end{align}
where we considered the extra matter at a single intermediate scale $M_I$ (with $\Delta b_i \equiv \sum_I b_i^I$), the generalization to multiple thresholds being straightforward. To obtain an estimate of the typical value of $A_{SD}$, one can consider only the contributions from $b^\textsc{sm}_i$ in the above equation and take $\alphaU = 1/25$, which gives $A_{SD} \approx 1.3$.

\subsection{Proton decay matrix elements}
A recent lattice QCD evaluation of the hadronic matrix elements in Eqs.~(\ref{eq:p-to-e}-\ref{eq:p-to-K}) gives~\cite{Aoki:2017puj}:
\begin{equation}
\begin{aligned}
& \langle \pi^0 | (ud)_R u_L |p\rangle ~= -0.131(4)(13)\,\text{GeV}^2\,, 
&\langle \pi^0 | (ud)_L u_L |p\rangle ~= 0.134(5)(16)\,\text{GeV}^2\,, \\
& \langle K^0 | (us)_R u_L |p\rangle ~= 0.103(3)(11)\,\text{GeV}^2\,, 
&\langle K^0 | (us)_L u_L |p\rangle ~= 0.057(2)(6)\,\text{GeV}^2\,, \\
& \langle \pi^+ | (du)_R d_L |p\rangle ~= -0.186(6)(18)\,\text{GeV}^2\,, & \\
& \langle K^+ | (us)_R d_L |p\rangle  ~= -0.049(2)(5)\,\text{GeV}^2\,,  
&\langle K^+ | (ud)_R s_L |p\rangle ~= -0.134(4)(14)\,\text{GeV}^2 \,.
\end{aligned}
\end{equation}

\subsection{Other possible contributions to proton decay}

As neutrinos are Majorana particles following the type II seesaw mechanism, the proton could also decay to a meson and a lepton (instead of an antilepton), thus breaking ${B-L}$. However, these processes are strongly suppressed by the small neutrino mass, as we will show below.

According to \eref{eq:typeIIint}, the leptoquark $\widetilde{R_2}$ can convert a down-type quark to a lepton and induce ${B-L}$ violating proton decay processes (such as $p \to \pi^+ \nu$) by interacting with other scalar fields via $\mu \phi_{\bf 5}^i \phi_{\bf 5}^j (\phi_{\bf 15}^*)_{ij}$. The resulting $d=7$ effective operator, which has been already discussed in Ref.~\cite{Dorsner:2005fq}, reads:
\begin{equation}
    {\cal O}_{d=7}=\frac{\mu Y_{d\ell}Y_{LQ}^{\dagger}}{M_T^2 M_{LQ}^2} \overline{U_R^c}\, D_R \, \overline{L_L}\,D_R\, H^*.
\end{equation}
where $M_T$ is the mass of the colour triplet $(\mathbf{\bar 3},{\bf 1},-1/3)$ in $\phi_{\textbf{5}}$ that also generates the standard $B-L$ conserving contributions to proton decay via the $d=6$ operator: 
\begin{equation}
    {\cal O}_{d=6}=\frac{Y_{d\ell}Y_{d\ell}^{\dagger}}{M_T^2} \overline{U_R^c}\, D_R \, \overline{L_L^c}\,Q_L.
\end{equation}
Hence, in order to make sure that the processes induced by ${\cal O}_{d=7}$ are as suppressed as the ones from ${\cal O}_{d=6}$, we need to require that $\mu\, v_{\text{EW}} Y_{LQ}/M_{LQ}^2 \lesssim Y_{d\ell}$.
Relating these parameters to the effective Majorana neutrino mass $\langle m_{\beta\beta} \rangle \sim Y_{\Delta} v_{\Delta}\sim  Y_{\Delta} \frac{\mu\, v_\text{EW}^2}{M_{\Delta}^2}\sim  Y_{LQ} \frac{\mu\, v_\text{EW}^2}{M_{\Delta}^2}$ and to the charged lepton masses $m_{\ell}\sim v_\text{EW} Y_{d\ell}$, the condition becomes: 
\begin{equation}
\label{condi}
    \langle m_{\beta\beta} \rangle \times \frac{M_{\Delta}^2}{ M_{LQ}^2}  \lesssim  m_{\ell} \,.
\end{equation}
According to the Bayesian analysis in \sref{sec:model2}, TeV-scale masses for both $\Delta$ and $\widetilde{R_2}$ are favoured, then Eq.~(\ref{condi}) is always verified. In other words, the smallness of the absolute neutrino mass further suppresses the $B-L$ violating processes relative to the ordinary proton decay induced by the colour triplet. Notice that, taking $\langle m_{\beta\beta} \rangle =0.1$~eV and $m_\ell = m_e$, the above condition is still fulfilled up to $M_\Delta \approx 2000\times M_{LQ}$.
For values of $M_\Delta$ larger than that, the induced proton decay could still evade the experimental bounds, considering that the processes mediated by the colour triplet with $M_T = M_\textsc{gut}$ are suppressed compared to those induced by the $SU(5)$ gauge bosons, since the Yukawa couplings in $Y_{d\ell}$ and $Y_{LQ}$ are much smaller than the unified gauge coupling strength\,---\,the latter couplings being severely constrained by LFV processes, as discussed in \sref{sec:lfv}.

%%%%%%%%%%%%%%%%%%%%%%%%%%%%%%%%%%%%%

\section{RGEs of triplet and leptoquark Yukawa couplings}
\label{app:LQrge}

The 1-loop RGEs for the \eref{eq:typeIIint} interactions related to type II seesaw are given by~\cite{Schmidt:2007nq,Bandyopadhyay:2021kue}:
\begin{align}
\label{eq:yrunning1}
(4\pi)^2 \frac{d Y_\Delta}{dt}~=~& \left[-\frac{9}{10}g_1^2-\frac{9}{2}g_2^2+\text{Tr}\left(Y_{\Delta}^{\dagger}Y_{\Delta} \right)\right] Y_\Delta + \nonumber\\
& \left[3 \,Y_{\Delta}Y_{\Delta}^{\dagger}Y_{\Delta}+\frac12\,Y_{\ell}^TY_{\ell}^*Y_{\Delta}+\frac12\,Y_{\Delta}Y_{\ell}^{\dagger}Y_{\ell} \right]\,, \\
    (4\pi)^2 \frac{d Y_{LQ}}{dt} ~=~& \left[-\frac{13}{20}g_1^2-\frac{9}{4}g_2^2-4g_3^2+\text{Tr}\left(Y_{LQ}^{\dagger}Y_{LQ} \right)\right] Y_{LQ} + \nonumber\\
    & \left[\frac{5}{2}\, Y_{LQ}Y_{LQ}^{\dagger}Y_{LQ}+Y_{d}Y_{d}^{\dagger}Y_{LQ}+\frac12\,Y_{LQ}Y_{\ell}^{\dagger}Y_{\ell} \right]\,, 
\label{eq:yrunning2}
\end{align}
where $t\equiv \ln (\mu /m_Z)$.
As one can see, $Y_{LQ}$ runs more than $Y_{\Delta}$ due to the term $\propto g_3^2$, reflecting the fact that the leptoquark is strongly interacting while the triplet $\Delta$ is colour neutral. As a consequence the GUT relation $Y_{LQ} = \sqrt2 \,Y_\Delta$ does not hold at lower energies.
On the other hand, notice that the flavour structure of the matrices can be only changed by the cubic terms $\propto Y_{\Delta}Y_{\Delta}^{\dagger}Y_{\Delta}$ and $Y_{LQ}Y_{LQ}^{\dagger}Y_{LQ}$. For the light spectra we are interested in, LFV processes constrain the entries of $Y_\Delta$ and $Y_{LQ}$ to be rather small, as discussed in \sref{sec:results}, hence the effect of these terms is negligible and the flavour structure of the two matrices will remain the same at all scales to a very good approximation.

Solving the above RGEs for a typical Model 2 spectrum, $M_{\varrho_3}=M_{\Delta}=M_{LQ}=M_{\varphi_8}=1~\text{TeV}$ (implying $M_{\varrho_8}\approx1.3\times10^{15}~\text{GeV}$, $\mgut\approx2.5\times10^{15}~\text{GeV}$, 
$\alphaU^{-1}\approx 33$), one obtains $Y_{LQ}(M_{LQ})/Y_{LQ}(\mgut)\approx2.4$ and $Y_{\Delta}(M_{\Delta})/Y_{\Delta}(\mgut)\approx1.6$, which results in the low-energy relation $Y_{LQ}\approx 2.1\, Y_{\Delta}$. 
Since this result is not much affected by details of the spectrum, we employed this constant factor in the numerical analysis of \sref{sec:lfv}.

%%%%%%%%%%%%%%%%%%%%%%%%%%%%%%%%%%%%%

\section{Dependence of the LFV rates on neutrino parameters}
\label{app:nupar}

\begin{figure}[t!]
    \centering
    \includegraphics[width=0.9\textwidth]{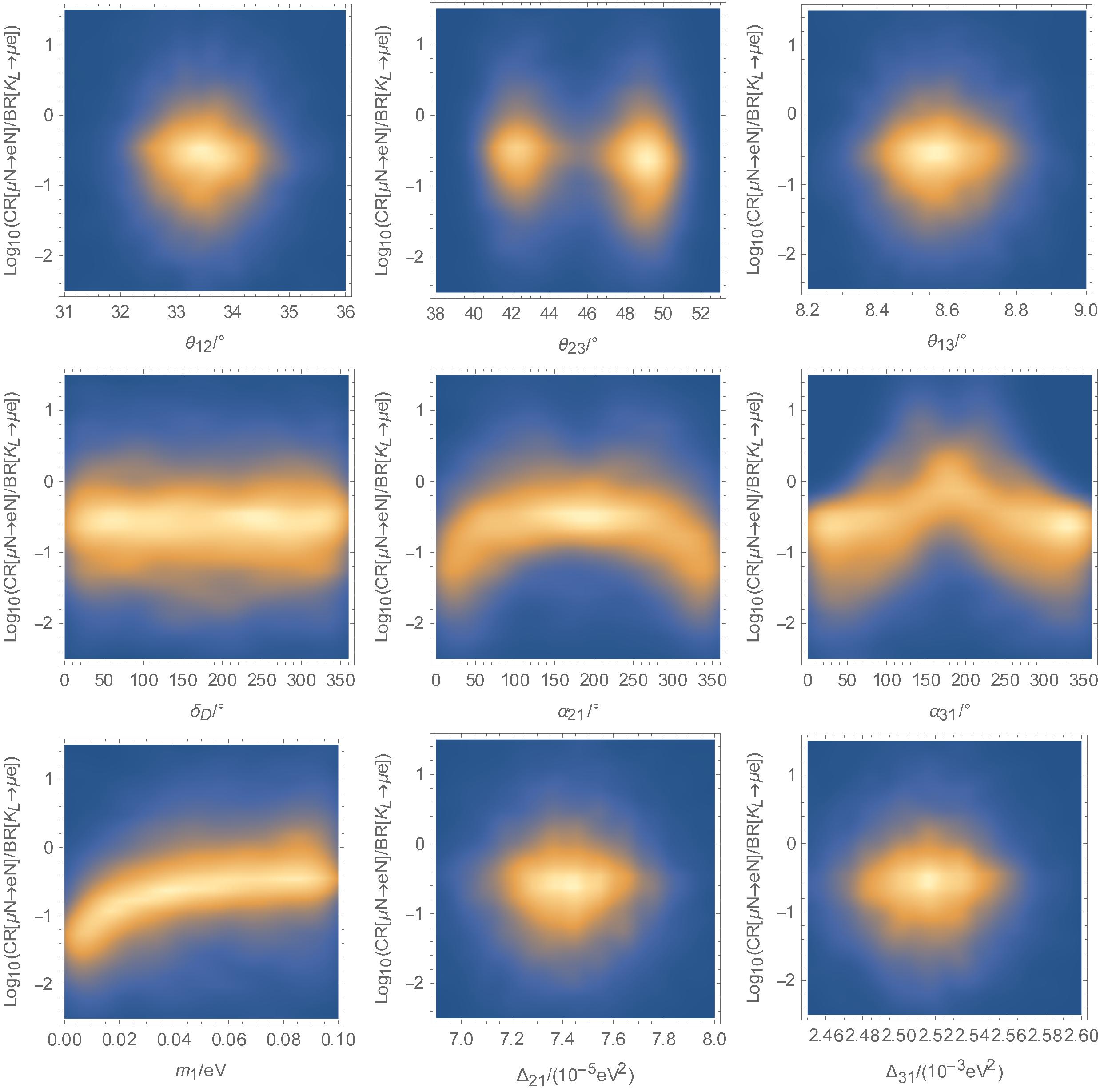}
    \caption{Dependence of $\text{CR}(\mu \,\textrm{Al}\rightarrow e\,\,\textrm{Al})/\text{BR}(K_L\rightarrow \mu e)$ on each the 9 neutrino sector parameters defined in \aref{app:nupar} marginalised over the other parameters for the NO case. The results of the fit for the three oscillation angles and the two mass splittings reported in~\cite{Esteban:2020cvm,nufit} have been employed here.}
    \label{allparameters}
\end{figure}

\begin{figure}[t!]
    \centering
    \includegraphics[width=0.9\textwidth]{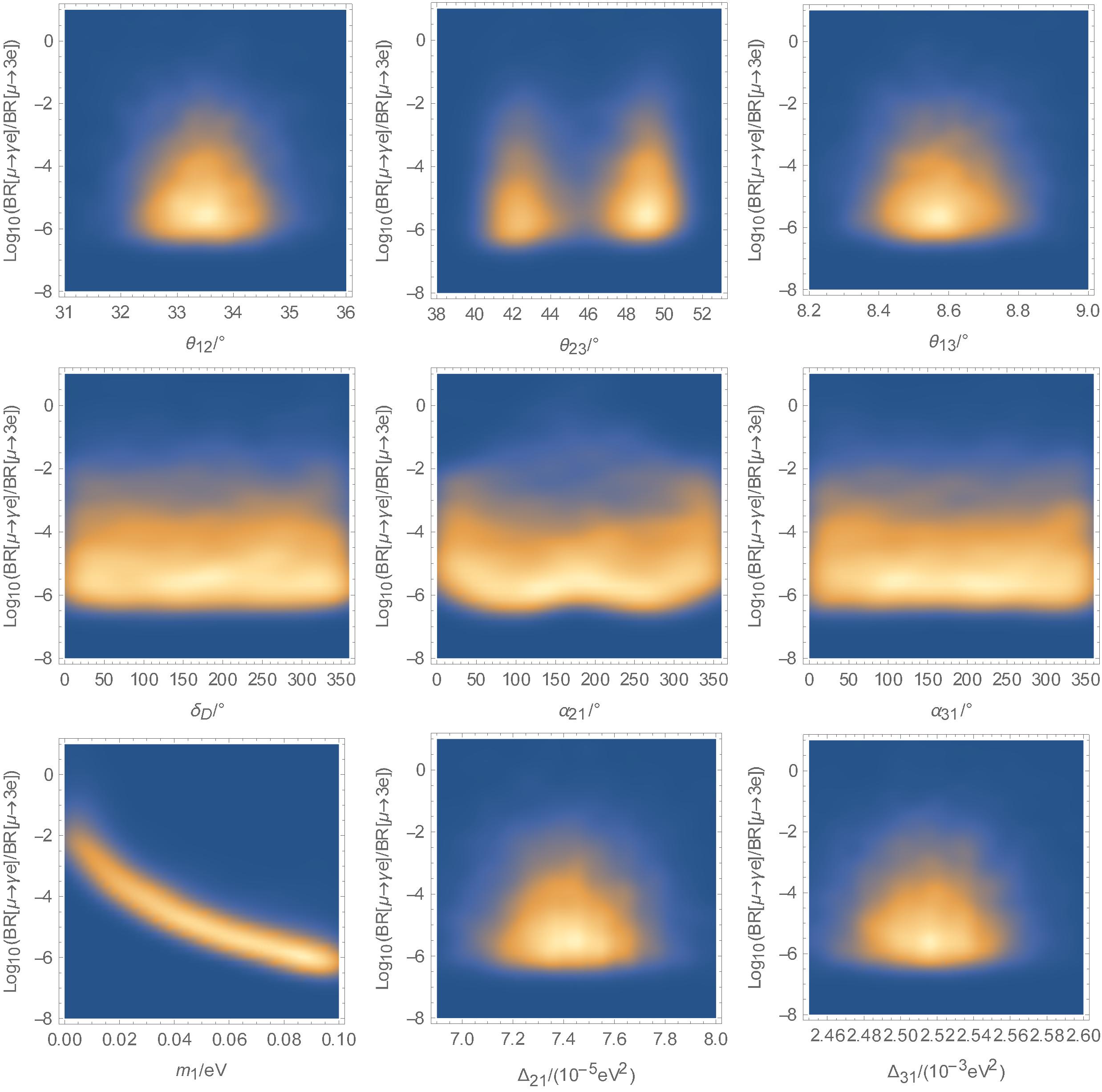}
    \caption{Same as \fref{allparameters} for $\textrm{BR}(\mu\to e\gamma)/\textrm{BR}(\mu\to eee)$.}
    \label{allparameters2}
\end{figure}

The PMNS matrix in \eref{diag} reads: 
\begin{equation}
\label{PMNS}
  U_\textsc{pmns}= 
  \left(
\begin{array}{ccc}
 1 & 0 & 0 \\
 0 & c_{23} & s_{23} \\
 0 & -s_{23} & c_{23} \\
\end{array}
\right)
\left(
\begin{array}{ccc}
 c_{13} & 0 & e^{-i \delta _D} s_{13} \\
 0 & 1 & 0 \\
 -e^{i \delta _D} s_{13} & 0 & c_{13} \\
\end{array}
\right) 
\left(
\begin{array}{ccc}
 c_{12} & s_{12} & 0 \\
 -s_{12} & c_{12} & 0 \\
 0 & 0 & 1 \\
\end{array}
\right) \cdot P\,,
\end{equation}    
where $s_{ij}\equiv\sin\theta_{ij},~c_{ij}\equiv \cos\theta_{ij}$
and $P$ is the matrix containing the Majorana phases:
\begin{equation}
    P \equiv \textrm{diag}(1,e^{- i \alpha_{21}/2},e^{-i \alpha_{31}/2})\,.
\end{equation}

Besides the six parameters in the PMNS, the neutrino sector comprises three mass parameters, namely the mass splittings $\Delta_{21}$ and $\Delta_{31}$ (with $\Delta_{ij}\equiv m_i^2 - m_j^2$) and the absolute mass $m_\textrm{min} = m_1~[\textrm{NO}],~m_3~[\textrm{IO}]$.

In \fref{allparameters}, we show the dependence of the ratio $\text{CR}(\mu \,\textrm{Al}\rightarrow e\,\,\textrm{Al})/\text{BR}(K_L\rightarrow \mu e)$ on the above-defined parameters for the normal ordering case and the oscillation parameters resulting from the fit in Refs.~\cite{Esteban:2020cvm,nufit}. Pronounced effects are observed only in the case of the Majorana phases and the absolute neutrino mass.

For completeness, we show in \fref{allparameters2} the same analysis for $\textrm{BR}(\mu\to e\gamma)/\textrm{BR}(\mu\to eee)$, which exhibits a strong sensitivity on $m_1$ but not a very prominent one on the phases.
It is interesting to notice that \fref{allparameters2} displays general results for type II seesaw, independent of our specific GUT models.
This shows that, in case of a positive signal for $\mu\to eee$ at Mu3e~\cite{Blondel:2013ia}, a future experiment able to go substantially beyond the sensitivity of MEGII~\cite{Baldini:2018nnn} on $\mu\to e\gamma$ (for ideas in this sense see Ref.~\cite{CGroup:2022tli}) would be particularly sensitive to a very light absolute mass.

%%%%%%%%%%%%%%%%%%%%%%%%%%%%%%%%%%%%%
\bibliographystyle{JHEP} 
\bibliography{gutbib.bib}% Produces the bibliography via BibTeX.

\providecommand{\href}[2]{#2}\begingroup\raggedright\begin{thebibliography}{10}

\bibitem{Georgi:1974sy}
H.~Georgi and S.~L. Glashow, \emph{{Unity of All Elementary Particle Forces}},
  \href{https://doi.org/10.1103/PhysRevLett.32.438}{\emph{Phys. Rev. Lett.}
  {\bfseries 32} (1974) 438--441}.

\bibitem{Fritzsch:1974nn}
H.~Fritzsch and P.~Minkowski, \emph{{Unified Interactions of Leptons and
  Hadrons}}, \href{https://doi.org/10.1016/0003-4916(75)90211-0}{\emph{Annals
  Phys.} {\bfseries 93} (1975) 193--266}.

\bibitem{Croon:2019kpe}
D.~Croon, T.~E. Gonzalo, L.~Graf, N.~Ko\v{s}nik and G.~White, \emph{{GUT
  Physics in the era of the LHC}},
  \href{https://doi.org/10.3389/fphy.2019.00076}{\emph{Front. in Phys.}
  {\bfseries 7} (2019) 76}, [\href{https://arxiv.org/abs/1903.04977}{{\ttfamily
  1903.04977}}].

\bibitem{abada2007low}
A.~Abada, C.~Biggio, F.~Bonnet, M.~B. Gavela and T.~Hambye, \emph{{Low energy
  effects of neutrino masses}},
  \href{https://doi.org/10.1088/1126-6708/2007/12/061}{\emph{JHEP} {\bfseries
  12} (2007) 061}, [\href{https://arxiv.org/abs/0707.4058}{{\ttfamily
  0707.4058}}].

\bibitem{Cai:2017jrq}
Y.~Cai, J.~Herrero-Garc\'\i{}a, M.~A. Schmidt, A.~Vicente and R.~R. Volkas,
  \emph{{From the trees to the forest: a review of radiative neutrino mass
  models}}, \href{https://doi.org/10.3389/fphy.2017.00063}{\emph{Front. in
  Phys.} {\bfseries 5} (2017) 63},
  [\href{https://arxiv.org/abs/1706.08524}{{\ttfamily 1706.08524}}].

\bibitem{Magg:1980ut}
M.~Magg and C.~Wetterich, \emph{{Neutrino Mass Problem and Gauge Hierarchy}},
  \href{https://doi.org/10.1016/0370-2693(80)90825-4}{\emph{Phys. Lett. B}
  {\bfseries 94} (1980) 61--64}.

\bibitem{Lazarides:1980nt}
G.~Lazarides, Q.~Shafi and C.~Wetterich, \emph{{Proton Lifetime and Fermion
  Masses in an SO(10) Model}},
  \href{https://doi.org/10.1016/0550-3213(81)90354-0}{\emph{Nucl. Phys. B}
  {\bfseries 181} (1981) 287--300}.

\bibitem{Mohapatra:1980yp}
R.~N. Mohapatra and G.~Senjanovic, \emph{{Neutrino Masses and Mixings in Gauge
  Models with Spontaneous Parity Violation}},
  \href{https://doi.org/10.1103/PhysRevD.23.165}{\emph{Phys. Rev. D} {\bfseries
  23} (1981) 165}.

\bibitem{Schechter:1980gr}
J.~Schechter and J.~W.~F. Valle, \emph{{Neutrino Masses in SU(2) x U(1)
  Theories}}, \href{https://doi.org/10.1103/PhysRevD.22.2227}{\emph{Phys. Rev.
  D} {\bfseries 22} (1980) 2227}.

\bibitem{Dorsner:2005fq}
I.~Dorsner and P.~Fileviez~Perez, \emph{{Unification without supersymmetry:
  Neutrino mass, proton decay and light leptoquarks}},
  \href{https://doi.org/10.1016/j.nuclphysb.2005.06.016}{\emph{Nucl. Phys. B}
  {\bfseries 723} (2005) 53--76},
  [\href{https://arxiv.org/abs/hep-ph/0504276}{{\ttfamily hep-ph/0504276}}].

\bibitem{Dorsner:2005ii}
I.~Dorsner, P.~Fileviez~Perez and R.~Gonzalez~Felipe, \emph{{Phenomenological
  and cosmological aspects of a minimal GUT scenario}},
  \href{https://doi.org/10.1016/j.nuclphysb.2006.05.006}{\emph{Nucl. Phys. B}
  {\bfseries 747} (2006) 312--327},
  [\href{https://arxiv.org/abs/hep-ph/0512068}{{\ttfamily hep-ph/0512068}}].

\bibitem{Dorsner:2006dj}
I.~Dorsner and P.~Fileviez~Perez, \emph{{Unification versus proton decay in
  SU(5)}}, \href{https://doi.org/10.1016/j.physletb.2006.09.034}{\emph{Phys.
  Lett. B} {\bfseries 642} (2006) 248--252},
  [\href{https://arxiv.org/abs/hep-ph/0606062}{{\ttfamily hep-ph/0606062}}].

\bibitem{Dorsner:2006hw}
I.~Dorsner, P.~Fileviez~Perez and G.~Rodrigo, \emph{{Fermion masses and the UV
  cutoff of the minimal realistic SU(5)}},
  \href{https://doi.org/10.1103/PhysRevD.75.125007}{\emph{Phys. Rev. D}
  {\bfseries 75} (2007) 125007},
  [\href{https://arxiv.org/abs/hep-ph/0607208}{{\ttfamily hep-ph/0607208}}].

\bibitem{Dorsner:2007fy}
I.~Dorsner and I.~Mocioiu, \emph{{Predictions from type II see-saw mechanism in
  SU(5)}}, \href{https://doi.org/10.1016/j.nuclphysb.2007.12.004}{\emph{Nucl.
  Phys. B} {\bfseries 796} (2008) 123--136},
  [\href{https://arxiv.org/abs/0708.3332}{{\ttfamily 0708.3332}}].

\bibitem{FileviezPerez:2008dw}
P.~Fileviez~Perez, T.~Han, T.~Li and M.~J. Ramsey-Musolf, \emph{{Leptoquarks
  and Neutrino Masses at the LHC}},
  \href{https://doi.org/10.1016/j.nuclphysb.2009.04.009}{\emph{Nucl. Phys. B}
  {\bfseries 819} (2009) 139--176},
  [\href{https://arxiv.org/abs/0810.4138}{{\ttfamily 0810.4138}}].

\bibitem{Antusch:2022afk}
S.~Antusch, K.~Hinze and S.~Saad, \emph{{Viable quark-lepton Yukawa ratios and
  nucleon decay predictions in $SU(5)$ GUTs with type-II seesaw}},
  \href{https://arxiv.org/abs/2205.01120}{{\ttfamily 2205.01120}}.

\bibitem{Bajc:2006ia}
B.~Bajc and G.~Senjanovic, \emph{{Seesaw at LHC}},
  \href{https://doi.org/10.1088/1126-6708/2007/08/014}{\emph{JHEP} {\bfseries
  08} (2007) 014}, [\href{https://arxiv.org/abs/hep-ph/0612029}{{\ttfamily
  hep-ph/0612029}}].

\bibitem{Dorsner:2006fx}
I.~Dorsner and P.~Fileviez~Perez, \emph{{Upper Bound on the Mass of the Type
  III Seesaw Triplet in an SU(5) Model}},
  \href{https://doi.org/10.1088/1126-6708/2007/06/029}{\emph{JHEP} {\bfseries
  06} (2007) 029}, [\href{https://arxiv.org/abs/hep-ph/0612216}{{\ttfamily
  hep-ph/0612216}}].

\bibitem{Bajc:2007zf}
B.~Bajc, M.~Nemevsek and G.~Senjanovic, \emph{{Probing seesaw at LHC}},
  \href{https://doi.org/10.1103/PhysRevD.76.055011}{\emph{Phys. Rev. D}
  {\bfseries 76} (2007) 055011},
  [\href{https://arxiv.org/abs/hep-ph/0703080}{{\ttfamily hep-ph/0703080}}].

\bibitem{FileviezPerez:2007bcw}
P.~Fileviez~Perez, \emph{{Renormalizable adjoint SU(5)}},
  \href{https://doi.org/10.1016/j.physletb.2007.07.075}{\emph{Phys. Lett. B}
  {\bfseries 654} (2007) 189--193},
  [\href{https://arxiv.org/abs/hep-ph/0702287}{{\ttfamily hep-ph/0702287}}].

\bibitem{Arhrib:2009mz}
A.~Arhrib, B.~Bajc, D.~K. Ghosh, T.~Han, G.-Y. Huang, I.~Puljak et~al.,
  \emph{{Collider Signatures for Heavy Lepton Triplet in Type I+III Seesaw}},
  \href{https://doi.org/10.1103/PhysRevD.82.053004}{\emph{Phys. Rev. D}
  {\bfseries 82} (2010) 053004},
  [\href{https://arxiv.org/abs/0904.2390}{{\ttfamily 0904.2390}}].

\bibitem{perez2018seesaw}
P.~Fileviez~P\'erez, A.~Gross and C.~Murgui, \emph{{Seesaw scale, unification,
  and proton decay}},
  \href{https://doi.org/10.1103/PhysRevD.98.035032}{\emph{Phys. Rev. D}
  {\bfseries 98} (2018) 035032},
  [\href{https://arxiv.org/abs/1804.07831}{{\ttfamily 1804.07831}}].

\bibitem{Olivas:2021nft}
U.~C. Olivas, K.~Kowalska and D.~Kumar, \emph{{Road map through the desert with
  scalars}},  \href{https://arxiv.org/abs/2112.11742}{{\ttfamily 2112.11742}}.

\bibitem{Calibbi:2017uvl}
L.~Calibbi and G.~Signorelli, \emph{{Charged Lepton Flavour Violation: An
  Experimental and Theoretical Introduction}},
  \href{https://doi.org/10.1393/ncr/i2018-10144-0}{\emph{Riv. Nuovo Cim.}
  {\bfseries 41} (2018) 71--174},
  [\href{https://arxiv.org/abs/1709.00294}{{\ttfamily 1709.00294}}].

\bibitem{Barrie:2021mwi}
N.~D. Barrie, C.~Han and H.~Murayama, \emph{{Affleck-Dine Leptogenesis from
  Higgs Inflation}},
  \href{https://doi.org/10.1103/PhysRevLett.128.141801}{\emph{Phys. Rev. Lett.}
  {\bfseries 128} (2022) 141801},
  [\href{https://arxiv.org/abs/2106.03381}{{\ttfamily 2106.03381}}].

\bibitem{Barrie:2022cub}
N.~D. Barrie, C.~Han and H.~Murayama, \emph{{Type II Seesaw leptogenesis}},
  \href{https://doi.org/10.1007/JHEP05(2022)160}{\emph{JHEP} {\bfseries 05}
  (2022) 160}, [\href{https://arxiv.org/abs/2204.08202}{{\ttfamily
  2204.08202}}].

\bibitem{Dorsner:2016wpm}
I.~Dor\v{s}ner, S.~Fajfer, A.~Greljo, J.~F. Kamenik and N.~Ko\v{s}nik,
  \emph{{Physics of leptoquarks in precision experiments and at particle
  colliders}}, \href{https://doi.org/10.1016/j.physrep.2016.06.001}{\emph{Phys.
  Rept.} {\bfseries 641} (2016) 1--68},
  [\href{https://arxiv.org/abs/1603.04993}{{\ttfamily 1603.04993}}].

\bibitem{Huang:2020hdv}
G.-y. Huang and S.~Zhou, \emph{{Precise Values of Running Quark and Lepton
  Masses in the Standard Model}},
  \href{https://doi.org/10.1103/PhysRevD.103.016010}{\emph{Phys. Rev. D}
  {\bfseries 103} (2021) 016010},
  [\href{https://arxiv.org/abs/2009.04851}{{\ttfamily 2009.04851}}].

\bibitem{Antusch:2013jca}
S.~Antusch and V.~Maurer, \emph{{Running quark and lepton parameters at various
  scales}}, \href{https://doi.org/10.1007/JHEP11(2013)115}{\emph{JHEP}
  {\bfseries 11} (2013) 115},
  [\href{https://arxiv.org/abs/1306.6879}{{\ttfamily 1306.6879}}].

\bibitem{Ellis:1979fg}
J.~R. Ellis and M.~K. Gaillard, \emph{{Fermion Masses and Higgs Representations
  in SU(5)}}, \href{https://doi.org/10.1016/0370-2693(79)90476-3}{\emph{Phys.
  Lett. B} {\bfseries 88} (1979) 315--319}.

\bibitem{Witten:1979nr}
E.~Witten, \emph{{Neutrino Masses in the Minimal O(10) Theory}},
  \href{https://doi.org/10.1016/0370-2693(80)90666-8}{\emph{Phys. Lett. B}
  {\bfseries 91} (1980) 81--84}.

\bibitem{Dorsner:2014wva}
I.~Dorsner, S.~Fajfer and I.~Mustac, \emph{{Light vector-like fermions in a
  minimal SU(5) setup}},
  \href{https://doi.org/10.1103/PhysRevD.89.115004}{\emph{Phys. Rev. D}
  {\bfseries 89} (2014) 115004},
  [\href{https://arxiv.org/abs/1401.6870}{{\ttfamily 1401.6870}}].

\bibitem{Babu:2012pb}
K.~S. Babu, B.~Bajc and Z.~Tavartkiladze, \emph{{Realistic Fermion Masses and
  Nucleon Decay Rates in SUSY SU(5) with Vector-Like Matter}},
  \href{https://doi.org/10.1103/PhysRevD.86.075005}{\emph{Phys. Rev. D}
  {\bfseries 86} (2012) 075005},
  [\href{https://arxiv.org/abs/1207.6388}{{\ttfamily 1207.6388}}].

\bibitem{giveon19915}
A.~Giveon, L.~J. Hall and U.~Sarid, \emph{{SU(5) unification revisited}},
  \href{https://doi.org/10.1016/0370-2693(91)91289-8}{\emph{Phys. Lett. B}
  {\bfseries 271} (1991) 138--144}.

\bibitem{Machacek:1983tz}
M.~E. Machacek and M.~T. Vaughn, \emph{{Two Loop Renormalization Group
  Equations in a General Quantum Field Theory. 1. Wave Function
  Renormalization}},
  \href{https://doi.org/10.1016/0550-3213(83)90610-7}{\emph{Nucl. Phys. B}
  {\bfseries 222} (1983) 83--103}.

\bibitem{Zyla:2020zbs}
{\scshape Particle Data Group} collaboration, P.~Zyla et~al., \emph{{Review of
  Particle Physics}}, \href{https://doi.org/10.1093/ptep/ptaa104}{\emph{PTEP}
  {\bfseries 2020} (2020) 083C01}.

\bibitem{Alciati:2005ur}
M.~L. Alciati, F.~Feruglio, Y.~Lin and A.~Varagnolo, \emph{{Proton lifetime
  from SU(5) unification in extra dimensions}},
  \href{https://doi.org/10.1088/1126-6708/2005/03/054}{\emph{JHEP} {\bfseries
  03} (2005) 054}, [\href{https://arxiv.org/abs/hep-ph/0501086}{{\ttfamily
  hep-ph/0501086}}].

\bibitem{Super-Kamiokande:2020wjk}
{\scshape Super-Kamiokande} collaboration, A.~Takenaka et~al., \emph{{Search
  for proton decay via $p\to e^+\pi^0$ and $p\to \mu^+\pi^0$ with an enlarged
  fiducial volume in Super-Kamiokande I-IV}},
  \href{https://doi.org/10.1103/PhysRevD.102.112011}{\emph{Phys. Rev. D}
  {\bfseries 102} (2020) 112011},
  [\href{https://arxiv.org/abs/2010.16098}{{\ttfamily 2010.16098}}].

\bibitem{Super-Kamiokande:2005lev}
{\scshape Super-Kamiokande} collaboration, K.~Kobayashi et~al., \emph{{Search
  for nucleon decay via modes favored by supersymmetric grand unification
  models in Super-Kamiokande-I}},
  \href{https://doi.org/10.1103/PhysRevD.72.052007}{\emph{Phys. Rev. D}
  {\bfseries 72} (2005) 052007},
  [\href{https://arxiv.org/abs/hep-ex/0502026}{{\ttfamily hep-ex/0502026}}].

\bibitem{Super-Kamiokande:2022egr}
{\scshape Super-Kamiokande} collaboration, R.~Matsumoto et~al., \emph{{Search
  for proton decay via $p\rightarrow \mu^+K^0$ in 0.37 megaton-years exposure
  of Super-Kamiokande}},  \href{https://arxiv.org/abs/2208.13188}{{\ttfamily
  2208.13188}}.

\bibitem{Super-Kamiokande:2013rwg}
{\scshape Super-Kamiokande} collaboration, K.~Abe et~al., \emph{{Search for
  Nucleon Decay via $n \to \bar{\nu} \pi^{0}$ and $p \to \bar{\nu} \pi^{+}$ in
  Super-Kamiokande}},
  \href{https://doi.org/10.1103/PhysRevLett.113.121802}{\emph{Phys. Rev. Lett.}
  {\bfseries 113} (2014) 121802},
  [\href{https://arxiv.org/abs/1305.4391}{{\ttfamily 1305.4391}}].

\bibitem{Super-Kamiokande:2014otb}
{\scshape Super-Kamiokande} collaboration, K.~Abe et~al., \emph{{Search for
  proton decay via $p\to\nu K^+$ using 260 kiloton\textperiodcentered{}year
  data of Super-Kamiokande}},
  \href{https://doi.org/10.1103/PhysRevD.90.072005}{\emph{Phys. Rev. D}
  {\bfseries 90} (2014) 072005},
  [\href{https://arxiv.org/abs/1408.1195}{{\ttfamily 1408.1195}}].

\bibitem{Nath:2006ut}
P.~Nath and P.~Fileviez~Perez, \emph{{Proton stability in grand unified
  theories, in strings and in branes}},
  \href{https://doi.org/10.1016/j.physrep.2007.02.010}{\emph{Phys. Rept.}
  {\bfseries 441} (2007) 191--317},
  [\href{https://arxiv.org/abs/hep-ph/0601023}{{\ttfamily hep-ph/0601023}}].

\bibitem{Dorsner:2004xa}
I.~Dorsner and P.~Fileviez~Perez, \emph{{How long could we live?}},
  \href{https://doi.org/10.1016/j.physletb.2005.08.039}{\emph{Phys. Lett. B}
  {\bfseries 625} (2005) 88--95},
  [\href{https://arxiv.org/abs/hep-ph/0410198}{{\ttfamily hep-ph/0410198}}].

\bibitem{CDF:2022hxs}
{\scshape CDF} collaboration, T.~Aaltonen et~al., \emph{{High-precision
  measurement of the W boson mass with the CDF II detector}},
  \href{https://doi.org/10.1126/science.abk1781}{\emph{Science} {\bfseries 376}
  (2022) 170--176}.

\bibitem{DiLuzio:2022xns}
L.~Di~Luzio, R.~Gr\"ober and P.~Paradisi, \emph{{Higgs physics confronts the
  $M_W$ anomaly}},  \href{https://arxiv.org/abs/2204.05284}{{\ttfamily
  2204.05284}}.

\bibitem{Evans:2022dgq}
J.~L. Evans, T.~T. Yanagida and N.~Yokozaki, \emph{{W boson mass anomaly and
  grand unification}},  \href{https://arxiv.org/abs/2205.03877}{{\ttfamily
  2205.03877}}.

\bibitem{Senjanovic:2022zwy}
G.~Senjanovi\'c and M.~Zantedeschi, \emph{{$SU(5)$ grand unification and
  $W$-boson mass}},  \href{https://arxiv.org/abs/2205.05022}{{\ttfamily
  2205.05022}}.

\bibitem{JUNO:2015sjr}
{\scshape JUNO} collaboration, Z.~Djurcic et~al., \emph{{JUNO Conceptual Design
  Report}},  \href{https://arxiv.org/abs/1508.07166}{{\ttfamily 1508.07166}}.

\bibitem{DUNE:2015lol}
{\scshape DUNE} collaboration, R.~Acciarri et~al., \emph{{Long-Baseline
  Neutrino Facility (LBNF) and Deep Underground Neutrino Experiment (DUNE)}:
  {Conceptual Design Report, Volume 2: The Physics Program for DUNE at LBNF}},
  \href{https://arxiv.org/abs/1512.06148}{{\ttfamily 1512.06148}}.

\bibitem{Hyper-Kamiokande:2018ofw}
{\scshape Hyper-Kamiokande} collaboration, K.~Abe et~al.,
  \emph{{Hyper-Kamiokande Design Report}},
  \href{https://arxiv.org/abs/1805.04163}{{\ttfamily 1805.04163}}.

\bibitem{Ashanujjaman:2021txz}
S.~Ashanujjaman and K.~Ghosh, \emph{{Revisiting type-II see-saw: present limits
  and future prospects at LHC}},
  \href{https://doi.org/10.1007/JHEP03(2022)195}{\emph{JHEP} {\bfseries 03}
  (2022) 195}, [\href{https://arxiv.org/abs/2108.10952}{{\ttfamily
  2108.10952}}].

\bibitem{ATLAS:2017xqs}
{\scshape ATLAS} collaboration, M.~Aaboud et~al., \emph{{Search for doubly
  charged Higgs boson production in multi-lepton final states with the ATLAS
  detector using proton\textendash{}proton collisions at $\sqrt{s}=13\,\text
  {TeV}$}}, \href{https://doi.org/10.1140/epjc/s10052-018-5661-z}{\emph{Eur.
  Phys. J. C} {\bfseries 78} (2018) 199},
  [\href{https://arxiv.org/abs/1710.09748}{{\ttfamily 1710.09748}}].

\bibitem{ATLAS:2021jol}
{\scshape ATLAS} collaboration, G.~Aad et~al., \emph{{Search for doubly and
  singly charged Higgs bosons decaying into vector bosons in multi-lepton final
  states with the ATLAS detector using proton-proton collisions at $
  \sqrt{\mathrm{s}} $ = 13 TeV}},
  \href{https://doi.org/10.1007/JHEP06(2021)146}{\emph{JHEP} {\bfseries 06}
  (2021) 146}, [\href{https://arxiv.org/abs/2101.11961}{{\ttfamily
  2101.11961}}].

\bibitem{ATLAS:2020dsk}
{\scshape ATLAS} collaboration, G.~Aad et~al., \emph{{Search for pairs of
  scalar leptoquarks decaying into quarks and electrons or muons in $ \sqrt{s}
  $ = 13 TeV $pp$ collisions with the ATLAS detector}},
  \href{https://doi.org/10.1007/JHEP10(2020)112}{\emph{JHEP} {\bfseries 10}
  (2020) 112}, [\href{https://arxiv.org/abs/2006.05872}{{\ttfamily
  2006.05872}}].

\bibitem{manohar2006flavor}
A.~V. Manohar and M.~B. Wise, \emph{{Flavor changing neutral currents, an
  extended scalar sector, and the Higgs production rate at the CERN LHC}},
  \href{https://doi.org/10.1103/PhysRevD.74.035009}{\emph{Phys. Rev. D}
  {\bfseries 74} (2006) 035009},
  [\href{https://arxiv.org/abs/hep-ph/0606172}{{\ttfamily hep-ph/0606172}}].

\bibitem{darme2018cornering}
L.~Darm\'e, B.~Fuks and M.~Goodsell, \emph{{Cornering sgluons with
  four-top-quark events}},
  \href{https://doi.org/10.1016/j.physletb.2018.08.001}{\emph{Phys. Lett. B}
  {\bfseries 784} (2018) 223--228},
  [\href{https://arxiv.org/abs/1805.10835}{{\ttfamily 1805.10835}}].

\bibitem{Miralles:2019uzg}
V.~Miralles and A.~Pich, \emph{{LHC bounds on colored scalars}},
  \href{https://doi.org/10.1103/PhysRevD.100.115042}{\emph{Phys. Rev. D}
  {\bfseries 100} (2019) 115042},
  [\href{https://arxiv.org/abs/1910.07947}{{\ttfamily 1910.07947}}].

\bibitem{Cacciapaglia:2020vyf}
G.~Cacciapaglia, A.~Deandrea, T.~Flacke and A.~M. Iyer, \emph{{Gluon-Photon
  Signatures for color octet at the LHC (and beyond)}},
  \href{https://doi.org/10.1007/JHEP05(2020)027}{\emph{JHEP} {\bfseries 05}
  (2020) 027}, [\href{https://arxiv.org/abs/2002.01474}{{\ttfamily
  2002.01474}}].

\bibitem{CMS:2018mgb}
{\scshape CMS} collaboration, A.~M. Sirunyan et~al., \emph{{Search for narrow
  and broad dijet resonances in proton-proton collisions at $ \sqrt{s}=13 $ TeV
  and constraints on dark matter mediators and other new particles}},
  \href{https://doi.org/10.1007/JHEP08(2018)130}{\emph{JHEP} {\bfseries 08}
  (2018) 130}, [\href{https://arxiv.org/abs/1806.00843}{{\ttfamily
  1806.00843}}].

\bibitem{Kanemura:2022ahw}
S.~Kanemura and K.~Yagyu, \emph{{Implication of the $W$ boson mass anomaly at
  CDF II in the Higgs triplet model with a mass difference}},
  \href{https://arxiv.org/abs/2204.07511}{{\ttfamily 2204.07511}}.

\bibitem{TheMEG:2016wtm}
{\scshape MEG} collaboration, A.~M. Baldini et~al., \emph{{Search for the
  lepton flavour violating decay $\mu ^+ \rightarrow \mathrm {e}^+ \gamma $
  with the full dataset of the MEG experiment}},
  \href{https://doi.org/10.1140/epjc/s10052-016-4271-x}{\emph{Eur. Phys. J. C}
  {\bfseries 76} (2016) 434},
  [\href{https://arxiv.org/abs/1605.05081}{{\ttfamily 1605.05081}}].

\bibitem{Baldini:2018nnn}
{\scshape MEG II} collaboration, A.~Baldini et~al., \emph{{The design of the
  MEG II experiment}},
  \href{https://doi.org/10.1140/epjc/s10052-018-5845-6}{\emph{Eur. Phys. J. C}
  {\bfseries 78} (2018) 380},
  [\href{https://arxiv.org/abs/1801.04688}{{\ttfamily 1801.04688}}].

\bibitem{Bellgardt:1987du}
{\scshape SINDRUM} collaboration, U.~Bellgardt et~al., \emph{{Search for the
  Decay $\mu^+\to e^+e^+e^-$}},
  \href{https://doi.org/10.1016/0550-3213(88)90462-2}{\emph{Nucl. Phys. B}
  {\bfseries 299} (1988) 1--6}.

\bibitem{Blondel:2013ia}
A.~Blondel et~al., \emph{{Research Proposal for an Experiment to Search for the
  Decay $\mu \to eee$}},  \href{https://arxiv.org/abs/1301.6113}{{\ttfamily
  1301.6113}}.

\bibitem{Bertl:2006up}
{\scshape SINDRUM II} collaboration, W.~H. Bertl et~al., \emph{{A Search for
  muon to electron conversion in muonic gold}},
  \href{https://doi.org/10.1140/epjc/s2006-02582-x}{\emph{Eur. Phys. J. C}
  {\bfseries 47} (2006) 337--346}.

\bibitem{Bartoszek:2014mya}
{\scshape Mu2e} collaboration, L.~Bartoszek et~al., \emph{{Mu2e Technical
  Design Report}},  \href{https://arxiv.org/abs/1501.05241}{{\ttfamily
  1501.05241}}.

\bibitem{Kuno:2013mha}
{\scshape COMET} collaboration, Y.~Kuno, \emph{{A search for muon-to-electron
  conversion at J-PARC: The COMET experiment}},
  \href{https://doi.org/10.1093/ptep/pts089}{\emph{PTEP} {\bfseries 2013}
  (2013) 022C01}.

\bibitem{BNL:1998apv}
{\scshape BNL} collaboration, D.~Ambrose et~al., \emph{{New limit on muon and
  electron lepton number violation from K0(L) ---\ensuremath{>} mu+- e-+
  decay}}, \href{https://doi.org/10.1103/PhysRevLett.81.5734}{\emph{Phys. Rev.
  Lett.} {\bfseries 81} (1998) 5734--5737},
  [\href{https://arxiv.org/abs/hep-ex/9811038}{{\ttfamily hep-ex/9811038}}].

\bibitem{Goudzovski:2022vbt}
E.~Goudzovski et~al., \emph{{New Physics Searches at Kaon and Hyperon
  Factories}},  \href{https://arxiv.org/abs/2201.07805}{{\ttfamily
  2201.07805}}.

\bibitem{KTeV:2007cvy}
{\scshape KTeV} collaboration, E.~Abouzaid et~al., \emph{{Search for lepton
  flavor violating decays of the neutral kaon}},
  \href{https://doi.org/10.1103/PhysRevLett.100.131803}{\emph{Phys. Rev. Lett.}
  {\bfseries 100} (2008) 131803},
  [\href{https://arxiv.org/abs/0711.3472}{{\ttfamily 0711.3472}}].

\bibitem{Sher:2005sp}
A.~Sher et~al., \emph{{An Improved upper limit on the decay K+
  ---\ensuremath{>} pi+ mu+ e-}},
  \href{https://doi.org/10.1103/PhysRevD.72.012005}{\emph{Phys. Rev. D}
  {\bfseries 72} (2005) 012005},
  [\href{https://arxiv.org/abs/hep-ex/0502020}{{\ttfamily hep-ex/0502020}}].

\bibitem{Appel:2000tc}
R.~Appel et~al., \emph{{Search for lepton flavor violation in K+ decays}},
  \href{https://doi.org/10.1103/PhysRevLett.85.2877}{\emph{Phys. Rev. Lett.}
  {\bfseries 85} (2000) 2877--2880},
  [\href{https://arxiv.org/abs/hep-ex/0006003}{{\ttfamily hep-ex/0006003}}].

\bibitem{Kitano:2002mt}
R.~Kitano, M.~Koike and Y.~Okada, \emph{{Detailed calculation of lepton flavor
  violating muon electron conversion rate for various nuclei}},
  \href{https://doi.org/10.1103/PhysRevD.76.059902}{\emph{Phys. Rev. D}
  {\bfseries 66} (2002) 096002},
  [\href{https://arxiv.org/abs/hep-ph/0203110}{{\ttfamily hep-ph/0203110}}].

\bibitem{Heeck:2022wer}
J.~Heeck, R.~Szafron and Y.~Uesaka, \emph{{Isotope dependence of
  muon-to-electron conversion}},
  \href{https://arxiv.org/abs/2203.00702}{{\ttfamily 2203.00702}}.

\bibitem{bevcirevic2016lepton}
D.~Be\v{c}irevi\'c, O.~Sumensari and R.~Zukanovich~Funchal, \emph{{Lepton
  flavor violation in exclusive $b\rightarrow s$ decays}},
  \href{https://doi.org/10.1140/epjc/s10052-016-3985-0}{\emph{Eur. Phys. J. C}
  {\bfseries 76} (2016) 134},
  [\href{https://arxiv.org/abs/1602.00881}{{\ttfamily 1602.00881}}].

\bibitem{Cirigliano:2011ny}
V.~Cirigliano, G.~Ecker, H.~Neufeld, A.~Pich and J.~Portoles, \emph{{Kaon
  Decays in the Standard Model}},
  \href{https://doi.org/10.1103/RevModPhys.84.399}{\emph{Rev. Mod. Phys.}
  {\bfseries 84} (2012) 399},
  [\href{https://arxiv.org/abs/1107.6001}{{\ttfamily 1107.6001}}].

\bibitem{KTeV:2004ozu}
{\scshape KTeV} collaboration, T.~Alexopoulos et~al., \emph{{Measurements of
  semileptonic K(L) decay form-factors}},
  \href{https://doi.org/10.1103/PhysRevD.70.092007}{\emph{Phys. Rev. D}
  {\bfseries 70} (2004) 092007},
  [\href{https://arxiv.org/abs/hep-ex/0406003}{{\ttfamily hep-ex/0406003}}].

\bibitem{FlavourLatticeAveragingGroup:2019iem}
{\scshape Flavour Lattice Averaging Group} collaboration, S.~Aoki et~al.,
  \emph{{FLAG Review 2019: Flavour Lattice Averaging Group (FLAG)}},
  \href{https://doi.org/10.1140/epjc/s10052-019-7354-7}{\emph{Eur. Phys. J. C}
  {\bfseries 80} (2020) 113},
  [\href{https://arxiv.org/abs/1902.08191}{{\ttfamily 1902.08191}}].

\bibitem{Esteban:2020cvm}
I.~Esteban, M.~C. Gonzalez-Garcia, M.~Maltoni, T.~Schwetz and A.~Zhou,
  \emph{{The fate of hints: updated global analysis of three-flavor neutrino
  oscillations}}, \href{https://doi.org/10.1007/JHEP09(2020)178}{\emph{JHEP}
  {\bfseries 09} (2020) 178},
  [\href{https://arxiv.org/abs/2007.14792}{{\ttfamily 2007.14792}}].

\bibitem{nufit}
I.~Esteban, M.~C. Gonzalez-Garcia, M.~Maltoni, T.~Schwetz and A.~Zhou, ``{NuFIT
  5.1 (2021)}.'' \url{http://www.nu-fit.org.}

\bibitem{Zhou:2015qua}
S.~Zhou, \emph{{Update on two-zero textures of the Majorana neutrino mass
  matrix in light of recent T2K, Super-Kamiokande and NO$\nu$A results}},
  \href{https://doi.org/10.1088/1674-1137/40/3/033102}{\emph{Chin. Phys. C}
  {\bfseries 40} (2016) 033102},
  [\href{https://arxiv.org/abs/1509.05300}{{\ttfamily 1509.05300}}].

\bibitem{Agostini:2022zub}
M.~Agostini, G.~Benato, J.~A. Detwiler, J.~Men\'endez and F.~Vissani,
  \emph{{Toward the discovery of matter creation with neutrinoless double-beta
  decay}},  \href{https://arxiv.org/abs/2202.01787}{{\ttfamily 2202.01787}}.

\bibitem{Guth:1981uk}
A.~H. Guth and E.~J. Weinberg, \emph{{Cosmological Consequences of a First
  Order Phase Transition in the SU(5) Grand Unified Model}},
  \href{https://doi.org/10.1103/PhysRevD.23.876}{\emph{Phys. Rev. D} {\bfseries
  23} (1981) 876}.

\bibitem{Kumericki:2017sfc}
K.~Kumericki, T.~Mede and I.~Picek, \emph{{Renormalizable SU(5) Completions of
  a Zee-type Neutrino Mass Model}},
  \href{https://doi.org/10.1103/PhysRevD.97.055012}{\emph{Phys. Rev. D}
  {\bfseries 97} (2018) 055012},
  [\href{https://arxiv.org/abs/1712.05246}{{\ttfamily 1712.05246}}].

\bibitem{Branco:2011iw}
G.~C. Branco, P.~M. Ferreira, L.~Lavoura, M.~N. Rebelo, M.~Sher and J.~P.
  Silva, \emph{{Theory and phenomenology of two-Higgs-doublet models}},
  \href{https://doi.org/10.1016/j.physrep.2012.02.002}{\emph{Phys. Rept.}
  {\bfseries 516} (2012) 1--102},
  [\href{https://arxiv.org/abs/1106.0034}{{\ttfamily 1106.0034}}].

\bibitem{Aoki:2017puj}
Y.~Aoki, T.~Izubuchi, E.~Shintani and A.~Soni, \emph{{Improved lattice
  computation of proton decay matrix elements}},
  \href{https://doi.org/10.1103/PhysRevD.96.014506}{\emph{Phys. Rev. D}
  {\bfseries 96} (2017) 014506},
  [\href{https://arxiv.org/abs/1705.01338}{{\ttfamily 1705.01338}}].

\bibitem{Schmidt:2007nq}
M.~A. Schmidt, \emph{{Renormalization group evolution in the type I+ II seesaw
  model}}, \href{https://doi.org/10.1103/PhysRevD.76.073010}{\emph{Phys. Rev.
  D} {\bfseries 76} (2007) 073010},
  [\href{https://arxiv.org/abs/0705.3841}{{\ttfamily 0705.3841}}].

\bibitem{Bandyopadhyay:2021kue}
P.~Bandyopadhyay, S.~Jangid and A.~Karan, \emph{{Constraining Scalar Doublet
  and Triplet Leptoquarks with Vacuum Stability and Perturbativity}},
  \href{https://arxiv.org/abs/2111.03872}{{\ttfamily 2111.03872}}.

\bibitem{CGroup:2022tli}
M.~Aoki et~al., \emph{{A New Charged Lepton Flavor Violation Program at
  Fermilab}},  in \emph{{2022 Snowmass Summer Study}}, 3, 2022,
  \href{https://arxiv.org/abs/2203.08278}{{\ttfamily 2203.08278}}.

\end{thebibliography}\endgroup
%%%%%%%%%%%%%%%%%%%%%%%%%%%%%%%%%%%%%

\end{document}